\documentclass[
 aip,
 amsmath,amssymb,
preprint,%
]{revtex4-1}

\newcommand{\Eqref}[1]{Eq.~\eqref{#1}}

\newcommand{\Figref}[1]{Fig.~\ref{#1}}
\newcommand{\Figsref}[1]{Figs.~\ref{#1}}
\newcommand{\Secref}[1]{Sec.~\ref{#1}}

\newcommand{\taud}{\ensuremath{\tau_\text{d}}}
\newcommand{\tauw}{\ensuremath{\tau_\text{w}}}
\newcommand{\Aave}{\ensuremath{\langle{A}\rangle}}
\newcommand{\Psiave}{\ensuremath{\langle{\Psi}\rangle}}
\newcommand{\Psirms}{\ensuremath{\Psi_\text{rms}}}

\usepackage{graphicx}
\usepackage{dcolumn}
\usepackage{bm}

\usepackage[utf8]{inputenc}
\usepackage[T1]{fontenc}
\usepackage{mathptmx}
\usepackage{xcolor}

\DeclareUnicodeCharacter{0308}{} 

\begin{document}

\title{Numerical turbulence simulations of intermittent fluctuations in the scrape-off layer of magnetized plasmas}

\author{G.~Decristoforo}
	\altaffiliation{Department of Physics and Technology, UiT The Arctic University of Norway, NO-9037 Troms{\o}, Norway}
\author{A.~Theodorsen}
	\altaffiliation{Department of Physics and Technology, UiT The Arctic University of Norway, NO-9037 Troms{\o}, Norway}
\author{J.~Omotani}
	\altaffiliation{United Kingdom Atomic Energy Authority, Culham Centre for Fusion Energy, Culham Science Centre, Abingdon, Oxon, OX14 3DB, UK}
\author{T.~Nicholas}
	\altaffiliation{York Plasma Institute, Department of Physics, University of York, Heslington, York YO10 5DD, UK}
\author{O.~E.~Garcia}
	\altaffiliation{Department of Physics and Technology, UiT The Arctic University of Norway, NO-9037 Troms{\o}, Norway}

\date{\today}

\begin{abstract}
Intermittent fluctuations in the boundary of magnetically confined plasmas are investigated by numerical turbulence simulations of a reduced fluid model describing the evolution of the plasma density and electric drift vorticity in the two-dimensional plane perpendicular to the magnetic field. Two different cases are considered, one describing resistive drift waves in the edge region and another including only the interchange instability due to unfavorable magnetic field curvature in the scrape-off layer. Analysis of long data time series obtained by single-point recordings are compared to predictions of a stochastic model describing the plasma fluctuations as a super-position of uncorrelated pulses. For both cases investigated, the radial particle density profile in the scrape-off layer is exponential with a radially constant scale length. The probability density function for the particle density fluctuations in the far scrape-off layer has an exponential tail. Radial motion of blob-like structures leads to large-amplitude bursts with an exponential distribution of peak amplitudes and the waiting times between them. The average burst shape is well described by a two-sided exponential function. The frequency power spectral density of the particle density is simply that of the average burst shape and is the same for all radial positions in the scrape-off layer. The fluctuation statistics obtained from the numerical simulations are in excellent agreement with recent experimental measurements on magnetically confined plasmas. The statistical framework defines a new validation metric for boundary turbulence simulations.
\end{abstract}

\maketitle

\section{Introduction}\label{intro}

At the boundary of magnetically confined plasma, turbulent transport of particles and heat in the outermost region enhances plasma interactions with the material surfaces. This can become a serious issue for future fusion experiments and reactors.\cite{pitts,lipschultz,loarte} A complete description of the physical mechanisms underlying the cross-field plasma and heat transport in the scrape-off layer (SOL) and its effects on plasma--wall interactions is necessary if reliable predictions for reactor relevant devices are to be obtained. Unfortunately, such an understanding is at present still not fully achieved and predictions and extrapolations are often based on empirical scaling laws or highly simplified transport modelling with limited theoretical foundation.\cite{loarte,wiesen,bonnin}

Fluctuations and turbulent motions in the boundary region of magnetized plasmas have been extensively investigated both experimentally and theoretically. It is recognized that in the SOL radial motion of blob-like filament structures is the dominant mechanism for cross-field transport of particles and heat.\cite{d2004blob,d2004blob,krasheninnikov2008recent,garcia2009blob,dippolito-11} This leads to broadening and flattening of radial profiles and high average particle density in the SOL that increases plasma--wall interactions.\cite{asakura,labombard2001,lipschultz2002,garcia2005tcv,rudakov2005,labombard2005,garcia2007jnm,garcia2007tcvnf,garcia2007tcvppcf,carralero2014,carralero2015,militello2016nf,vianello2017,vianello2020} Experimental measurements using Langmuir probes and gas puff imaging have revealed highly intermittent fluctuations of the particle density in the far SOL. Interestingly, measurements across a variety of magnetic geometries, including conventional tokamaks, spherical tokamaks, reversed field pinches and stellarators have shown similar fluctuation characteristics.\cite{liewer1985measurements,endler1995turbulent,carreras2005plasma,zweben2007edge} Recent statistical analysis of exceptionally long fluctuation data time series from several tokamak devices has shown that the fluctuations are well described as a super-position of uncorrelated exponential pulses with fixed duration, arriving according to a Poisson process and with exponentially distributed pulse amplitudes.\cite{garcia2013intermittent,garcia2013intermittent2,garcia2015tcv,theodorsen2016scrape,garcia2017sol,theodorsen2017cmod,walkden,walkden2017,kube2018intermittent,garcia2018intermittent,theodorsen2018universality,benze,kube2019statistical,kuang,kube2020intermittent} A statistical framework based on filtered Poisson processes has proven an accurate description of both average radial profiles and fluctuations in the boundary of magnetically confined plasma.\cite{garcia2012stochastic,kube2015pop,theodorsen2016pop,garcia2016stochastic,militello2016,mo2016,theodorsen2017ps,garcia2017pop,theodorsen2018pre,militello2018,theodorsen2018ppcf}

So far, this stochastic model has not been utilized to analyze fluctuation data from numerical turbulence simulations of the boundary region of magnetized plasmas. In order to obtain statistically significant results, long simulation data time series or a large ensemble are required, equivalent to several hundred milliseconds in experiments with medium-sized magnetically confined plasma. Since most turbulence simulation studies have been focused on the dynamics of individual blob structures or on the effects of specific physical mechanisms on turbulence and transport, the simulations have likely not produced time series data of sufficient duration in order to analyze them in the same manner as the experimental measurements.\cite{garcia2013intermittent,garcia2013intermittent2,garcia2015tcv,theodorsen2016scrape,garcia2017sol,theodorsen2017cmod,walkden,walkden2017,kube2018intermittent,garcia2018intermittent,theodorsen2018universality,benze,kube2019statistical,kuang,kube2020intermittent} In this paper we present the first results from applying the same statistical framework on numerical simulation data as has recently been done on experimental measurements. By using a simplified turbulence model describing the fluctuations in the two-dimensional plane perpendicular to the magnetic field, we have obtained data time series sufficiently long to allow unambiguous identification of the fluctuation statistics. The main goal of this study is to clarify these statistical properties and compare them with that found from experimental measurements. This is considered an essential step towards validation of turbulence simulation codes.\cite{terry,greenwald,holland}

A recent analysis of fluctuation data time series obtained from numerical simulations of turbulent Rayleigh–B{\'e}nard-convection in two dimensions has given some illuminating results.\cite{decristoforo2020intermittent} This model has frequently been used as a simplified description of the non-linear interchange dynamics in the SOL of magnetically confined plasmas.\cite{pogutse,sugama,beyer,horton,berning,garcia2003rb,garciabian2003,garcia2003ps,wilczynski} In Ref.~\onlinecite{decristoforo2020intermittent} it was found that the fluctuation time series are well described as a super-position of Lorentzian pulses, resulting in an exponential frequency power spectral density. In the present study, more sophisticated models for SOL turbulence are investigated, including sheath dissipation due to losses along magnetic field lines intersecting material surfaces as well as drift wave dynamics in the edge region.\cite{benkadda,garcia-2001,sarazin1998theoretical,ghendrih2003,sarazin2003theoretical,garcia-esel-prl,garcia-esel-php,garcia-esel-ps,myra2008,russell,myra2013,bisai2004,bisai2005edge,bisai2005edge2,nielsen2015,nielsen2017,olsen2018} The resulting far SOL data time series are shown to be dominated by large-amplitude bursts with a two-sided exponential pulse shape and fluctuation statistics that compare favorably with those found in experimental measurements.\cite{garcia2013intermittent,garcia2013intermittent2,garcia2015tcv,theodorsen2016scrape,garcia2017sol,theodorsen2017cmod,walkden,walkden2017,kube2018intermittent,garcia2018intermittent,theodorsen2018universality,benze,kube2019statistical,kuang,kube2020intermittent}

In this contribution we present a detailed statistical analysis of fluctuation data time series from numerical simulations of a two-dimensional reduced fluid model describing the evolution of the electron density and electric drift vorticity. The paper is structured as follows. The reduced fluid model equations, normalization and parameters are discussed in \Secref{sec:model}. A brief introduction to the stochastic model is also presented here. We present the results for the time-averaged profiles and probability densities in \Secref{sec:av_prof} and for the fluctuation statistics in \Secref{sec:fluc_stat}. A discussion of the results and the conclusions are finally presented in \Secref{sec:concl}.

\section{Model equations}\label{sec:model}

The reduced fluid model investigated here is motivated by previous simulation studies performed by Sarazin \textit{et al.},\cite{sarazin1998theoretical,ghendrih2003,sarazin2003theoretical} Garcia \textit{et al.}\cite{garcia-esel-prl,garcia-esel-php,garcia-esel-ps}, Myra \textit{et al.}\cite{myra2008,russell,myra2013}, Bisai \textit{et al.}\cite{bisai2004,bisai2005edge,bisai2005edge2} and Nielsen \textit{et al}.\cite{nielsen2015,nielsen2017,olsen2018} One particular case of the model is equivalent to that used in Ref.~\onlinecite{sarazin2003theoretical} and simulates SOL conditions in the entire simulation domain where a particle source is located close to the inner boundary. The particle density profile results from a balance between the plasma source, the sheath dissipation and radial transport due to the interchange instability. Another case of the model is similar to that used in Ref.~\onlinecite{bisai2005edge} and features a simulation domain separating an edge region corresponding to plasma dynamics on closed magnetic flux surfaces and a SOL region where sheath dissipation balances the interchange drive. The source term is located in the plasma edge region where parallel resistivity gives rise to unstable drift waves. Despite these two fundamentally different descriptions of the primary instability mechanism underlying the SOL turbulence, the resulting fluctuations are remarkably similar as will be shown in the following.

We use two-field fluid model equations describing the plasma evolution in the edge and SOL regions for a quasi-neutral plasma, neglecting electron inertia and assuming for simplicity isothermal electrons and negligibly small ion temperature. We make these simplifying assumptions in order to obtain long fluctuation data time series from the numerical simulations. We choose a slab geometry where $x$ refers to the radial direction and $y$ to the binormal or poloidal direction. The reduced electron continuity and electron drift vorticity equations are given by
\begin{subequations} \label{modeleqs}
\begin{gather}
\frac{\text{d} n}{\text{d} t} + g \left(\frac{\partial n}{\partial y} -n\frac{\partial \phi}{\partial y} \right) = S_n + D_\perp \nabla_\perp^2 n + \Bigg \langle \frac{1}{L_\parallel} \nabla_\parallel J_{\parallel\text{e}} \Bigg \rangle_\parallel ,  \label{3D_n}
\\
\frac{\text{d} \nabla_\perp^2 \phi}{\text{d} t} + \frac{g}{n}\frac{\partial n}{\partial y} = \nu_\perp \nabla_\perp^4 \phi + \Bigg \langle \frac{1}{n L_\parallel} \nabla_\parallel J_{\parallel} \Bigg \rangle_\parallel ,  \label{3D_vort}
\end{gather}
\end{subequations}
where $n$ represents the normalized electron density, $\phi$ is the normalized electric potential, $g$ is normalized effective gravity (that is, drive from unfavorable magnetic curvature), $S_n$ is the plasma source term, and $D_\perp$ and $\nu_\perp$ are the normalized particle and vorticity diffusion coefficients. We use the standard Bohm normalization as previously used and discussed in Refs.~\onlinecite{garcia-2001,benkadda,sarazin1998theoretical,ghendrih2003,sarazin2003theoretical,garcia-esel-prl,garcia-esel-php,garcia-esel-ps,myra2008,russell,myra2013,bisai2004,bisai2004,bisai2005edge,bisai2005edge2}. In addition we have the advective derivative $\text{d}/\text{d}t = \partial / \partial t + \mathbf{V}_\text{E}\cdot \nabla_\perp$, where $\mathbf{V}_\text{E} = \hat{\mathbf{z}}\times\nabla\phi$ is the electric drift. The plasma source term is given by $S_n(x)=S_0 \,\text{exp}(-(x - x_0)^2/\lambda_\text{s}^2)$, where $S_0$ is the maximum amplitude of the source, $x_0$ is the source location and $\lambda_\text{s}$ is the $e$-folding length for the source.

\begin{figure}[t]
	\centering
	\includegraphics[width=8cm]{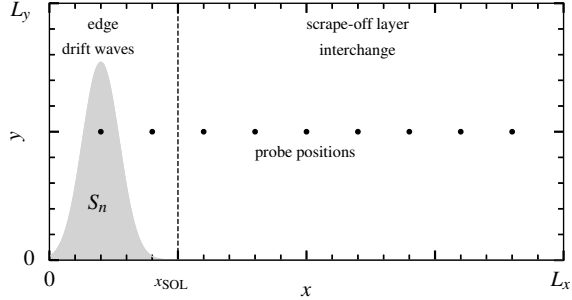}
	\caption{Schematic illustration of the simulation domain for the $x_\text{SOL}=50$ case. The position of the plasma source term (gray shaded) and the border between edge and SOL (dashed vertical line) are indicated.}
	\label{domain}
\end{figure}

Equations \eqref{modeleqs} are averaged along the magnetic field lines, with the contribution from the normalized parallel electron $J_{\parallel\text{e}}$ and total plasma currents $J_\parallel$ in the sheath connected regime given by 
\begin{subequations}
\begin{gather}
\Bigg \langle \frac{1}{L_\parallel} \nabla_\parallel J_{\parallel\text{e}} \Bigg \rangle_\parallel = - \sigma n\,\exp(\Lambda-\phi) + \chi( \widehat{\phi} - \widehat{n} ) ,
\\
\Bigg \langle \frac{1}{n L_\parallel} \nabla_\parallel J_{\parallel} \Bigg \rangle_\parallel =  \sigma \left[ 1-\exp(\Lambda-\phi) \right] + \chi( \widehat{\phi} - \widehat{n} ) .
\end{gather}
\end{subequations}
Here $\Lambda$ is the sheath potential, $\sigma$ the normalized sheath dissipation and $\chi$ the normalized parallel plasma conductivity. Like in several previous investigations, these parameters are taken to be a function of the radial position in the boundary region.\cite{garcia-esel-prl,garcia-esel-php,garcia-esel-ps,myra2008,russell,myra2013,bisai2004,bisai2005edge,bisai2005edge2,nielsen2015,nielsen2017,olsen2018} In particular, the sheath dissipation coefficient $\sigma$ is finite in the SOL region ($x>x_\text{SOL}$) and vanishes in the edge ($x<x_\text{SOL}$), which corresponds to the region with closed magnetic flux surfaces,
\begin{equation}
\sigma(x) = 
\begin{cases}
0 , & 0 \leq x < x_\text{SOL} ,
\\
\sigma_0 , & x_\text{SOL} \leq x \leq L_x .
\end{cases}
\end{equation}
Similarly, the plasma conductivity $\chi$ is neglected in the SOL and is finite in the edge region,
\begin{equation}
\chi(x) = 
\begin{cases}
\chi_0 , & 0 \leq x < x_\text{SOL} ,
\\
0 , & x_\text{SOL} \leq x \leq L_x .
\end{cases}
\end{equation}
The simulation domain is sketched in \Figref{domain}, showing the location of the plasma source and the separation between the edge and SOL regions. Furthermore, the spatially fluctuating electron density $\widehat{n}$ and plasma potential $\widehat{\phi}$ are defined as $\widehat{n}=n-\langle{n}\rangle_y$ and $\widehat{\phi}=\phi-\langle{\phi}\rangle_y$ where $\langle{\cdot}\rangle_y$ refers to the flux surface average. This leads to the final reduced electron continuity and electric drift vorticity equations,
\begin{subequations}
\begin{gather}
\frac{\text{d} n}{\text{d} t} + g \left(\frac{\partial n}{\partial y} - n\frac{\partial \phi}{\partial y} \right) = S_n(x) + D_\perp \nabla_\perp^2 n - \sigma(x)\,n\,\text{exp}(\Lambda-\phi) + \chi(x)( \widehat{\phi} - \widehat{n} ) ,
\\
\frac{\text{d} \nabla_\perp^2 \phi}{\text{d} t} + \frac{g}{n}\frac{\partial n}{\partial y} = \nu_\perp \nabla_\perp^4 \phi + \sigma(x) \left[1-\text{exp}(\Lambda-\phi)\right] + \chi(x)( \widehat{\phi} - \widehat{n} ) .
\end{gather}
\end{subequations}
In the following we present results from numerical simulations of this model for two different cases. In the first case, the domain is split into two regions, effectively the edge and the SOL regions, by taking $x_\text{SOL}=50$. In the second case, a pure SOL plasma is considered with $x_\text{SOL}=0$, thus plasma conductivity $\chi$ is not present in the simulation domain.

The input parameters have been chosen to be similar to that used in previous publications based on this model.\cite{bisai2005edge} For all runs presented here, the simulation domain lengths are chosen to be $L_x=200$ and $L_y=100$, with the border between the edge and the SOL at $x_\text{SOL}=50$ for the two-region case. It has been verified that a change of the size of the simulation domain does not influence the fluctuation statistics. The simulation code is implemented in BOUT++\cite{dudson2009bout++} utilizing the STORM branch,\cite{militello2017interaction} which uses a finite difference scheme in the $x$-direction and a spectral scheme in the $y$-direction. Time integration is performed with  the PVODE solver.\cite{byrne1999pvode} We use a resolution of $512\times256$ grid points for all runs. We further take $D_\perp=\nu_\perp=10^{-2}$, $g=10^{-3}$,  $\chi=6\times10^{-4}$, $S_0=11/2000$, $\sigma_0=5\times10^{-4}$, $\Lambda=0.5\text{ln}(2\pi m_\text{i}/m_\text{e})$ with deuterium ions, $x_0=20$ and $\lambda_\text{s}=10$. We apply periodic boundary conditions in the poloidal direction and zero gradient boundary conditions in the radial direction for both the electron density and vorticity fields. For the plasma potential we use zero gradient boundary conditions at the outer boundary and fixed boundary conditions $\phi(x=0)=0$ at the inner boundary.

\begin{figure}[t]
	\centering
	\includegraphics[width=12cm]{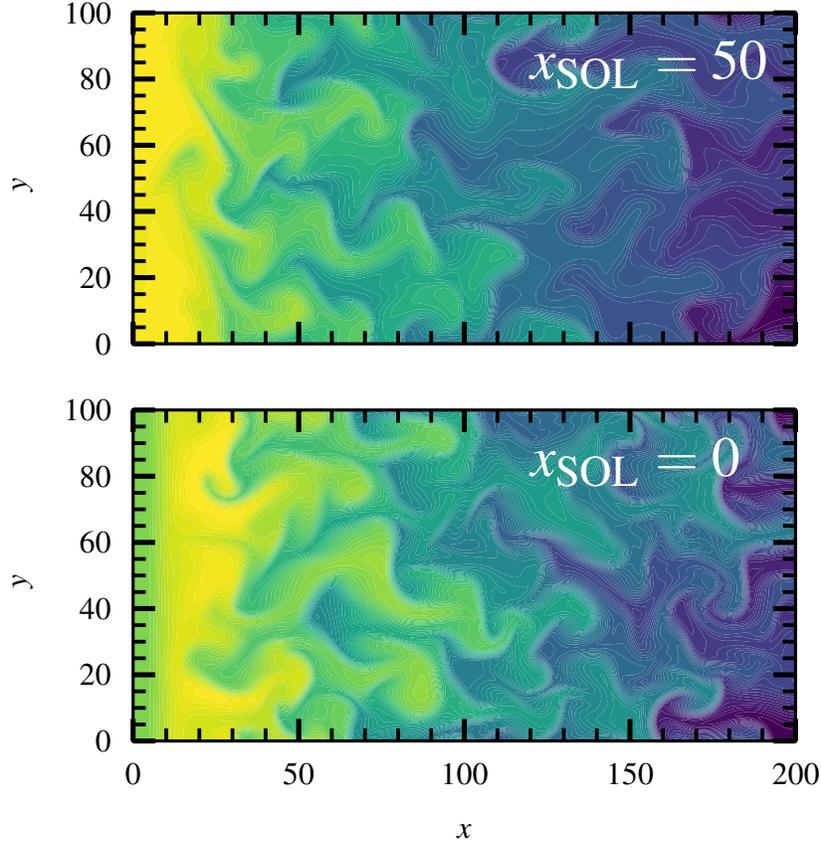}
	\caption{Contour plots of $\mathrm{log}(n)$ in the turbulent state for the $x_{\text{SOL}}=0$ and $x_{\text{SOL}}=50$ cases showing the presence of mushroom-shaped blob-like structures in the SOL.}
	\label{langmuir_positions}
\end{figure}

During the simulations, the plasma parameters at 9 different radial positions in the simulation domain are recorded with a sampling frequency of one in normalized time units. The location of these probes are presented in \Figref{domain}. This corresponds to single-point measurements in the experiments, and the simulation data will be analyzed in the same manner as has previously been done for experimental measurement data. The contour plots of the electron density in both simulation cases presented in \Figref{langmuir_positions} show several blob-like structures with the familiar mushroom-shape typical for strongly non-linear interchange motions.\cite{garcia2003ps}

\begin{figure}[t]
	\centering
	\includegraphics[width=8cm]{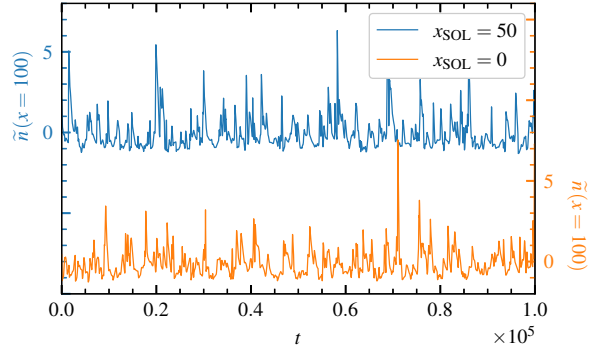}
	\caption{A short part of the normalized electron density time series recorded at $x=100$ for the $x_{\text{SOL}}=0$ and $x_{\text{SOL}}=50$ simulation cases.}
	\label{time_series}
\end{figure}

Time series of the plasma parameters with a duration of $2\times10^6$ time units have been obtained under statistically stationary conditions, that is, excluding initial transients in the turbulence simulations. 10 simulation runs with this duration time are performed for the two-region model and 7 for the one region model. The fluctuation statistics to be presented in \Secref{sec:fluc_stat} are based on these ensembles of simulation data. In the following analysis we will frequently consider plasma parameters normalized such as to have vanishing mean and unit standard deviation, for example. For the electron density we define
\begin{equation}
\widetilde{n} = \frac{n - \langle n \rangle }{n_\text{rms}} ,
\end{equation} 
where the angular brackets denote a time average and $n_\text{rms}$ is the root mean square value calculated from the time series. A short part of the normalized electron density time series are presented in \Figref{time_series} for both simulation cases, showing frequent appearance of large-amplitude bursts due to the high density blob-like structures moving radially outwards. The radial variation of the lowest order moments of these fluctuations are presented and discussed in \Secref{sec:av_prof}.

In the following, the numerical simulation data will be compared to predictions of a stochastic model which describes the fluctuations as a super-position of uncorrelated pulses with fixed shape and constant duration. This is written as\cite{garcia2012stochastic,kube2015pop,theodorsen2016pop,garcia2016stochastic,militello2016,mo2016,theodorsen2017ps,garcia2017pop,theodorsen2018pre,militello2018,theodorsen2018ppcf,garcia2017pop}
\begin{equation}
\Psi_K(t) = \sum_{k=1}^{K(T)} A_k \psi\left( \frac{t-t_k}{\taud} \right) ,
\end{equation}
where $\psi$ is the pulse function, $\taud$ is the pulse duration time, $K(T)$ is the number of pulses for a realization of duration $T$, and for the event labelled $k$ the pulse amplitude is $A_k$ and the arrival time $t_k$. The mean value of the random variable $\Psi_K$ is $\Psiave=(\taud/\tauw)\Aave$, where $\Aave$ is the average pulse amplitude and $\tauw$ is the average pulse waiting time. We will assume pulses arriving according to a Poisson process, which implies independent and exponentially distributed waiting times and independent arrival times uniformly distributed on the realization. We further assume independently and exponentially distributed amplitudes, $P_A(A)=\exp(-A/\Aave)/\Aave$, and
we will consider the case of a two-sided exponential pulse function,\cite{garcia2017pop}
\begin{equation} \label{twosidedexp}
\psi(\theta;\lambda) = \begin{cases}
\exp( \theta/\lambda ) , & \theta < 0 ,
\\
\exp( - \theta/(1-\lambda) ) , & \theta \geq 0 ,
\end{cases}
\end{equation}
where the pulse asymmetry parameter $\lambda$ is restricted to the range $0<\lambda<1$. For $\lambda<1/2$, the pulse rise time is faster that than the decay time, while the pulse shape is symmetric in the case $\lambda=1/2$. The frequency power spectral density for this process is just the spectrum of the pulse function,\cite{garcia2017pop}
\begin{equation} \label{lspectrum}
\Omega_{\widetilde{\Psi}}(\omega) = \frac{2\taud}{[1+(1-\lambda)^2(\taud\omega)^2][1+\lambda^2(\taud\omega)^2]} ,
\end{equation}
where $\omega$ is the angular frequency. Note that the power spectral density of $\widetilde{\Psi}$ is independent of the amplitude distribution. From this it follows that the frequency power spectral density can be used to estimate the pulse parameters $\taud$ and $\lambda$, which will be done in the following analysis of the numerical simulations.

The stationary probability density function (PDF) for the random variable $\Psi_K$ can be shown to be a Gamma distribution,\cite{theodorsen2018ppcf}
\begin{equation}
\Psiave P_{\Psi}(\Psi) = \frac{\gamma}{\Gamma(\gamma)} \left( \frac{\gamma\Psi}{\Psiave} \right)^{\gamma-1} \exp\left( \frac{\gamma\Psi}{\Psiave} \right) ,
\end{equation}
with shape parameter $\gamma=\taud/\tauw$, that is, the ratio of the pulse duration and the average pulse waiting time $\tauw$. This parameter describes the degree of pulse overlap, which determines the level of intermittency in the process. From the Gamma distribution it follows that the skewness moment is $S_\Psi=\langle(\Psi-\Psiave)^3\rangle/\Psirms^3=2/\gamma^{1/2}$ and the flatness moment is $F_\Psi=\langle(\Psi-\Psiave)^4\rangle/\Psirms^4=3+6/\gamma$. Accordingly, there is a parabolic relationship between these moments given by $F_\Psi=3+3S_\Psi^2/2$. For strong pulse overlap and large $\gamma$, the probability density function approaches a normal distribution and the skewness $S_\Psi$ and excess flatness $F_\Psi-3$ moments vanish.

\section{Profiles and distributions}\label{sec:av_prof}

\begin{figure}[t]
	\centering
	\includegraphics[width=8cm]{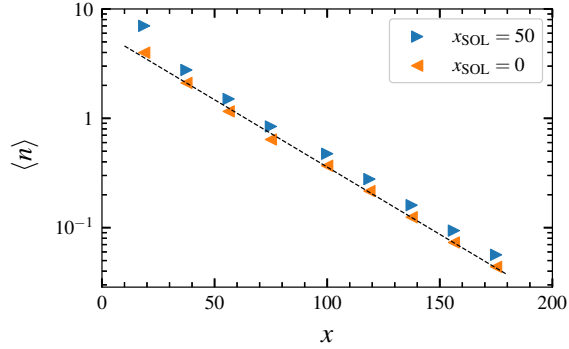}
	\caption{Time-averaged electron density profile for the $x_{\text{SOL}}=0$ and $x_{\text{SOL}}=50$ cases. The broken line is the best fit of an exponential function with a scale length of 35.5.}
	\label{profiles}
\end{figure}

\begin{figure}[t]
	\centering
	\includegraphics[width=8cm]{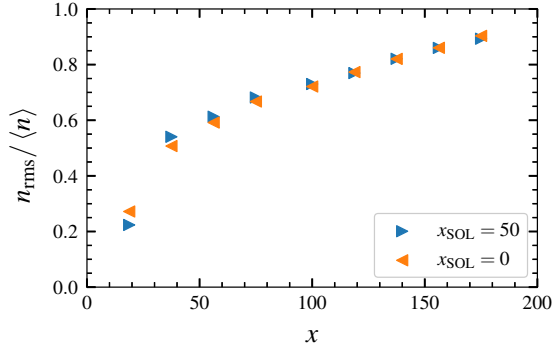}
	\caption{The relative fluctuation level of the electron density at different positions in the SOL for the $x_{\text{SOL}}=0$ and $x_{\text{SOL}}=50$ cases.}
	\label{fluc}
\end{figure}

The time-averaged electron density profiles in the turbulence simulations are presented in \Figref{profiles}. Since the $x_{\text{SOL}}=50$ case does not include any sheath dissipation in the edge region, the average density is higher here than for the $x_{\text{SOL}}=0$ case. Throughout the entire SOL region, we observe that the electron density decreases exponentially with a radially constant scale length of 35.5. This is to be compared with the equilibrium SOL profile scale length in the absence of turbulence given by $(D_\perp/\sigma_0)=\sqrt{20}$ for the simulation parameters used here. Interestingly, both the scale length and the absolute density are very similar for the two simulation cases investigated. We further show the relative fluctuation level at different radial positions for both cases in \Figref{fluc}. The normalized fluctuation level is very high, increases radially outwards and is roughly similar for the two simulation cases.

\begin{figure}[t]
	\centering
	\includegraphics[width=8cm]{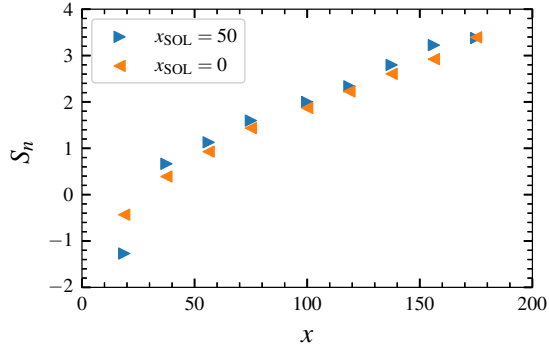}
	\caption{Skewness of the electron density fluctuations at different radial positions for the $x_{\text{SOL}}=0$ and $x_{\text{SOL}}=50$ cases.}
	\label{skewness}
\end{figure}

\begin{figure}[t]
	\centering
	\includegraphics[width=8cm]{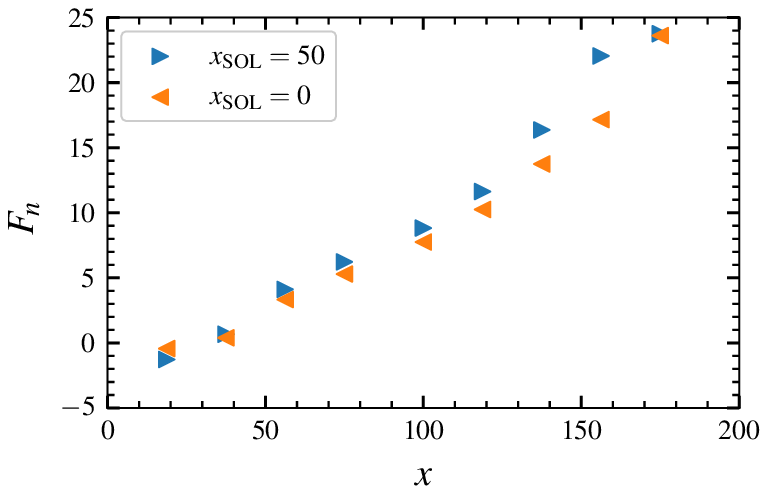}
	\caption{Flatness of the electron density fluctuations at different radial positions for the $x_{\text{SOL}}=0$ and $x_{\text{SOL}}=50$ cases.}
	\label{kurtosis}
\end{figure}

\begin{figure}[t]
	\centering
	\includegraphics[width=8cm]{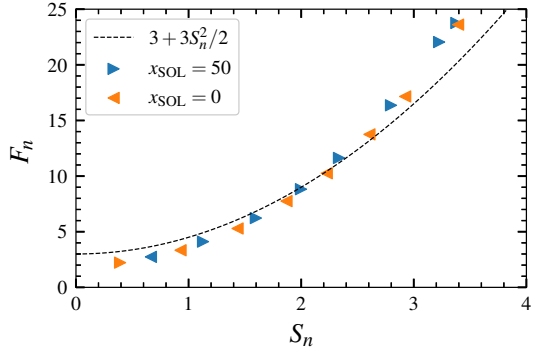}
	\caption{Flatness plotted versus skewness for the electron density fluctuations in the SOL. The broken line shows the parabolic relationship $F_n=3+3S_n^2/2$ for comparison.}
	\label{K_vs_S}
\end{figure}

The radial variation of the skewness and flatness moments of the electron density fluctuations are presented in \Figsref{skewness} and \ref{kurtosis}, respectively. From these figures it is clear that the intermittency of the fluctuations increases radially outwards in the SOL, qualitatively similar for the $x_\text{SOL}=0$ and $x_\text{SOL}=50$ cases. By plotting the flatness moment versus the skewness, presented in \Figref{K_vs_S}, it is seen that for both simulation cases there is a nearly parabolic relationship between these higher order moments. Such a parabolic relationship is predicted by the stochastic model describing the fluctuations as a super-position of uncorrelated pulses,\cite{garcia2012stochastic,kube2015pop,theodorsen2016pop,garcia2016stochastic} which can be related to blob-like structures moving radially outwards in the SOL as seen in \Figref{langmuir_positions}.

\begin{figure}[t]
	\centering
	\includegraphics[width=8cm]{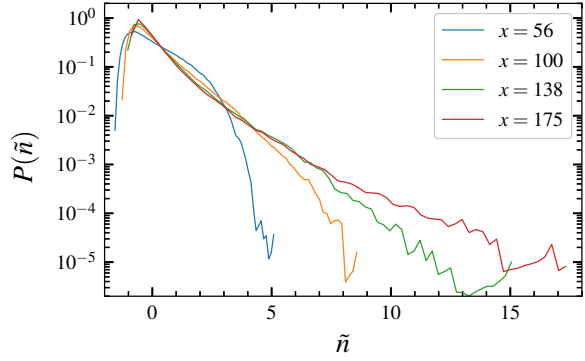}
	\caption{Probability density functions of the normalized electron density recorded at different radial positions for the $x_\text{SOL}=0$ case.}
	\label{PDF_sarazin}
\end{figure}

\begin{figure}[t]
	\centering
	\includegraphics[width=8cm]{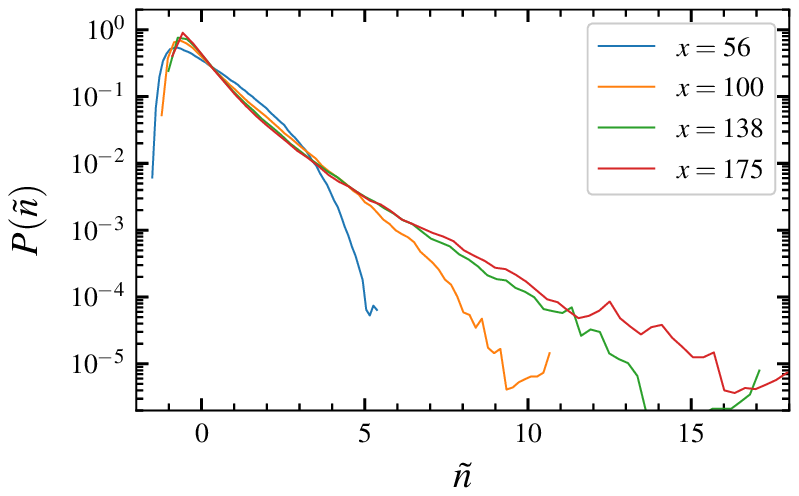}
	\caption{Probability density functions of the normalized electron density recorded at different radial positions for the $x_\text{SOL}=50$ case.}
	\label{PDF_bisai}
\end{figure}

\begin{figure}[t]
	\centering
	\includegraphics[width=8cm]{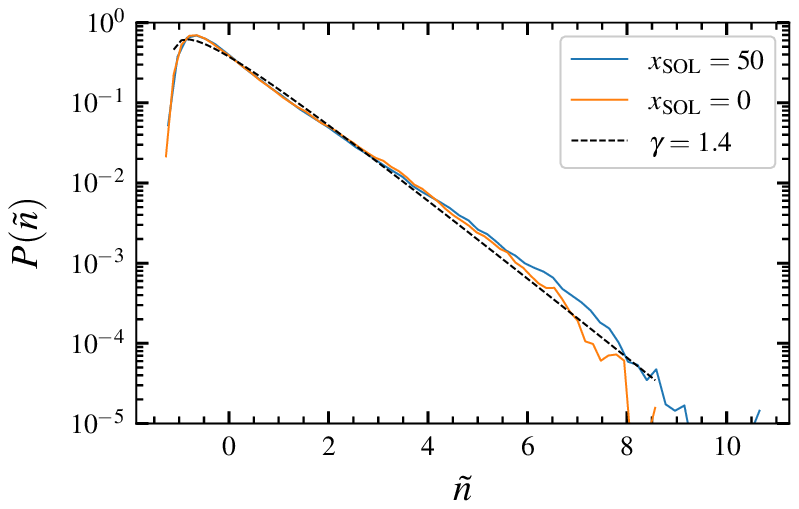}
	\caption{Probability density functions of the normalized electron density recorded at $x=100$ for both simulations cases compared to a Gamma distribution with shape parameter $\gamma=1.4$ shown with the dashed black line.}
	\label{PDF}
\end{figure}

The PDFs for the normalized electron density fluctuations at different radial positions are presented in \Figsref{PDF_sarazin} and \ref{PDF_bisai} for the $x_\text{SOL}=0$ and $x_\text{SOL}=50$ cases, respectively. The PDFs change from a narrow and nearly symmetric distribution in the edge/near SOL region to a distribution with an exponential tail for large fluctuation amplitudes in the far SOL. In \Figref{PDF} we further compare the PDFs of the electron density time series recorded in the far SOL at $x=100$ for both simulation cases with a Gamma distribution with a shape parameter of $1.4$. Such a Gamma distribution is predicted by the stochastic model describing the fluctuations as a super-position of uncorrelated exponential pulses. The Gamma distribution is clearly an excellent description of the PDF for the electron density fluctuations in the simulations. A similar change in the shape of the PDF radially outwards in the SOL has also been reported from previous turbulence simulations.\cite{garcia-esel-prl,garcia-esel-php,garcia-esel-ps}

\section{Fluctuation statistics}\label{sec:fluc_stat}

\begin{figure}[t]
	\centering
	\includegraphics[width=8cm]{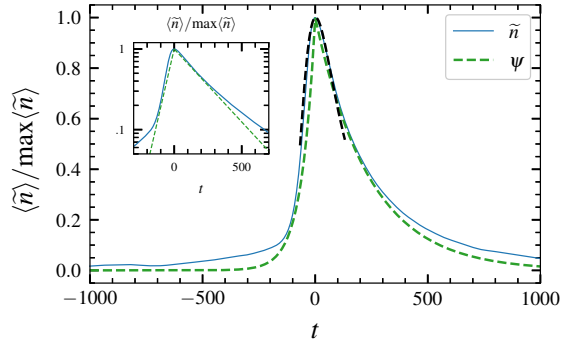}
	\caption{Conditionally averaged burst shape at $x=100$ of the $x_\text{SOL}=50$ case (full blue line) compared to a two-sided exponential pulse (dashed orange line), as well as a skewed Lorentzian pulse for short time lags (dashed black line). The conditional average is normalized by its peak amplitude.}
	\label{CA}
\end{figure}

In this section we present a detailed analysis of the electron density fluctuations recorded at $x=100$. In order to reveal the typical shape of large-amplitude bursts in the time series, a conditional averaging method which allows for overlapping events is applied. This identifies a total of 3128 conditional events with peak amplitudes larger than 2.5 times the root mean square value above the mean for the $x_\text{SOL}=50$ case and 1701 conditional events for the $x_\text{SOL}=0$ case. The average burst structures are presented in \Figref{CA} and shows an asymmetric shape with a fast rise and a slower decay. The is compared to an asymmetric, two-sided exponential function given by \Eqref{twosidedexp} with duration time $\taud=300$ and asymmetry parameter $\lambda=0.2$. The conditional burst shape is shown with semi-logarithmic axes in the inset in \Figref{CA}, showing that the decay of the conditional pulse shape is approximately exponential. However, the two-sided exponential function obviously fails to describe the smooth peak of the average burst shape in the simulations. As shown for short time lags in \Figref{CA}, this is better described using a skewed Lorentzian pulse as a fit function with duration $80$ and skewness parameter $1$ for the $x_\text{SOL}=50$ case.\cite{maggs2011,garcia2017l,garcia2018l,garcia2018l2} The slightly elevated tails of the conditional burst shape is likely due to finite pulse overlap in the turbulence simulations.

\begin{figure}[t]
	\centering
	\includegraphics[width=8cm]{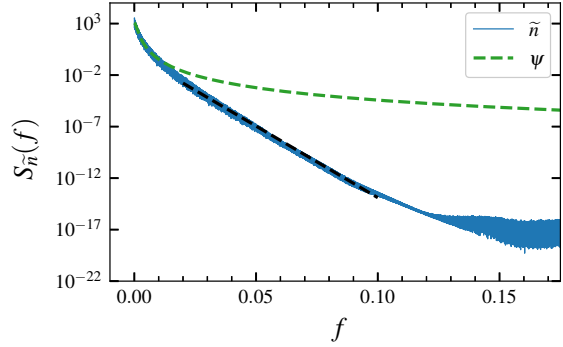}
	\caption{Frequency power spectral density for the electron density fluctuations recorded at $x=100$ for the $x_\text{SOL}=50$ case (full line). This is compared to the predictions of a stochastic model describing the fluctuations as a super-position of uncorrelated, two-sided exponential pulses (dashed orange line), as well as an exponential function for the high frequency part (dashed black line).}
	\label{PSD}
\end{figure}

The frequency power spectral density of the electron density fluctuations recorded at $x=100$ is presented with semi-logarithmic axes in \Figref{PSD} for the $x_\text{SOL}=50$ case. This shows an exponential decrease of power with frequency for high frequencies. This exponential fall off is attributed to the smooth peak of the large-amplitude bursts in the simulations. In agreement with the fit of a Lorentzian function to the peak of the conditionally averaged burst shape in \Figref{CA}, the power spectral density decreases exponentially with a characteristic scale given by the duration of the Lorentzian-shaped peak.\cite{garcia2018l} The flattening of the power spectral density at low powers and high frequencies is due to  the noise floor implied by round off errors in the computations.\cite{theodorsen2017ps}

\begin{figure}[t]
	\centering
	\includegraphics[width=8cm]{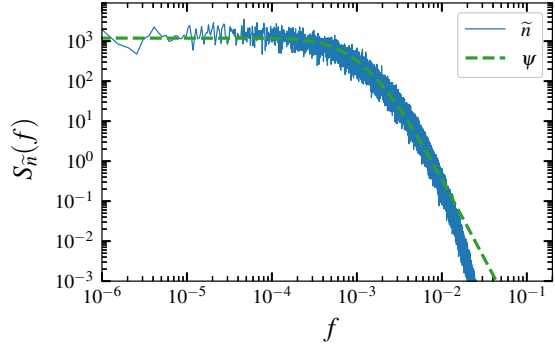}
	\caption{Frequency power spectral density of the electron density fluctuations recorded at $x=100$ for the $x_\text{SOL}=50$ case (full blue line). This is compared to the predictions of a stochastic model describing the fluctuations as a super-position of uncorrelated, two-sided exponential pulses with duration time $\tau_\text{d}=300$ and asymmetry parameter $\lambda=0.2$ (dashed orange line).}
	\label{PSD_noise}
\end{figure}

The frequency power spectral density due to a super-position of uncorrelated exponential pulses is clearly not a good description of the simulation data for high frequencies. However, presenting the power spectrum with double-logarithmic axes shows that the spectrum given by \Eqref{lspectrum} gives excellent agreement for high powers and low frequencies. This is clearly shown in \Figref{PSD_noise} for the case $x_\text{SOL}=50$. The exponential decay of the power at high frequencies is clearly due to the smooth peak of the large-amplitude bursts in the time series. This is consistent with the conditionally averaged burst shape presented in \Figref{CA}. Similar results for conditional averaging and frequency power spectra are found for the case $x_\text{SOL}=0$ but with slightly different pulse parameters.

\begin{figure}[t]
	\centering
	\includegraphics[width=8cm]{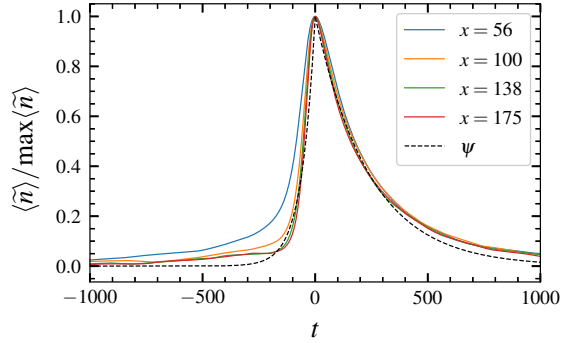}
	\caption{Conditionally averaged burst shapes at different radial positions for the $x_\text{SOL}=50$ case. The conditional averages are normalized by their peak amplitude.}
	\label{CA_radial_pos}
\end{figure}

\begin{figure}[t]
	\centering
	\includegraphics[width=8cm]{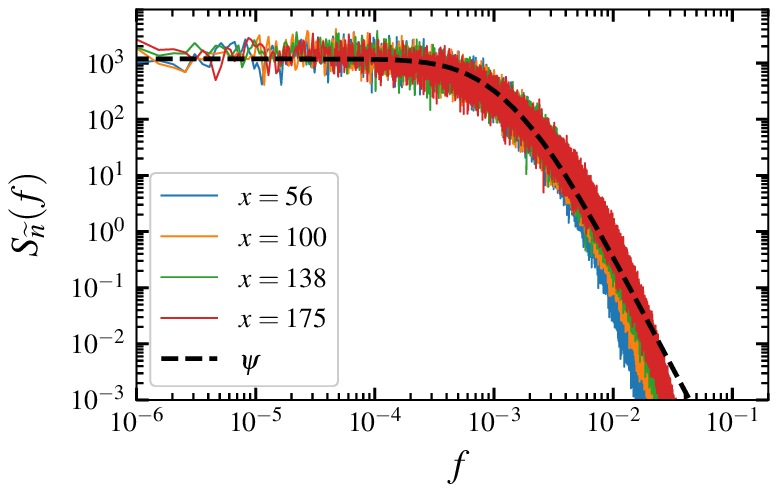}
	\caption{Frequency power spectral densities of the electron density fluctuation recorded at different radial positions for the $x_\text{SOL}=50$ model. The dashed line shows the spectrum due to a super-position of uncorrelated, two-sided exponential pulses with duration time $\taud$.}
	\label{PSD_radial_pos}
\end{figure}

\begin{figure}[t]
	\centering
	\includegraphics[width=8cm]{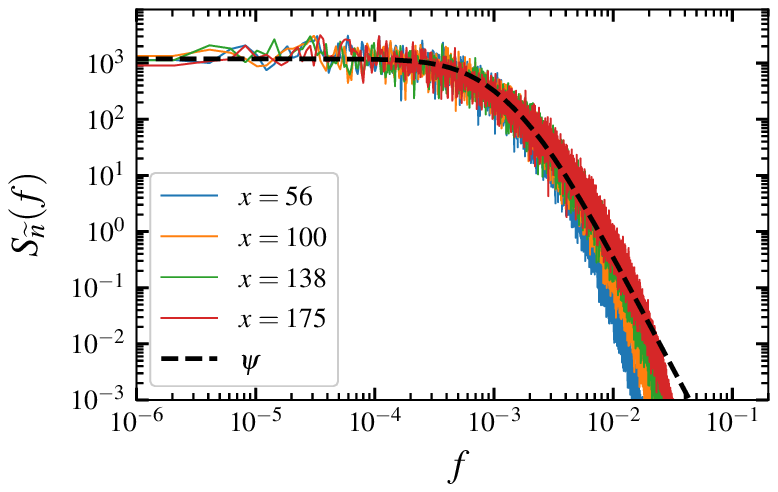}
	\caption{Frequency power spectral densities of the electron density fluctuation recorded at different radial positions for the $x_\text{SOL}=0$ model.  The dashed line shows the spectrum due to a super-position of uncorrelated two-sided exponential pulses.}
	\label{PSD_radial_pos_sarazin}
\end{figure}

The conditionally averaged burst shape is presented in \Figref{CA_radial_pos} for different radial positions in the SOL for the $x_\text{SOL}=50$ case. Here it is seen that the burst shape in the far SOL region is the same for all radial positions, despite the fact that the relative fluctuation amplitude increases radially outwards. Accordingly, as predicted by the stochastic model, the frequency power spectral density has the same shape for all these different radial positions, as is shown in \Figsref{PSD_radial_pos} and \ref{PSD_radial_pos_sarazin} for both the one- and two-region cases. The spectra are well described by that of a two-sided exponential pulse function, shown by the dashed black line in the figures.

\begin{figure}[t]
	\centering
	\includegraphics[width=8cm]{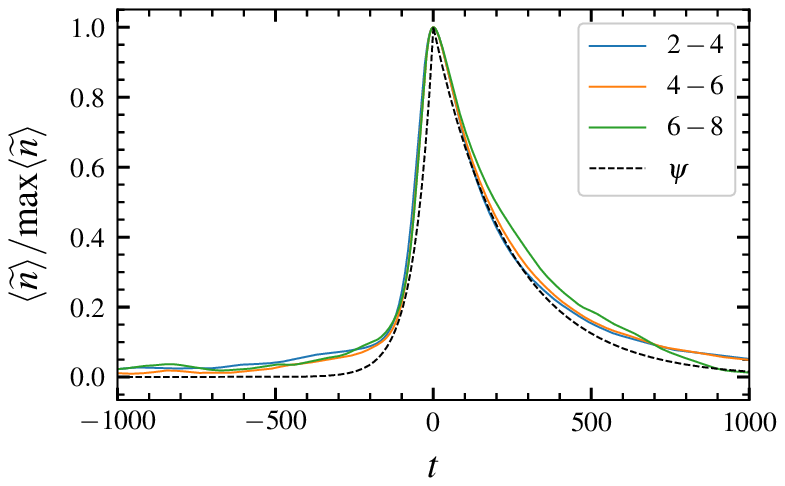}
	\caption{Conditionally averaged burst shape at $x=100$ of the $x_\text{SOL}=50$ case for different conditional amplitude threshold intervals. The conditional averages are normalized by their peak amplitudes.}
	\label{CA_different_levels}
\end{figure}

Restricting the peak amplitude of conditional events in the electron density to be within a range of 2--4, 4--6 and 6--8 times the rms value, the appropriately scaled conditional burst shapes are presented in \Figref{CA_different_levels}. This reveals that the average burst shape and duration do not depend on the burst amplitude and is again well described by a two-sided exponential function except for the smooth peak. This supports the assumption of fixed pulse duration in the stochastic model describing the fluctuations as a super-positions of pulses.

\begin{figure}[t]
	\centering
	\includegraphics[width=8cm]{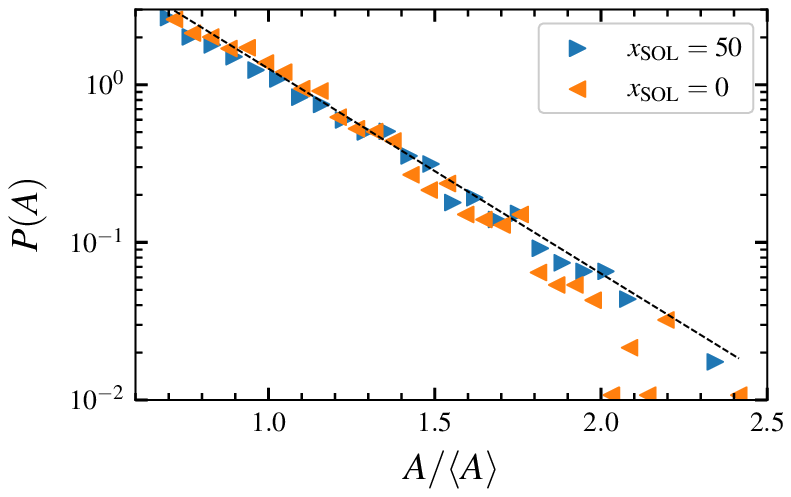}
	\caption{Probability density functions of conditional burst amplitudes of the electron density time series recorded at $x=100$.}
	\label{PDF_peak}
\end{figure}

\begin{figure}[t]
	\centering
	\includegraphics[width=8cm]{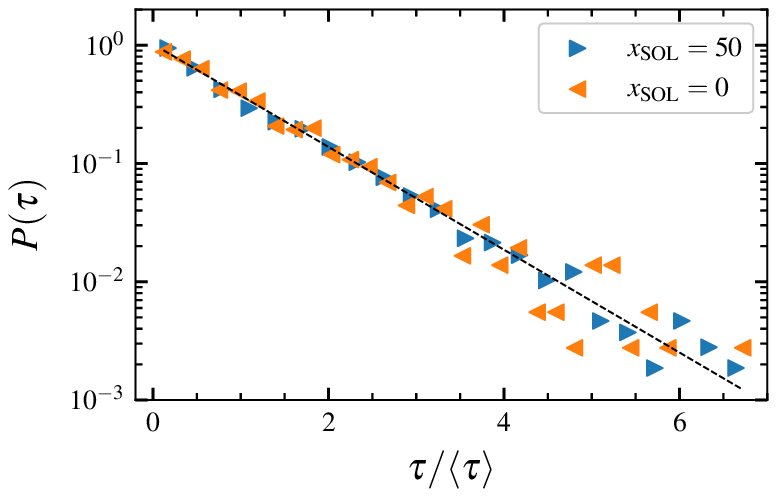}
	\caption{Probability density functions of waiting times between consecutive large-amplitude burst in the electron density time series recorded at $x=100$.}
	\label{PDF_wait}
\end{figure}

From the conditional averaging we further obtain the peak amplitudes of conditional events and the waiting times between them. The PDFs of these are presented in \Figsref{PDF_peak} and \ref{PDF_wait}, respectively. The distributions are similar for both simulation cases and are clearly well described by an exponential distribution as shown by the dashed black line in the plots. This is in agreement with the assumptions for the stochastic model presented in \Secref{sec:model}. In particular, the exponential waiting time distribution supports the hypothesis that the events are uncorrelated and arrive according to a Poisson process.

\section{Discussion and conclusions}\label{sec:concl}

The abundant experimental evidence for universal statistical properties of fluctuations in the SOL of magnetically confined fusion plasmas sets high requirements for validation of turbulence simulation codes for the boundary region.\cite{terry,greenwald,holland}  In this context, we have examined the statistical properties of the electron density fluctuations in the SOL by numerical simulations of plasma turbulence in the two-dimensional plane perpendicular to the magnetic field. Two model cases have been considered, one describing resistive drift waves in the edge region and another including only the interchange instability due to unfavorable magnetic field curvature. For both cases, mushroom-shaped blob-like structures move radially outwards, resulting in large-amplitude fluctuations and high average particle densities in the SOL. The numerical simulations show that the time-averaged radial profile decreases exponentially with radial distance into the SOL with the same characteristic length scale for both simulation cases. Moreover, the fluctuation statistics in the SOL are the same for both cases. This is despite the different linear instability mechanisms driving the fluctuations in the edge/near SOL region in the two simulation cases. It appears that any drift-ordered instability mechanism will lead to formation of filament structures when coupled to a SOL region with unfavorable magnetic field curvature.

According to a stochastic model describing the profile as due to radial motion of filament structures, the profile scale length is given by the product of the radial filament velocity and the parallel transit time.\cite{garcia2016stochastic,militello2016,mo2016} This suggests that typical filament velocities are the same in both simulation cases. Future work will investigate the distribution of filament sizes and velocities by analysis of the velocity fluctuations and applying a blob tracking algorithm as described in Ref.~\onlinecite{decristoforo2020blob}.

The relative fluctuation level increases radially outwards, nearly reaching unity in the far SOL for the plasma parameters investigated here. Similarly, the skewness and flatness moments also increase into the SOL, and these higher order moments closely follow a quadratic dependence as predicted by the stochastic model describing the fluctuations as a super-position of uncorrelated pulses. The PDF of the electron density fluctuations changes from a nearly Gaussian distribution in the edge/near SOL region to a distribution with an exponential tail for large amplitudes in the far SOL. In the far SOL region, the PDFs are well described by a Gamma distribution with the shape parameter given by the ratio of the pulse duration and average waiting time. The increase of this intermittency parameter with radial distance into the SOL suggests that only the most coherent and large-amplitude blob structures are able to move through the entire SOL region before they disperse and break up due to secondary instabilities.

A conditional averaging analysis has revealed that the shape of large-amplitude bursts in single-point recordings in the far SOL is well described by a two-sided exponential pulse, as has previously been found in experimental measurements.
However, the high resolution and smoothness of the solution from the numerical computations implies that the burst structure has a rounded peak as opposed the to break point in experimental measurements due to their much lower sampling rate and additional measurement noise. The smooth peak is well described by a skewed Lorentzian pulse function. This is further supported by the frequency power spectral density, which is well described by that of a two-sided exponential pulse for high powers and low frequencies. However, for low powers and high frequencies, the frequency power spectral density has an exponential decay which obviously can be attributed to the smooth, Lorentzian shaped peak of large-amplitude fluctuations in the numerical simulations. In experimental measurements, this exponential tail in the spectrum may readily be masked by low sampling rates, limiting the highest frequencies resolved, or by additive measurement noise, limiting the lowest power resolved.\cite{theodorsen2017ps,garcia2017pop}

In summary, it is here demonstrated that a simple but self-consistent model for turbulent fluctuations in the scrape-off layer reveals the same statistical properties of large-amplitude events as found in the experiments. This includes exponentially distributed pulse amplitudes and waiting times, the latter supporting the assumption of Poisson events.\cite{garcia2017sol,theodorsen2017cmod,kube2018intermittent,garcia2018intermittent,kube2019statistical,kube2020intermittent} The simulation data also agree with predictions of the stochastic model, namely an exponential average profile, Gamma distributed fluctuation amplitudes and a frequency power spectral density determined by the average shape of large-amplitude bursts. It is concluded that the filtered Poisson process, describing the fluctuations in single-point recordings as a super-position of uncorrelated pulses with fixed duration, is an excellent description of the SOL plasma fluctuations in the turbulence simulations investigated here. 

The simple turbulence model used in this study does not include finite ion temperature effects, X-point physics, parallel collisional conductivity in the scrape-off layer, or any effect of interactions with neutral particles. Numerous SOL turbulence models and codes are now being extended to include these features.\cite{halpern,tamain,dudson,held,wiesenberger,stegmeir,zhang,rms,giacomin} The statistical framework with super-position of filaments can be used for analysis and interpretation of simulation results in these more advanced models, similar to what has been done here and previously for experimental measurements. As such, this work sets a new standard for validation of turbulence simulation codes.\cite{terry,greenwald,holland}

\section*{Acknowledgements}

This work was supported by the UiT Aurora Centre Program, UiT The Arctic University of Norway (2020). AT was supported by a Troms{\o} Science Foundation Starting Grant under grant number 19$\_$SG$\_$AT. GD acknowledges the generous hospitality of the Culham Centre for Fusion Energy (CCFE) where parts of this work were conducted. Two supercomputers provided by the Norwegian Metacenter for Computational Science (NOTUR) were used for the computational work, the Fram and Stallo clusters at the University of Troms{\o} under project nn9348k. The MARCONI supercomputer was used for parts of the computational work under the Project No.\ FUA34$\_$SOLBOUT4. The storm2d project has been used as a template for the presented physical models. This work was funded in part by the RCUK Energy Programme (Grant number EP/T012250/1) and the EPSRC Centre for Doctoral Training in the Science and Technology of Fusion Energy (Grant number EP/L01663X/1), as well as an iCASE award from CCFE.

\section*{Data Availability}

The data that support the findings of this study are available from the corresponding author upon reasonable request.

\nocite{*}
\bibliography{sources}

\begin{thebibliography}{100}%
\makeatletter
\providecommand \@ifxundefined [1]{%
 \@ifx{#1\undefined}
}%
\providecommand \@ifnum [1]{%
 \ifnum #1\expandafter \@firstoftwo
 \else \expandafter \@secondoftwo
 \fi
}%
\providecommand \@ifx [1]{%
 \ifx #1\expandafter \@firstoftwo
 \else \expandafter \@secondoftwo
 \fi
}%
\providecommand \natexlab [1]{#1}%
\providecommand \enquote  [1]{``#1''}%
\providecommand \bibnamefont  [1]{#1}%
\providecommand \bibfnamefont [1]{#1}%
\providecommand \citenamefont [1]{#1}%
\providecommand \href@noop [0]{\@secondoftwo}%
\providecommand \href [0]{\begingroup \@sanitize@url \@href}%
\providecommand \@href[1]{\@@startlink{#1}\@@href}%
\providecommand \@@href[1]{\endgroup#1\@@endlink}%
\providecommand \@sanitize@url [0]{\catcode `\\12\catcode `\$12\catcode
  `\&12\catcode `\#12\catcode `\^12\catcode `\_12\catcode `\%12\relax}%
\providecommand \@@startlink[1]{}%
\providecommand \@@endlink[0]{}%
\providecommand \url  [0]{\begingroup\@sanitize@url \@url }%
\providecommand \@url [1]{\endgroup\@href {#1}{\urlprefix }}%
\providecommand \urlprefix  [0]{URL }%
\providecommand \Eprint [0]{\href }%
\providecommand \doibase [0]{http://dx.doi.org/}%
\providecommand \selectlanguage [0]{\@gobble}%
\providecommand \bibinfo  [0]{\@secondoftwo}%
\providecommand \bibfield  [0]{\@secondoftwo}%
\providecommand \translation [1]{[#1]}%
\providecommand \BibitemOpen [0]{}%
\providecommand \bibitemStop [0]{}%
\providecommand \bibitemNoStop [0]{.\EOS\space}%
\providecommand \EOS [0]{\spacefactor3000\relax}%
\providecommand \BibitemShut  [1]{\csname bibitem#1\endcsname}%
\let\auto@bib@innerbib\@empty
\bibitem [{\citenamefont {Pitts}\ \emph {et~al.}(2005)\citenamefont {Pitts},
  \citenamefont {Coad}, \citenamefont {Coster}, \citenamefont {Federici},
  \citenamefont {Fundamenski}, \citenamefont {Horacek}, \citenamefont
  {Krieger}, \citenamefont {Kukushkin}, \citenamefont {Likonen}, \citenamefont
  {Matthews}, \citenamefont {Rubel}, \citenamefont {Strachan},\ and\
  \citenamefont {{JET-EFDA contributors}}}]{pitts}%
  \BibitemOpen
  \bibfield  {author} {\bibinfo {author} {\bibfnamefont {R.~A.}\ \bibnamefont
  {Pitts}}, \bibinfo {author} {\bibfnamefont {J.~P.}\ \bibnamefont {Coad}},
  \bibinfo {author} {\bibfnamefont {D.~P.}\ \bibnamefont {Coster}}, \bibinfo
  {author} {\bibfnamefont {G.}~\bibnamefont {Federici}}, \bibinfo {author}
  {\bibfnamefont {W.}~\bibnamefont {Fundamenski}}, \bibinfo {author}
  {\bibfnamefont {J.}~\bibnamefont {Horacek}}, \bibinfo {author} {\bibfnamefont
  {K.}~\bibnamefont {Krieger}}, \bibinfo {author} {\bibfnamefont
  {A.}~\bibnamefont {Kukushkin}}, \bibinfo {author} {\bibfnamefont
  {J.}~\bibnamefont {Likonen}}, \bibinfo {author} {\bibfnamefont {G.~F.}\
  \bibnamefont {Matthews}}, \bibinfo {author} {\bibfnamefont {M.}~\bibnamefont
  {Rubel}}, \bibinfo {author} {\bibfnamefont {J.~D.}\ \bibnamefont {Strachan}},
  \ and\ \bibinfo {author} {\bibnamefont {{JET-EFDA contributors}}},\
  }\bibfield  {title} {\enquote {\bibinfo {title} {Material erosion and
  migration in tokamaks},}\ }\href@noop {} {\bibfield  {journal} {\bibinfo
  {journal} {Plasma Physics and Controlled Fusion}\ }\textbf {\bibinfo {volume}
  {47}},\ \bibinfo {pages} {B303} (\bibinfo {year} {2005})}\BibitemShut
  {NoStop}%
\bibitem [{\citenamefont {Lipschultz}\ \emph {et~al.}(2007)\citenamefont
  {Lipschultz}, \citenamefont {Bonnin}, \citenamefont {Counsell}, \citenamefont
  {Kallenbach}, \citenamefont {Kukushkin}, \citenamefont {Krieger},
  \citenamefont {Leonard}, \citenamefont {Loarte}, \citenamefont {Neu},
  \citenamefont {Pitts}, \citenamefont {Rognlien}, \citenamefont {Roth},
  \citenamefont {Skinner}, \citenamefont {Terry}, \citenamefont {Tsitrone},
  \citenamefont {Whyte}, \citenamefont {Zweben}, \citenamefont {Asakura},
  \citenamefont {Coster}, \citenamefont {Doerner}, \citenamefont {Dux},
  \citenamefont {Federici}, \citenamefont {Fenstermacher}, \citenamefont
  {Fundamenski}, \citenamefont {Ghendrih}, \citenamefont {Herrmann},
  \citenamefont {Hu}, \citenamefont {Krasheninnikov}, \citenamefont {Kirnev},
  \citenamefont {Kreter}, \citenamefont {Kurnaev}, \citenamefont {La{B}ombard},
  \citenamefont {Lisgo}, \citenamefont {Nakano}, \citenamefont {Ohno},
  \citenamefont {Pacher}, \citenamefont {Paley}, \citenamefont {Pan},
  \citenamefont {Pautasso}, \citenamefont {Philipps}, \citenamefont {Rohde},
  \citenamefont {Rudakov}, \citenamefont {Stangeby}, \citenamefont {Takamura},
  \citenamefont {Tanabe}, \citenamefont {Yang},\ and\ \citenamefont
  {Zhu}}]{lipschultz}%
  \BibitemOpen
  \bibfield  {author} {\bibinfo {author} {\bibfnamefont {B.}~\bibnamefont
  {Lipschultz}}, \bibinfo {author} {\bibfnamefont {X.}~\bibnamefont {Bonnin}},
  \bibinfo {author} {\bibfnamefont {G.}~\bibnamefont {Counsell}}, \bibinfo
  {author} {\bibfnamefont {A.}~\bibnamefont {Kallenbach}}, \bibinfo {author}
  {\bibfnamefont {A.}~\bibnamefont {Kukushkin}}, \bibinfo {author}
  {\bibfnamefont {K.}~\bibnamefont {Krieger}}, \bibinfo {author} {\bibfnamefont
  {A.}~\bibnamefont {Leonard}}, \bibinfo {author} {\bibfnamefont
  {A.}~\bibnamefont {Loarte}}, \bibinfo {author} {\bibfnamefont
  {R.}~\bibnamefont {Neu}}, \bibinfo {author} {\bibfnamefont {R.}~\bibnamefont
  {Pitts}}, \bibinfo {author} {\bibfnamefont {T.}~\bibnamefont {Rognlien}},
  \bibinfo {author} {\bibfnamefont {J.}~\bibnamefont {Roth}}, \bibinfo {author}
  {\bibfnamefont {C.}~\bibnamefont {Skinner}}, \bibinfo {author} {\bibfnamefont
  {J.}~\bibnamefont {Terry}}, \bibinfo {author} {\bibfnamefont
  {E.}~\bibnamefont {Tsitrone}}, \bibinfo {author} {\bibfnamefont
  {D.}~\bibnamefont {Whyte}}, \bibinfo {author} {\bibfnamefont
  {S.}~\bibnamefont {Zweben}}, \bibinfo {author} {\bibfnamefont
  {N.}~\bibnamefont {Asakura}}, \bibinfo {author} {\bibfnamefont
  {D.}~\bibnamefont {Coster}}, \bibinfo {author} {\bibfnamefont
  {R.}~\bibnamefont {Doerner}}, \bibinfo {author} {\bibfnamefont
  {R.}~\bibnamefont {Dux}}, \bibinfo {author} {\bibfnamefont {G.}~\bibnamefont
  {Federici}}, \bibinfo {author} {\bibfnamefont {M.}~\bibnamefont
  {Fenstermacher}}, \bibinfo {author} {\bibfnamefont {W.}~\bibnamefont
  {Fundamenski}}, \bibinfo {author} {\bibfnamefont {P.}~\bibnamefont
  {Ghendrih}}, \bibinfo {author} {\bibfnamefont {A.}~\bibnamefont {Herrmann}},
  \bibinfo {author} {\bibfnamefont {J.}~\bibnamefont {Hu}}, \bibinfo {author}
  {\bibfnamefont {S.}~\bibnamefont {Krasheninnikov}}, \bibinfo {author}
  {\bibfnamefont {G.}~\bibnamefont {Kirnev}}, \bibinfo {author} {\bibfnamefont
  {A.}~\bibnamefont {Kreter}}, \bibinfo {author} {\bibfnamefont
  {V.}~\bibnamefont {Kurnaev}}, \bibinfo {author} {\bibfnamefont
  {B.}~\bibnamefont {La{B}ombard}}, \bibinfo {author} {\bibfnamefont
  {S.}~\bibnamefont {Lisgo}}, \bibinfo {author} {\bibfnamefont
  {T.}~\bibnamefont {Nakano}}, \bibinfo {author} {\bibfnamefont
  {N.}~\bibnamefont {Ohno}}, \bibinfo {author} {\bibfnamefont {H.}~\bibnamefont
  {Pacher}}, \bibinfo {author} {\bibfnamefont {J.}~\bibnamefont {Paley}},
  \bibinfo {author} {\bibfnamefont {Y.}~\bibnamefont {Pan}}, \bibinfo {author}
  {\bibfnamefont {G.}~\bibnamefont {Pautasso}}, \bibinfo {author}
  {\bibfnamefont {V.}~\bibnamefont {Philipps}}, \bibinfo {author}
  {\bibfnamefont {V.}~\bibnamefont {Rohde}}, \bibinfo {author} {\bibfnamefont
  {D.}~\bibnamefont {Rudakov}}, \bibinfo {author} {\bibfnamefont
  {P.}~\bibnamefont {Stangeby}}, \bibinfo {author} {\bibfnamefont
  {S.}~\bibnamefont {Takamura}}, \bibinfo {author} {\bibfnamefont
  {T.}~\bibnamefont {Tanabe}}, \bibinfo {author} {\bibfnamefont
  {Y.}~\bibnamefont {Yang}}, \ and\ \bibinfo {author} {\bibfnamefont
  {S.}~\bibnamefont {Zhu}},\ }\bibfield  {title} {\enquote {\bibinfo {title}
  {Plasma–surface interaction, scrape-off layer and divertor physics:
  implications for {ITER}},}\ }\href@noop {} {\bibfield  {journal} {\bibinfo
  {journal} {Nuclear Fusion}\ }\textbf {\bibinfo {volume} {47}},\ \bibinfo
  {pages} {1189} (\bibinfo {year} {2007})}\BibitemShut {NoStop}%
\bibitem [{\citenamefont {Loarte}\ \emph {et~al.}(2007)\citenamefont {Loarte},
  \citenamefont {Lipschultz}, \citenamefont {Kukushkin}, \citenamefont
  {Matthews}, \citenamefont {Stangeby}, \citenamefont {Asakura}, \citenamefont
  {Counsell}, \citenamefont {Federici}, \citenamefont {Kallenbach},
  \citenamefont {Krieger}, \citenamefont {Mahdavi}, \citenamefont {Philipps},
  \citenamefont {Reiter}, \citenamefont {Roth}, \citenamefont {Strachan},
  \citenamefont {Whyte}, \citenamefont {Doerner}, \citenamefont {Eich},
  \citenamefont {Fundamenski}, \citenamefont {Herrmann}, \citenamefont
  {Fenstermacher}, \citenamefont {Ghendrih}, \citenamefont {Groth},
  \citenamefont {Kirschner}, \citenamefont {Konoshima}, \citenamefont
  {La{B}ombard}, \citenamefont {Lang}, \citenamefont {Leonard}, \citenamefont
  {Monier-Garbet}, \citenamefont {Neu}, \citenamefont {Pacher}, \citenamefont
  {Pegourie}, \citenamefont {Pitts}, \citenamefont {Takamura}, \citenamefont
  {Terry}, \citenamefont {Tsitrone},\ and\ \citenamefont {{the {ITPA}
  {S}crape-off {L}ayer and {D}ivertor {P}hysics {T}opical {G}roup}}}]{loarte}%
  \BibitemOpen
  \bibfield  {author} {\bibinfo {author} {\bibfnamefont {A.}~\bibnamefont
  {Loarte}}, \bibinfo {author} {\bibfnamefont {B.}~\bibnamefont {Lipschultz}},
  \bibinfo {author} {\bibfnamefont {A.}~\bibnamefont {Kukushkin}}, \bibinfo
  {author} {\bibfnamefont {G.}~\bibnamefont {Matthews}}, \bibinfo {author}
  {\bibfnamefont {P.}~\bibnamefont {Stangeby}}, \bibinfo {author}
  {\bibfnamefont {N.}~\bibnamefont {Asakura}}, \bibinfo {author} {\bibfnamefont
  {G.}~\bibnamefont {Counsell}}, \bibinfo {author} {\bibfnamefont
  {G.}~\bibnamefont {Federici}}, \bibinfo {author} {\bibfnamefont
  {A.}~\bibnamefont {Kallenbach}}, \bibinfo {author} {\bibfnamefont
  {K.}~\bibnamefont {Krieger}}, \bibinfo {author} {\bibfnamefont
  {A.}~\bibnamefont {Mahdavi}}, \bibinfo {author} {\bibfnamefont
  {V.}~\bibnamefont {Philipps}}, \bibinfo {author} {\bibfnamefont
  {D.}~\bibnamefont {Reiter}}, \bibinfo {author} {\bibfnamefont
  {J.}~\bibnamefont {Roth}}, \bibinfo {author} {\bibfnamefont {J.}~\bibnamefont
  {Strachan}}, \bibinfo {author} {\bibfnamefont {D.}~\bibnamefont {Whyte}},
  \bibinfo {author} {\bibfnamefont {R.}~\bibnamefont {Doerner}}, \bibinfo
  {author} {\bibfnamefont {T.}~\bibnamefont {Eich}}, \bibinfo {author}
  {\bibfnamefont {W.}~\bibnamefont {Fundamenski}}, \bibinfo {author}
  {\bibfnamefont {A.}~\bibnamefont {Herrmann}}, \bibinfo {author}
  {\bibfnamefont {M.}~\bibnamefont {Fenstermacher}}, \bibinfo {author}
  {\bibfnamefont {P.}~\bibnamefont {Ghendrih}}, \bibinfo {author}
  {\bibfnamefont {M.}~\bibnamefont {Groth}}, \bibinfo {author} {\bibfnamefont
  {A.}~\bibnamefont {Kirschner}}, \bibinfo {author} {\bibfnamefont
  {S.}~\bibnamefont {Konoshima}}, \bibinfo {author} {\bibfnamefont
  {B.}~\bibnamefont {La{B}ombard}}, \bibinfo {author} {\bibfnamefont
  {P.}~\bibnamefont {Lang}}, \bibinfo {author} {\bibfnamefont {A.}~\bibnamefont
  {Leonard}}, \bibinfo {author} {\bibfnamefont {P.}~\bibnamefont
  {Monier-Garbet}}, \bibinfo {author} {\bibfnamefont {R.}~\bibnamefont {Neu}},
  \bibinfo {author} {\bibfnamefont {H.}~\bibnamefont {Pacher}}, \bibinfo
  {author} {\bibfnamefont {B.}~\bibnamefont {Pegourie}}, \bibinfo {author}
  {\bibfnamefont {R.}~\bibnamefont {Pitts}}, \bibinfo {author} {\bibfnamefont
  {S.}~\bibnamefont {Takamura}}, \bibinfo {author} {\bibfnamefont
  {J.}~\bibnamefont {Terry}}, \bibinfo {author} {\bibfnamefont
  {E.}~\bibnamefont {Tsitrone}}, \ and\ \bibinfo {author} {\bibnamefont {{the
  {ITPA} {S}crape-off {L}ayer and {D}ivertor {P}hysics {T}opical {G}roup}}},\
  }\bibfield  {title} {\enquote {\bibinfo {title} {Chapter 4: {P}ower and
  particle control},}\ }\href@noop {} {\bibfield  {journal} {\bibinfo
  {journal} {Nuclear Fusion}\ }\textbf {\bibinfo {volume} {47}},\ \bibinfo
  {pages} {S203} (\bibinfo {year} {2007})}\BibitemShut {NoStop}%
\bibitem [{\citenamefont {Wiesen}\ \emph {et~al.}(2015)\citenamefont {Wiesen},
  \citenamefont {Reiter}, \citenamefont {Kotov}, \citenamefont {Baelmans},
  \citenamefont {Dekeyser}, \citenamefont {Kukushkin}, \citenamefont {Lisgo},
  \citenamefont {Pitts}, \citenamefont {Rozhansky}, \citenamefont {Saibene},
  \citenamefont {Veselova},\ and\ \citenamefont {Voskoboynikov}}]{wiesen}%
  \BibitemOpen
  \bibfield  {author} {\bibinfo {author} {\bibfnamefont {S.}~\bibnamefont
  {Wiesen}}, \bibinfo {author} {\bibfnamefont {D.}~\bibnamefont {Reiter}},
  \bibinfo {author} {\bibfnamefont {V.}~\bibnamefont {Kotov}}, \bibinfo
  {author} {\bibfnamefont {M.}~\bibnamefont {Baelmans}}, \bibinfo {author}
  {\bibfnamefont {W.}~\bibnamefont {Dekeyser}}, \bibinfo {author}
  {\bibfnamefont {A.~S.}\ \bibnamefont {Kukushkin}}, \bibinfo {author}
  {\bibfnamefont {S.~W.}\ \bibnamefont {Lisgo}}, \bibinfo {author}
  {\bibfnamefont {R.~A.}\ \bibnamefont {Pitts}}, \bibinfo {author}
  {\bibfnamefont {V.}~\bibnamefont {Rozhansky}}, \bibinfo {author}
  {\bibfnamefont {G.}~\bibnamefont {Saibene}}, \bibinfo {author} {\bibfnamefont
  {I.}~\bibnamefont {Veselova}}, \ and\ \bibinfo {author} {\bibfnamefont
  {S.}~\bibnamefont {Voskoboynikov}},\ }\bibfield  {title} {\enquote {\bibinfo
  {title} {The new {SOLPS-ITER} code package},}\ }\href@noop {} {\bibfield
  {journal} {\bibinfo  {journal} {Journal of Nuclear Materials}\ }\textbf
  {\bibinfo {volume} {463}},\ \bibinfo {pages} {480} (\bibinfo {year}
  {2015})}\BibitemShut {NoStop}%
\bibitem [{\citenamefont {Bonnin}\ \emph {et~al.}(2016)\citenamefont {Bonnin},
  \citenamefont {Dekeyser}, \citenamefont {Pitts}, \citenamefont {Coster},
  \citenamefont {Voskoboynikov},\ and\ \citenamefont {Wiesen}}]{bonnin}%
  \BibitemOpen
  \bibfield  {author} {\bibinfo {author} {\bibfnamefont {X.}~\bibnamefont
  {Bonnin}}, \bibinfo {author} {\bibfnamefont {W.}~\bibnamefont {Dekeyser}},
  \bibinfo {author} {\bibfnamefont {R.}~\bibnamefont {Pitts}}, \bibinfo
  {author} {\bibfnamefont {D.}~\bibnamefont {Coster}}, \bibinfo {author}
  {\bibfnamefont {S.}~\bibnamefont {Voskoboynikov}}, \ and\ \bibinfo {author}
  {\bibfnamefont {S.}~\bibnamefont {Wiesen}},\ }\bibfield  {title} {\enquote
  {\bibinfo {title} {Presentation of the new {SOLPS-ITER} code package for
  tokamak plasma edge modelling},}\ }\href@noop {} {\bibfield  {journal}
  {\bibinfo  {journal} {Plasma and Fusion Research}\ }\textbf {\bibinfo
  {volume} {11}},\ \bibinfo {pages} {1403102} (\bibinfo {year}
  {2016})}\BibitemShut {NoStop}%
\bibitem [{\citenamefont {D'{I}ppolito}\ \emph {et~al.}(2004)\citenamefont
  {D'{I}ppolito}, \citenamefont {Myra}, \citenamefont {Krasheninnikov},
  \citenamefont {Yu},\ and\ \citenamefont {Pigarov}}]{d2004blob}%
  \BibitemOpen
  \bibfield  {author} {\bibinfo {author} {\bibfnamefont {D.~A.}\ \bibnamefont
  {D'{I}ppolito}}, \bibinfo {author} {\bibfnamefont {J.~R.}\ \bibnamefont
  {Myra}}, \bibinfo {author} {\bibfnamefont {S.~I.}\ \bibnamefont
  {Krasheninnikov}}, \bibinfo {author} {\bibfnamefont {G.~Q.}\ \bibnamefont
  {Yu}}, \ and\ \bibinfo {author} {\bibfnamefont {A.~Y.}\ \bibnamefont
  {Pigarov}},\ }\bibfield  {title} {\enquote {\bibinfo {title} {Blob transport
  in the tokamak scrape-off-layer},}\ }\href@noop {} {\bibfield  {journal}
  {\bibinfo  {journal} {Contributions to Plasma Physics}\ }\textbf {\bibinfo
  {volume} {44}},\ \bibinfo {pages} {205} (\bibinfo {year} {2004})}\BibitemShut
  {NoStop}%
\bibitem [{\citenamefont {Krasheninnikov}, \citenamefont {D'{I}ppolito},\ and\
  \citenamefont {Myra}(2008)}]{krasheninnikov2008recent}%
  \BibitemOpen
  \bibfield  {author} {\bibinfo {author} {\bibfnamefont {S.~I.}\ \bibnamefont
  {Krasheninnikov}}, \bibinfo {author} {\bibfnamefont {D.~A.}\ \bibnamefont
  {D'{I}ppolito}}, \ and\ \bibinfo {author} {\bibfnamefont {J.~R.}\
  \bibnamefont {Myra}},\ }\bibfield  {title} {\enquote {\bibinfo {title}
  {Recent theoretical progress in understanding coherent structures in edge and
  {SOL} turbulence},}\ }\href@noop {} {\bibfield  {journal} {\bibinfo
  {journal} {Journal of Plasma Physics}\ }\textbf {\bibinfo {volume} {74}},\
  \bibinfo {pages} {679} (\bibinfo {year} {2008})}\BibitemShut {NoStop}%
\bibitem [{\citenamefont {Garcia}(2009)}]{garcia2009blob}%
  \BibitemOpen
  \bibfield  {author} {\bibinfo {author} {\bibfnamefont {O.~E.}\ \bibnamefont
  {Garcia}},\ }\bibfield  {title} {\enquote {\bibinfo {title} {Blob transport
  in the plasma edge: a review},}\ }\href@noop {} {\bibfield  {journal}
  {\bibinfo  {journal} {Plasma and Fusion Research}\ }\textbf {\bibinfo
  {volume} {4}},\ \bibinfo {pages} {019} (\bibinfo {year} {2009})}\BibitemShut
  {NoStop}%
\bibitem [{\citenamefont {D'Ippolito}, \citenamefont {Myra},\ and\
  \citenamefont {Zweben}(2011)}]{dippolito-11}%
  \BibitemOpen
  \bibfield  {author} {\bibinfo {author} {\bibfnamefont {D.~A.}\ \bibnamefont
  {D'Ippolito}}, \bibinfo {author} {\bibfnamefont {J.~R.}\ \bibnamefont
  {Myra}}, \ and\ \bibinfo {author} {\bibfnamefont {S.~J.}\ \bibnamefont
  {Zweben}},\ }\bibfield  {title} {\enquote {\bibinfo {title} {Convective
  transport by intermittent blob-filaments: {C}omparison of theory and
  experiment},}\ }\href@noop {} {\bibfield  {journal} {\bibinfo  {journal}
  {Physics of Plasmas}\ }\textbf {\bibinfo {volume} {18}},\ \bibinfo {pages}
  {060501} (\bibinfo {year} {2011})}\BibitemShut {NoStop}%
\bibitem [{\citenamefont {Asakura}\ \emph {et~al.}(1997)\citenamefont
  {Asakura}, \citenamefont {Koide}, \citenamefont {Itami}, \citenamefont
  {Hosogan}, \citenamefont {Shimizu}, \citenamefont {Tsuji-Iio}, \citenamefont
  {Sakurai},\ and\ \citenamefont {Sakasai}}]{asakura}%
  \BibitemOpen
  \bibfield  {author} {\bibinfo {author} {\bibfnamefont {N.}~\bibnamefont
  {Asakura}}, \bibinfo {author} {\bibfnamefont {Y.}~\bibnamefont {Koide}},
  \bibinfo {author} {\bibfnamefont {K.}~\bibnamefont {Itami}}, \bibinfo
  {author} {\bibfnamefont {N.}~\bibnamefont {Hosogan}}, \bibinfo {author}
  {\bibfnamefont {K.}~\bibnamefont {Shimizu}}, \bibinfo {author} {\bibfnamefont
  {S.}~\bibnamefont {Tsuji-Iio}}, \bibinfo {author} {\bibfnamefont
  {S.}~\bibnamefont {Sakurai}}, \ and\ \bibinfo {author} {\bibfnamefont
  {A.}~\bibnamefont {Sakasai}},\ }\bibfield  {title} {\enquote {\bibinfo
  {title} {{SO}l plasma profiles under radiative and detached divertor
  conditions in {JT-60U}},}\ }\href@noop {} {\bibfield  {journal} {\bibinfo
  {journal} {Journal of Nuclear Materials}\ }\textbf {\bibinfo {volume}
  {241-243}},\ \bibinfo {pages} {559} (\bibinfo {year} {1997})}\BibitemShut
  {NoStop}%
\bibitem [{\citenamefont {La{B}ombard}\ \emph {et~al.}(2001)\citenamefont
  {La{B}ombard}, \citenamefont {Boivin}, \citenamefont {Greenwald},
  \citenamefont {Hughes}, \citenamefont {Lipschultz}, \citenamefont
  {Mossessian}, \citenamefont {Pitcher}, \citenamefont {Terry}, \citenamefont
  {Zweben},\ and\ \citenamefont {{Alcator Group}}}]{labombard2001}%
  \BibitemOpen
  \bibfield  {author} {\bibinfo {author} {\bibfnamefont {B.}~\bibnamefont
  {La{B}ombard}}, \bibinfo {author} {\bibfnamefont {R.~L.}\ \bibnamefont
  {Boivin}}, \bibinfo {author} {\bibfnamefont {M.}~\bibnamefont {Greenwald}},
  \bibinfo {author} {\bibfnamefont {J.}~\bibnamefont {Hughes}}, \bibinfo
  {author} {\bibfnamefont {B.}~\bibnamefont {Lipschultz}}, \bibinfo {author}
  {\bibfnamefont {D.}~\bibnamefont {Mossessian}}, \bibinfo {author}
  {\bibfnamefont {C.~S.}\ \bibnamefont {Pitcher}}, \bibinfo {author}
  {\bibfnamefont {J.~L.}\ \bibnamefont {Terry}}, \bibinfo {author}
  {\bibfnamefont {S.~J.}\ \bibnamefont {Zweben}}, \ and\ \bibinfo {author}
  {\bibnamefont {{Alcator Group}}},\ }\bibfield  {title} {\enquote {\bibinfo
  {title} {Particle transport in the scrape-off layer and its relationship to
  discharge density limit in {A}lcator {C}-{M}od},}\ }\href@noop {} {\bibfield
  {journal} {\bibinfo  {journal} {Physics of Plasmas}\ }\textbf {\bibinfo
  {volume} {8}},\ \bibinfo {pages} {2107} (\bibinfo {year} {2001})}\BibitemShut
  {NoStop}%
\bibitem [{\citenamefont {Lipschultz}\ \emph {et~al.}(2002)\citenamefont
  {Lipschultz}, \citenamefont {La{B}ombard}, \citenamefont {Pitcher},\ and\
  \citenamefont {Boivin}}]{lipschultz2002}%
  \BibitemOpen
  \bibfield  {author} {\bibinfo {author} {\bibfnamefont {B.}~\bibnamefont
  {Lipschultz}}, \bibinfo {author} {\bibfnamefont {B.}~\bibnamefont
  {La{B}ombard}}, \bibinfo {author} {\bibfnamefont {C.~S.}\ \bibnamefont
  {Pitcher}}, \ and\ \bibinfo {author} {\bibfnamefont {R.}~\bibnamefont
  {Boivin}},\ }\bibfield  {title} {\enquote {\bibinfo {title} {Investigation of
  the origin of neutrals in the main chamber of {A}lcator {C}-{M}od},}\
  }\href@noop {} {\bibfield  {journal} {\bibinfo  {journal} {Plasma Physics and
  Controlled Fusion}\ }\textbf {\bibinfo {volume} {44}},\ \bibinfo {pages}
  {733} (\bibinfo {year} {2002})}\BibitemShut {NoStop}%
\bibitem [{\citenamefont {Garcia}\ \emph
  {et~al.}(2005{\natexlab{a}})\citenamefont {Garcia}, \citenamefont {Horacek},
  \citenamefont {Pitts}, \citenamefont {Nielsen}, \citenamefont {Fundamenski},
  \citenamefont {Graves}, \citenamefont {Naulin},\ and\ \citenamefont
  {Rasmussen}}]{garcia2005tcv}%
  \BibitemOpen
  \bibfield  {author} {\bibinfo {author} {\bibfnamefont {O.~E.}\ \bibnamefont
  {Garcia}}, \bibinfo {author} {\bibfnamefont {J.}~\bibnamefont {Horacek}},
  \bibinfo {author} {\bibfnamefont {R.~A.}\ \bibnamefont {Pitts}}, \bibinfo
  {author} {\bibfnamefont {A.~H.}\ \bibnamefont {Nielsen}}, \bibinfo {author}
  {\bibfnamefont {W.}~\bibnamefont {Fundamenski}}, \bibinfo {author}
  {\bibfnamefont {J.~P.}\ \bibnamefont {Graves}}, \bibinfo {author}
  {\bibfnamefont {V.}~\bibnamefont {Naulin}}, \ and\ \bibinfo {author}
  {\bibfnamefont {J.~J.}\ \bibnamefont {Rasmussen}},\ }\bibfield  {title}
  {\enquote {\bibinfo {title} {Interchange turbulence in the {TCV} scrape-off
  layer},}\ }\href@noop {} {\bibfield  {journal} {\bibinfo  {journal} {Plasma
  Physics and Controlled Fusion}\ }\textbf {\bibinfo {volume} {48}},\ \bibinfo
  {pages} {L1} (\bibinfo {year} {2005}{\natexlab{a}})}\BibitemShut {NoStop}%
\bibitem [{\citenamefont {Rudakov}\ \emph {et~al.}(2005)\citenamefont
  {Rudakov}, \citenamefont {Boedo}, \citenamefont {Moyer}, \citenamefont
  {Stangeby}, \citenamefont {Watkins}, \citenamefont {Whyte}, \citenamefont
  {Zeng}, \citenamefont {Brooks}, \citenamefont {Doerner}, \citenamefont
  {Evans}, \citenamefont {Fenstermacher}, \citenamefont {Groth}, \citenamefont
  {Hollmann}, \citenamefont {Krasheninnikov}, \citenamefont {Lasnier},
  \citenamefont {Leonard}, \citenamefont {Mahdavi}, \citenamefont {Mc{K}ee},
  \citenamefont {Mc{L}ean}, \citenamefont {Pigarov}, \citenamefont {Wampler},
  \citenamefont {Wang}, \citenamefont {West},\ and\ \citenamefont
  {Wong}}]{rudakov2005}%
  \BibitemOpen
  \bibfield  {author} {\bibinfo {author} {\bibfnamefont {D.~L.}\ \bibnamefont
  {Rudakov}}, \bibinfo {author} {\bibfnamefont {J.~A.}\ \bibnamefont {Boedo}},
  \bibinfo {author} {\bibfnamefont {R.~A.}\ \bibnamefont {Moyer}}, \bibinfo
  {author} {\bibfnamefont {P.~C.}\ \bibnamefont {Stangeby}}, \bibinfo {author}
  {\bibfnamefont {J.~G.}\ \bibnamefont {Watkins}}, \bibinfo {author}
  {\bibfnamefont {D.~G.}\ \bibnamefont {Whyte}}, \bibinfo {author}
  {\bibfnamefont {L.}~\bibnamefont {Zeng}}, \bibinfo {author} {\bibfnamefont
  {N.~H.}\ \bibnamefont {Brooks}}, \bibinfo {author} {\bibfnamefont {R.~P.}\
  \bibnamefont {Doerner}}, \bibinfo {author} {\bibfnamefont {T.~E.}\
  \bibnamefont {Evans}}, \bibinfo {author} {\bibfnamefont {M.~E.}\ \bibnamefont
  {Fenstermacher}}, \bibinfo {author} {\bibfnamefont {M.}~\bibnamefont
  {Groth}}, \bibinfo {author} {\bibfnamefont {E.~M.}\ \bibnamefont {Hollmann}},
  \bibinfo {author} {\bibfnamefont {S.~I.}\ \bibnamefont {Krasheninnikov}},
  \bibinfo {author} {\bibfnamefont {C.~J.}\ \bibnamefont {Lasnier}}, \bibinfo
  {author} {\bibfnamefont {A.~W.}\ \bibnamefont {Leonard}}, \bibinfo {author}
  {\bibfnamefont {M.~A.}\ \bibnamefont {Mahdavi}}, \bibinfo {author}
  {\bibfnamefont {G.~R.}\ \bibnamefont {Mc{K}ee}}, \bibinfo {author}
  {\bibfnamefont {A.~G.}\ \bibnamefont {Mc{L}ean}}, \bibinfo {author}
  {\bibfnamefont {A.~Y.}\ \bibnamefont {Pigarov}}, \bibinfo {author}
  {\bibfnamefont {W.~R.}\ \bibnamefont {Wampler}}, \bibinfo {author}
  {\bibfnamefont {G.}~\bibnamefont {Wang}}, \bibinfo {author} {\bibfnamefont
  {W.~P.}\ \bibnamefont {West}}, \ and\ \bibinfo {author} {\bibfnamefont
  {C.~P.~C.}\ \bibnamefont {Wong}},\ }\bibfield  {title} {\enquote {\bibinfo
  {title} {Far {SOL} transport and main wall plasma interaction in {DIII-D}},}\
  }\href@noop {} {\bibfield  {journal} {\bibinfo  {journal} {Nuclear Fusion}\
  }\textbf {\bibinfo {volume} {45}},\ \bibinfo {pages} {1589} (\bibinfo {year}
  {2005})}\BibitemShut {NoStop}%
\bibitem [{\citenamefont {La{B}ombard}\ \emph {et~al.}(2005)\citenamefont
  {La{B}ombard}, \citenamefont {Hughes}, \citenamefont {Mossessian},
  \citenamefont {Greenwald}, \citenamefont {Lipschultz}, \citenamefont
  {Terry},\ and\ \citenamefont {{the Alcator C-Mod Team}}}]{labombard2005}%
  \BibitemOpen
  \bibfield  {author} {\bibinfo {author} {\bibfnamefont {B.}~\bibnamefont
  {La{B}ombard}}, \bibinfo {author} {\bibfnamefont {J.~W.}\ \bibnamefont
  {Hughes}}, \bibinfo {author} {\bibfnamefont {D.}~\bibnamefont {Mossessian}},
  \bibinfo {author} {\bibfnamefont {M.}~\bibnamefont {Greenwald}}, \bibinfo
  {author} {\bibfnamefont {B.}~\bibnamefont {Lipschultz}}, \bibinfo {author}
  {\bibfnamefont {J.~L.}\ \bibnamefont {Terry}}, \ and\ \bibinfo {author}
  {\bibnamefont {{the Alcator C-Mod Team}}},\ }\bibfield  {title} {\enquote
  {\bibinfo {title} {Evidence for electromagnetic fluid drift turbulence
  controlling the edge plasma state in the {A}lcator {C}-{M}od tokamak},}\
  }\href@noop {} {\bibfield  {journal} {\bibinfo  {journal} {Nuclear Fusion}\
  }\textbf {\bibinfo {volume} {45}},\ \bibinfo {pages} {1658} (\bibinfo {year}
  {2005})}\BibitemShut {NoStop}%
\bibitem [{\citenamefont {Garcia}\ \emph
  {et~al.}(2007{\natexlab{a}})\citenamefont {Garcia}, \citenamefont {Pitts},
  \citenamefont {Horacek}, \citenamefont {Nielsen}, \citenamefont
  {Fundamenski}, \citenamefont {Naulin},\ and\ \citenamefont
  {Rasmussen}}]{garcia2007jnm}%
  \BibitemOpen
  \bibfield  {author} {\bibinfo {author} {\bibfnamefont {O.~E.}\ \bibnamefont
  {Garcia}}, \bibinfo {author} {\bibfnamefont {R.~A.}\ \bibnamefont {Pitts}},
  \bibinfo {author} {\bibfnamefont {J.}~\bibnamefont {Horacek}}, \bibinfo
  {author} {\bibfnamefont {A.~H.}\ \bibnamefont {Nielsen}}, \bibinfo {author}
  {\bibfnamefont {W.}~\bibnamefont {Fundamenski}}, \bibinfo {author}
  {\bibfnamefont {V.}~\bibnamefont {Naulin}}, \ and\ \bibinfo {author}
  {\bibfnamefont {J.~J.}\ \bibnamefont {Rasmussen}},\ }\bibfield  {title}
  {\enquote {\bibinfo {title} {Turbulent transport in the {TCV} {SOL}},}\
  }\href@noop {} {\bibfield  {journal} {\bibinfo  {journal} {Journal of Nuclear
  Materials}\ }\textbf {\bibinfo {volume} {363–365}},\ \bibinfo {pages} {575}
  (\bibinfo {year} {2007}{\natexlab{a}})}\BibitemShut {NoStop}%
\bibitem [{\citenamefont {Garcia}\ \emph
  {et~al.}(2007{\natexlab{b}})\citenamefont {Garcia}, \citenamefont {Horacek},
  \citenamefont {Pitts}, \citenamefont {Nielsen}, \citenamefont {Fundamenski},
  \citenamefont {Naulin},\ and\ \citenamefont {Rasmussen}}]{garcia2007tcvnf}%
  \BibitemOpen
  \bibfield  {author} {\bibinfo {author} {\bibfnamefont {O.~E.}\ \bibnamefont
  {Garcia}}, \bibinfo {author} {\bibfnamefont {J.}~\bibnamefont {Horacek}},
  \bibinfo {author} {\bibfnamefont {R.~A.}\ \bibnamefont {Pitts}}, \bibinfo
  {author} {\bibfnamefont {A.~H.}\ \bibnamefont {Nielsen}}, \bibinfo {author}
  {\bibfnamefont {W.}~\bibnamefont {Fundamenski}}, \bibinfo {author}
  {\bibfnamefont {V.}~\bibnamefont {Naulin}}, \ and\ \bibinfo {author}
  {\bibfnamefont {J.~J.}\ \bibnamefont {Rasmussen}},\ }\bibfield  {title}
  {\enquote {\bibinfo {title} {Fluctuations and transport in the {TCV}
  scrape-off layer},}\ }\href@noop {} {\bibfield  {journal} {\bibinfo
  {journal} {Nuclear Fusion}\ }\textbf {\bibinfo {volume} {47}},\ \bibinfo
  {pages} {667} (\bibinfo {year} {2007}{\natexlab{b}})}\BibitemShut {NoStop}%
\bibitem [{\citenamefont {Garcia}\ \emph
  {et~al.}(2007{\natexlab{c}})\citenamefont {Garcia}, \citenamefont {Pitts},
  \citenamefont {Horacek}, \citenamefont {Madsen}, \citenamefont {Naulin},
  \citenamefont {Nielsen},\ and\ \citenamefont
  {Rasmussen}}]{garcia2007tcvppcf}%
  \BibitemOpen
  \bibfield  {author} {\bibinfo {author} {\bibfnamefont {O.~E.}\ \bibnamefont
  {Garcia}}, \bibinfo {author} {\bibfnamefont {R.~A.}\ \bibnamefont {Pitts}},
  \bibinfo {author} {\bibfnamefont {J.}~\bibnamefont {Horacek}}, \bibinfo
  {author} {\bibfnamefont {J.}~\bibnamefont {Madsen}}, \bibinfo {author}
  {\bibfnamefont {V.}~\bibnamefont {Naulin}}, \bibinfo {author} {\bibfnamefont
  {A.~H.}\ \bibnamefont {Nielsen}}, \ and\ \bibinfo {author} {\bibfnamefont
  {J.~J.}\ \bibnamefont {Rasmussen}},\ }\bibfield  {title} {\enquote {\bibinfo
  {title} {Collisionality dependent transport in {TCV} {SOL} plasmas},}\
  }\href@noop {} {\bibfield  {journal} {\bibinfo  {journal} {Plasma Physics and
  Controlled Fusion}\ }\textbf {\bibinfo {volume} {49}},\ \bibinfo {pages}
  {B47} (\bibinfo {year} {2007}{\natexlab{c}})}\BibitemShut {NoStop}%
\bibitem [{\citenamefont {Carralero}\ \emph {et~al.}(2014)\citenamefont
  {Carralero}, \citenamefont {Birkenmeier}, \citenamefont {M{̈\"u}ller},
  \citenamefont {Manz}, \citenamefont {de{M}arne}, \citenamefont
  {M{\"u}̈ller}, \citenamefont {Reimold}, \citenamefont {Stroth},
  \citenamefont {Wischmeier}, \citenamefont {Wolfrum},\ and\ \citenamefont
  {{The ASDEX Upgrade Team}}}]{carralero2014}%
  \BibitemOpen
  \bibfield  {author} {\bibinfo {author} {\bibfnamefont {D.}~\bibnamefont
  {Carralero}}, \bibinfo {author} {\bibfnamefont {G.}~\bibnamefont
  {Birkenmeier}}, \bibinfo {author} {\bibfnamefont {H.~W.}\ \bibnamefont
  {M{̈\"u}ller}}, \bibinfo {author} {\bibfnamefont {P.}~\bibnamefont {Manz}},
  \bibinfo {author} {\bibfnamefont {P.}~\bibnamefont {de{M}arne}}, \bibinfo
  {author} {\bibfnamefont {S.~H.}\ \bibnamefont {M{\"u}̈ller}}, \bibinfo
  {author} {\bibfnamefont {F.}~\bibnamefont {Reimold}}, \bibinfo {author}
  {\bibfnamefont {U.}~\bibnamefont {Stroth}}, \bibinfo {author} {\bibfnamefont
  {M.}~\bibnamefont {Wischmeier}}, \bibinfo {author} {\bibfnamefont
  {E.}~\bibnamefont {Wolfrum}}, \ and\ \bibinfo {author} {\bibnamefont {{The
  ASDEX Upgrade Team}}},\ }\bibfield  {title} {\enquote {\bibinfo {title} {An
  experimental investigation of the high density transition of the scrape-off
  layer transport in {ASDEX} {U}pgrade},}\ }\href@noop {} {\bibfield  {journal}
  {\bibinfo  {journal} {Nuclear Fusion}\ }\textbf {\bibinfo {volume} {54}},\
  \bibinfo {pages} {123005} (\bibinfo {year} {2014})}\BibitemShut {NoStop}%
\bibitem [{\citenamefont {Carralero}\ \emph {et~al.}(2015)\citenamefont
  {Carralero}, \citenamefont {M{\"u}ller}, \citenamefont {Groth}, \citenamefont
  {Komm}, \citenamefont {Adamek}, \citenamefont {Birkenmeier}, \citenamefont
  {Brix}, \citenamefont {Janky}, \citenamefont {Hacek}, \citenamefont {Marsen},
  \citenamefont {Reimold}, \citenamefont {Silva}, \citenamefont {Stroth},
  \citenamefont {Wischmeier}, \citenamefont {Wolfrum}, \citenamefont {{ASDEX
  Upgrade Team}}, \citenamefont {{COMPASS Team}},\ and\ \citenamefont
  {{JET-EFDA Contributors}}}]{carralero2015}%
  \BibitemOpen
  \bibfield  {author} {\bibinfo {author} {\bibfnamefont {D.}~\bibnamefont
  {Carralero}}, \bibinfo {author} {\bibfnamefont {H.~W.}\ \bibnamefont
  {M{\"u}ller}}, \bibinfo {author} {\bibfnamefont {M.}~\bibnamefont {Groth}},
  \bibinfo {author} {\bibfnamefont {M.}~\bibnamefont {Komm}}, \bibinfo {author}
  {\bibfnamefont {J.}~\bibnamefont {Adamek}}, \bibinfo {author} {\bibfnamefont
  {G.}~\bibnamefont {Birkenmeier}}, \bibinfo {author} {\bibfnamefont
  {M.}~\bibnamefont {Brix}}, \bibinfo {author} {\bibfnamefont {F.}~\bibnamefont
  {Janky}}, \bibinfo {author} {\bibfnamefont {P.}~\bibnamefont {Hacek}},
  \bibinfo {author} {\bibfnamefont {S.}~\bibnamefont {Marsen}}, \bibinfo
  {author} {\bibfnamefont {F.}~\bibnamefont {Reimold}}, \bibinfo {author}
  {\bibfnamefont {C.}~\bibnamefont {Silva}}, \bibinfo {author} {\bibfnamefont
  {U.}~\bibnamefont {Stroth}}, \bibinfo {author} {\bibfnamefont
  {M.}~\bibnamefont {Wischmeier}}, \bibinfo {author} {\bibfnamefont
  {E.}~\bibnamefont {Wolfrum}}, \bibinfo {author} {\bibnamefont {{ASDEX Upgrade
  Team}}}, \bibinfo {author} {\bibnamefont {{COMPASS Team}}}, \ and\ \bibinfo
  {author} {\bibnamefont {{JET-EFDA Contributors}}},\ }\bibfield  {title}
  {\enquote {\bibinfo {title} {Implications of high density operation on sol
  transport: {A} multimachine investigation},}\ }\href@noop {} {\bibfield
  {journal} {\bibinfo  {journal} {Journal of Nuclear Materials}\ }\textbf
  {\bibinfo {volume} {463}},\ \bibinfo {pages} {123} (\bibinfo {year}
  {2015})}\BibitemShut {NoStop}%
\bibitem [{\citenamefont {Militello}\ \emph {et~al.}(2016)\citenamefont
  {Militello}, \citenamefont {Garzotti}, \citenamefont {Harrison},
  \citenamefont {Omotani}, \citenamefont {Scannell}, \citenamefont {Allan},
  \citenamefont {Kirk}, \citenamefont {Lupelli}, \citenamefont {Thornton},\
  and\ \citenamefont {{the MAST team}}}]{militello2016nf}%
  \BibitemOpen
  \bibfield  {author} {\bibinfo {author} {\bibfnamefont {F.}~\bibnamefont
  {Militello}}, \bibinfo {author} {\bibfnamefont {L.}~\bibnamefont {Garzotti}},
  \bibinfo {author} {\bibfnamefont {J.}~\bibnamefont {Harrison}}, \bibinfo
  {author} {\bibfnamefont {J.~T.}\ \bibnamefont {Omotani}}, \bibinfo {author}
  {\bibfnamefont {R.}~\bibnamefont {Scannell}}, \bibinfo {author}
  {\bibfnamefont {S.}~\bibnamefont {Allan}}, \bibinfo {author} {\bibfnamefont
  {A.}~\bibnamefont {Kirk}}, \bibinfo {author} {\bibfnamefont {I.}~\bibnamefont
  {Lupelli}}, \bibinfo {author} {\bibfnamefont {A.~J.}\ \bibnamefont
  {Thornton}}, \ and\ \bibinfo {author} {\bibnamefont {{the MAST team}}},\
  }\bibfield  {title} {\enquote {\bibinfo {title} {Characterisation of the
  {L}-mode scrape off layer in {MAST}: decay lengths},}\ }\href@noop {}
  {\bibfield  {journal} {\bibinfo  {journal} {Nuclear Fusion}\ }\textbf
  {\bibinfo {volume} {56}},\ \bibinfo {pages} {016006} (\bibinfo {year}
  {2016})}\BibitemShut {NoStop}%
\bibitem [{\citenamefont {Vianello}\ \emph {et~al.}(2017)\citenamefont
  {Vianello}, \citenamefont {Tsui}, \citenamefont {Theiler}, \citenamefont
  {Allan}, \citenamefont {Boedo}, \citenamefont {Labit}, \citenamefont
  {Reimerdes}, \citenamefont {Verhaegh}, \citenamefont {Vijvers}, \citenamefont
  {Walkden}, \citenamefont {Costea}, \citenamefont {Kovacic}, \citenamefont
  {Ionita}, \citenamefont {Naulin}, \citenamefont {Nielsen}, \citenamefont
  {Rasmussen}, \citenamefont {Schneider}, \citenamefont {Schrittwieser},
  \citenamefont {Spolaore}, \citenamefont {Carralero}, \citenamefont {Madsen},
  \citenamefont {Lipschultz}, \citenamefont {Militello}, \citenamefont {{The
  TCV Team}},\ and\ \citenamefont {{The EUROfusion MST1 Team}}}]{vianello2017}%
  \BibitemOpen
  \bibfield  {author} {\bibinfo {author} {\bibfnamefont {N.}~\bibnamefont
  {Vianello}}, \bibinfo {author} {\bibfnamefont {C.}~\bibnamefont {Tsui}},
  \bibinfo {author} {\bibfnamefont {C.}~\bibnamefont {Theiler}}, \bibinfo
  {author} {\bibfnamefont {S.}~\bibnamefont {Allan}}, \bibinfo {author}
  {\bibfnamefont {J.}~\bibnamefont {Boedo}}, \bibinfo {author} {\bibfnamefont
  {B.}~\bibnamefont {Labit}}, \bibinfo {author} {\bibfnamefont
  {H.}~\bibnamefont {Reimerdes}}, \bibinfo {author} {\bibfnamefont
  {K.}~\bibnamefont {Verhaegh}}, \bibinfo {author} {\bibfnamefont {W.~A.~J.}\
  \bibnamefont {Vijvers}}, \bibinfo {author} {\bibfnamefont {N.}~\bibnamefont
  {Walkden}}, \bibinfo {author} {\bibfnamefont {S.}~\bibnamefont {Costea}},
  \bibinfo {author} {\bibfnamefont {J.}~\bibnamefont {Kovacic}}, \bibinfo
  {author} {\bibfnamefont {C.}~\bibnamefont {Ionita}}, \bibinfo {author}
  {\bibfnamefont {V.}~\bibnamefont {Naulin}}, \bibinfo {author} {\bibfnamefont
  {A.~H.}\ \bibnamefont {Nielsen}}, \bibinfo {author} {\bibfnamefont {J.~J.}\
  \bibnamefont {Rasmussen}}, \bibinfo {author} {\bibfnamefont {B.}~\bibnamefont
  {Schneider}}, \bibinfo {author} {\bibfnamefont {R.}~\bibnamefont
  {Schrittwieser}}, \bibinfo {author} {\bibfnamefont {M.}~\bibnamefont
  {Spolaore}}, \bibinfo {author} {\bibfnamefont {D.}~\bibnamefont {Carralero}},
  \bibinfo {author} {\bibfnamefont {J.}~\bibnamefont {Madsen}}, \bibinfo
  {author} {\bibfnamefont {B.}~\bibnamefont {Lipschultz}}, \bibinfo {author}
  {\bibfnamefont {F.}~\bibnamefont {Militello}}, \bibinfo {author}
  {\bibnamefont {{The TCV Team}}}, \ and\ \bibinfo {author} {\bibnamefont {{The
  EUROfusion MST1 Team}}},\ }\bibfield  {title} {\enquote {\bibinfo {title}
  {Modification of {SOL} profiles and fluctuations with line-average density
  and divertor flux expansion in {TCV}},}\ }\href@noop {} {\bibfield  {journal}
  {\bibinfo  {journal} {Nuclear Fusion}\ }\textbf {\bibinfo {volume} {57}},\
  \bibinfo {pages} {116014} (\bibinfo {year} {2017})}\BibitemShut {NoStop}%
\bibitem [{\citenamefont {Vianello}\ \emph {et~al.}(2020)\citenamefont
  {Vianello}, \citenamefont {Carralero}, \citenamefont {Tsui}, \citenamefont
  {Naulin}, \citenamefont {Agostini}, \citenamefont {Cziegler}, \citenamefont
  {Labit}, \citenamefont {Theiler}, \citenamefont {Wolfrum}, \citenamefont
  {Aguiam}, \citenamefont {Allan}, \citenamefont {Bernert}, \citenamefont
  {Boedo}, \citenamefont {Costea}, \citenamefont {{De Oliveira}}, \citenamefont
  {Fevrier}, \citenamefont {Galdon-{Q}uiroga1}, \citenamefont {Grenfell},
  \citenamefont {Hakola}, \citenamefont {Ionita}, \citenamefont {Isliker},
  \citenamefont {Karpushov}, \citenamefont {Kovacic}, \citenamefont
  {Lipschultz}, \citenamefont {Maurizio}, \citenamefont {Mc{C}lements},
  \citenamefont {Militello}, \citenamefont {Nielsen}, \citenamefont {Olsen},
  \citenamefont {Rasmussen}, \citenamefont {Ravensbergen}, \citenamefont
  {Reimerdes}, \citenamefont {Schneider}, \citenamefont {Schrittwieser},
  \citenamefont {Seliunin}, \citenamefont {Spolaore}, \citenamefont {Verhaegh},
  \citenamefont {Vicente}, \citenamefont {Walkden}, \citenamefont {Zhang},
  \citenamefont {{the ASDEX Upgrade Team}}, \citenamefont {{the TCV Team}},\
  and\ \citenamefont {{the EUROfusion MST1 Team}}}]{vianello2020}%
  \BibitemOpen
  \bibfield  {author} {\bibinfo {author} {\bibfnamefont {N.}~\bibnamefont
  {Vianello}}, \bibinfo {author} {\bibfnamefont {D.}~\bibnamefont {Carralero}},
  \bibinfo {author} {\bibfnamefont {C.~K.}\ \bibnamefont {Tsui}}, \bibinfo
  {author} {\bibfnamefont {V.}~\bibnamefont {Naulin}}, \bibinfo {author}
  {\bibfnamefont {M.}~\bibnamefont {Agostini}}, \bibinfo {author}
  {\bibfnamefont {I.}~\bibnamefont {Cziegler}}, \bibinfo {author}
  {\bibfnamefont {B.}~\bibnamefont {Labit}}, \bibinfo {author} {\bibfnamefont
  {C.}~\bibnamefont {Theiler}}, \bibinfo {author} {\bibfnamefont
  {E.}~\bibnamefont {Wolfrum}}, \bibinfo {author} {\bibfnamefont
  {D.}~\bibnamefont {Aguiam}}, \bibinfo {author} {\bibfnamefont
  {S.}~\bibnamefont {Allan}}, \bibinfo {author} {\bibfnamefont
  {M.}~\bibnamefont {Bernert}}, \bibinfo {author} {\bibfnamefont
  {J.}~\bibnamefont {Boedo}}, \bibinfo {author} {\bibfnamefont
  {S.}~\bibnamefont {Costea}}, \bibinfo {author} {\bibfnamefont
  {H.}~\bibnamefont {{De Oliveira}}}, \bibinfo {author} {\bibfnamefont
  {O.}~\bibnamefont {Fevrier}}, \bibinfo {author} {\bibfnamefont
  {J.}~\bibnamefont {Galdon-{Q}uiroga1}}, \bibinfo {author} {\bibfnamefont
  {G.}~\bibnamefont {Grenfell}}, \bibinfo {author} {\bibfnamefont
  {A.}~\bibnamefont {Hakola}}, \bibinfo {author} {\bibfnamefont
  {C.}~\bibnamefont {Ionita}}, \bibinfo {author} {\bibfnamefont
  {H.}~\bibnamefont {Isliker}}, \bibinfo {author} {\bibfnamefont
  {A.}~\bibnamefont {Karpushov}}, \bibinfo {author} {\bibfnamefont
  {J.}~\bibnamefont {Kovacic}}, \bibinfo {author} {\bibfnamefont
  {B.}~\bibnamefont {Lipschultz}}, \bibinfo {author} {\bibfnamefont
  {R.}~\bibnamefont {Maurizio}}, \bibinfo {author} {\bibfnamefont
  {K.}~\bibnamefont {Mc{C}lements}}, \bibinfo {author} {\bibfnamefont
  {F.}~\bibnamefont {Militello}}, \bibinfo {author} {\bibfnamefont {A.~H.}\
  \bibnamefont {Nielsen}}, \bibinfo {author} {\bibfnamefont {J.}~\bibnamefont
  {Olsen}}, \bibinfo {author} {\bibfnamefont {J.~J.}\ \bibnamefont
  {Rasmussen}}, \bibinfo {author} {\bibfnamefont {T.}~\bibnamefont
  {Ravensbergen}}, \bibinfo {author} {\bibfnamefont {H.}~\bibnamefont
  {Reimerdes}}, \bibinfo {author} {\bibfnamefont {B.}~\bibnamefont
  {Schneider}}, \bibinfo {author} {\bibfnamefont {R.}~\bibnamefont
  {Schrittwieser}}, \bibinfo {author} {\bibfnamefont {E.}~\bibnamefont
  {Seliunin}}, \bibinfo {author} {\bibfnamefont {M.}~\bibnamefont {Spolaore}},
  \bibinfo {author} {\bibfnamefont {K.}~\bibnamefont {Verhaegh}}, \bibinfo
  {author} {\bibfnamefont {J.}~\bibnamefont {Vicente}}, \bibinfo {author}
  {\bibfnamefont {N.}~\bibnamefont {Walkden}}, \bibinfo {author} {\bibfnamefont
  {W.}~\bibnamefont {Zhang}}, \bibinfo {author} {\bibnamefont {{the ASDEX
  Upgrade Team}}}, \bibinfo {author} {\bibnamefont {{the TCV Team}}}, \ and\
  \bibinfo {author} {\bibnamefont {{the EUROfusion MST1 Team}}},\ }\bibfield
  {title} {\enquote {\bibinfo {title} {Scrape-off layer transport and filament
  characteristics in high-density tokamak regimes},}\ }\href@noop {} {\bibfield
   {journal} {\bibinfo  {journal} {Nuclear Fusion}\ }\textbf {\bibinfo {volume}
  {60}},\ \bibinfo {pages} {016001} (\bibinfo {year} {2020})}\BibitemShut
  {NoStop}%
\bibitem [{\citenamefont {Liewer}(1985)}]{liewer1985measurements}%
  \BibitemOpen
  \bibfield  {author} {\bibinfo {author} {\bibfnamefont {P.~C.}\ \bibnamefont
  {Liewer}},\ }\bibfield  {title} {\enquote {\bibinfo {title} {Measurements of
  microturbulence in tokamaks and comparisons with theories of turbulence and
  anomalous transport},}\ }\href@noop {} {\bibfield  {journal} {\bibinfo
  {journal} {Nuclear Fusion}\ }\textbf {\bibinfo {volume} {25}},\ \bibinfo
  {pages} {543} (\bibinfo {year} {1985})}\BibitemShut {NoStop}%
\bibitem [{\citenamefont {Endler}(1999)}]{endler1995turbulent}%
  \BibitemOpen
  \bibfield  {author} {\bibinfo {author} {\bibfnamefont {M.}~\bibnamefont
  {Endler}},\ }\bibfield  {title} {\enquote {\bibinfo {title} {Turbulent {SOL}
  transport in stellarators and tokamaks},}\ }\href@noop {} {\bibfield
  {journal} {\bibinfo  {journal} {Journal of Nuclear Materials}\ }\textbf
  {\bibinfo {volume} {266-269}},\ \bibinfo {pages} {84} (\bibinfo {year}
  {1999})}\BibitemShut {NoStop}%
\bibitem [{\citenamefont {Carreras}(2005)}]{carreras2005plasma}%
  \BibitemOpen
  \bibfield  {author} {\bibinfo {author} {\bibfnamefont {B.~A.}\ \bibnamefont
  {Carreras}},\ }\bibfield  {title} {\enquote {\bibinfo {title} {Plasma edge
  cross-field transport: experiment and theory},}\ }\href@noop {} {\bibfield
  {journal} {\bibinfo  {journal} {Journal of Nuclear Materials}\ }\textbf
  {\bibinfo {volume} {337}},\ \bibinfo {pages} {315} (\bibinfo {year}
  {2005})}\BibitemShut {NoStop}%
\bibitem [{\citenamefont {Zweben}\ \emph {et~al.}(2007)\citenamefont {Zweben},
  \citenamefont {Boedo}, \citenamefont {Grulke}, \citenamefont {Hidalgo},
  \citenamefont {La{B}ombard}, \citenamefont {Maqueda}, \citenamefont
  {Scarin},\ and\ \citenamefont {Terry}}]{zweben2007edge}%
  \BibitemOpen
  \bibfield  {author} {\bibinfo {author} {\bibfnamefont {S.~J.}\ \bibnamefont
  {Zweben}}, \bibinfo {author} {\bibfnamefont {J.~A.}\ \bibnamefont {Boedo}},
  \bibinfo {author} {\bibfnamefont {O.}~\bibnamefont {Grulke}}, \bibinfo
  {author} {\bibfnamefont {C.}~\bibnamefont {Hidalgo}}, \bibinfo {author}
  {\bibfnamefont {B.}~\bibnamefont {La{B}ombard}}, \bibinfo {author}
  {\bibfnamefont {R.~J.}\ \bibnamefont {Maqueda}}, \bibinfo {author}
  {\bibfnamefont {P.}~\bibnamefont {Scarin}}, \ and\ \bibinfo {author}
  {\bibfnamefont {J.~L.}\ \bibnamefont {Terry}},\ }\bibfield  {title} {\enquote
  {\bibinfo {title} {Edge turbulence measurements in toroidal fusion
  devices},}\ }\href@noop {} {\bibfield  {journal} {\bibinfo  {journal} {Plasma
  Physics and Controlled Fusion}\ }\textbf {\bibinfo {volume} {49}},\ \bibinfo
  {pages} {S1} (\bibinfo {year} {2007})}\BibitemShut {NoStop}%
\bibitem [{\citenamefont {Garcia}\ \emph
  {et~al.}(2013{\natexlab{a}})\citenamefont {Garcia}, \citenamefont {Cziegler},
  \citenamefont {Kube}, \citenamefont {La{B}ombard},\ and\ \citenamefont
  {Terry}}]{garcia2013intermittent}%
  \BibitemOpen
  \bibfield  {author} {\bibinfo {author} {\bibfnamefont {O.~E.}\ \bibnamefont
  {Garcia}}, \bibinfo {author} {\bibfnamefont {I.}~\bibnamefont {Cziegler}},
  \bibinfo {author} {\bibfnamefont {R.}~\bibnamefont {Kube}}, \bibinfo {author}
  {\bibfnamefont {B.}~\bibnamefont {La{B}ombard}}, \ and\ \bibinfo {author}
  {\bibfnamefont {J.~L.}\ \bibnamefont {Terry}},\ }\bibfield  {title} {\enquote
  {\bibinfo {title} {Burst statistics in {A}lcator {C}-{M}od {SOL}
  turbulence},}\ }\href@noop {} {\bibfield  {journal} {\bibinfo  {journal}
  {Journal of Nuclear Materials}\ }\textbf {\bibinfo {volume} {438}},\ \bibinfo
  {pages} {S180} (\bibinfo {year} {2013}{\natexlab{a}})}\BibitemShut {NoStop}%
\bibitem [{\citenamefont {Garcia}\ \emph
  {et~al.}(2013{\natexlab{b}})\citenamefont {Garcia}, \citenamefont {Fritzner},
  \citenamefont {Kube}, \citenamefont {Cziegler}, \citenamefont {LaBombard},\
  and\ \citenamefont {Terry}}]{garcia2013intermittent2}%
  \BibitemOpen
  \bibfield  {author} {\bibinfo {author} {\bibfnamefont {O.~E.}\ \bibnamefont
  {Garcia}}, \bibinfo {author} {\bibfnamefont {S.~M.}\ \bibnamefont
  {Fritzner}}, \bibinfo {author} {\bibfnamefont {R.}~\bibnamefont {Kube}},
  \bibinfo {author} {\bibfnamefont {I.}~\bibnamefont {Cziegler}}, \bibinfo
  {author} {\bibfnamefont {B.}~\bibnamefont {LaBombard}}, \ and\ \bibinfo
  {author} {\bibfnamefont {J.~L.}\ \bibnamefont {Terry}},\ }\bibfield  {title}
  {\enquote {\bibinfo {title} {Intermittent fluctuations in the {A}lcator
  {C}-{M}od scrape-off layer},}\ }\href@noop {} {\bibfield  {journal} {\bibinfo
   {journal} {Physics of Plasmas}\ }\textbf {\bibinfo {volume} {20}},\ \bibinfo
  {pages} {055901} (\bibinfo {year} {2013}{\natexlab{b}})}\BibitemShut
  {NoStop}%
\bibitem [{\citenamefont {Garcia}, \citenamefont {Horacek},\ and\ \citenamefont
  {Pitts}(2015)}]{garcia2015tcv}%
  \BibitemOpen
  \bibfield  {author} {\bibinfo {author} {\bibfnamefont {O.~E.}\ \bibnamefont
  {Garcia}}, \bibinfo {author} {\bibfnamefont {J.}~\bibnamefont {Horacek}}, \
  and\ \bibinfo {author} {\bibfnamefont {R.~A.}\ \bibnamefont {Pitts}},\
  }\bibfield  {title} {\enquote {\bibinfo {title} {Intermittent fluctuations in
  the {TCV} scrape-off layer},}\ }\href@noop {} {\bibfield  {journal} {\bibinfo
   {journal} {Nuclear Fusion}\ }\textbf {\bibinfo {volume} {55}},\ \bibinfo
  {pages} {062002} (\bibinfo {year} {2015})}\BibitemShut {NoStop}%
\bibitem [{\citenamefont {Theodorsen}\ \emph {et~al.}(2016)\citenamefont
  {Theodorsen}, \citenamefont {Garcia}, \citenamefont {Horacek}, \citenamefont
  {Kube},\ and\ \citenamefont {Pitts}}]{theodorsen2016scrape}%
  \BibitemOpen
  \bibfield  {author} {\bibinfo {author} {\bibfnamefont {A.}~\bibnamefont
  {Theodorsen}}, \bibinfo {author} {\bibfnamefont {O.~E.}\ \bibnamefont
  {Garcia}}, \bibinfo {author} {\bibfnamefont {J.}~\bibnamefont {Horacek}},
  \bibinfo {author} {\bibfnamefont {R.}~\bibnamefont {Kube}}, \ and\ \bibinfo
  {author} {\bibfnamefont {R.~A.}\ \bibnamefont {Pitts}},\ }\bibfield  {title}
  {\enquote {\bibinfo {title} {Scrape-off layer turbulence in {TCV}: Evidence
  in support of stochastic modelling},}\ }\href@noop {} {\bibfield  {journal}
  {\bibinfo  {journal} {Plasma Physics and Controlled Fusion}\ }\textbf
  {\bibinfo {volume} {58}},\ \bibinfo {pages} {044006} (\bibinfo {year}
  {2016})}\BibitemShut {NoStop}%
\bibitem [{\citenamefont {Garcia}\ \emph {et~al.}(2017)\citenamefont {Garcia},
  \citenamefont {Kube}, \citenamefont {Theodorsen}, \citenamefont {Bak},
  \citenamefont {Hong}, \citenamefont {Kim}, \citenamefont {Pitts},\ and\
  \citenamefont {{{KSTAR} {T}eam}}}]{garcia2017sol}%
  \BibitemOpen
  \bibfield  {author} {\bibinfo {author} {\bibfnamefont {O.~E.}\ \bibnamefont
  {Garcia}}, \bibinfo {author} {\bibfnamefont {R.}~\bibnamefont {Kube}},
  \bibinfo {author} {\bibfnamefont {A.}~\bibnamefont {Theodorsen}}, \bibinfo
  {author} {\bibfnamefont {J.-G.}\ \bibnamefont {Bak}}, \bibinfo {author}
  {\bibfnamefont {S.-H.}\ \bibnamefont {Hong}}, \bibinfo {author}
  {\bibfnamefont {H.-S.}\ \bibnamefont {Kim}}, \bibinfo {author} {\bibfnamefont
  {R.~A.}\ \bibnamefont {Pitts}}, \ and\ \bibinfo {author} {\bibnamefont
  {{{KSTAR} {T}eam}}},\ }\bibfield  {title} {\enquote {\bibinfo {title} {{SOL}
  width and intermittent fluctuations in {KSTAR}},}\ }\href@noop {} {\bibfield
  {journal} {\bibinfo  {journal} {Nuclear Materials and Energy}\ }\textbf
  {\bibinfo {volume} {12}},\ \bibinfo {pages} {36} (\bibinfo {year}
  {2017})}\BibitemShut {NoStop}%
\bibitem [{\citenamefont {Theodorsen}\ \emph {et~al.}(2017)\citenamefont
  {Theodorsen}, \citenamefont {Garcia}, \citenamefont {Kube}, \citenamefont
  {La{B}ombard},\ and\ \citenamefont {Terry}}]{theodorsen2017cmod}%
  \BibitemOpen
  \bibfield  {author} {\bibinfo {author} {\bibfnamefont {A.}~\bibnamefont
  {Theodorsen}}, \bibinfo {author} {\bibfnamefont {O.~E.}\ \bibnamefont
  {Garcia}}, \bibinfo {author} {\bibfnamefont {R.}~\bibnamefont {Kube}},
  \bibinfo {author} {\bibfnamefont {B.}~\bibnamefont {La{B}ombard}}, \ and\
  \bibinfo {author} {\bibfnamefont {J.~L.}\ \bibnamefont {Terry}},\ }\bibfield
  {title} {\enquote {\bibinfo {title} {Relationship between frequency power
  spectra and intermittent, large-amplitude bursts in the {A}lcator {C}-{M}od
  scrape-off layer},}\ }\href@noop {} {\bibfield  {journal} {\bibinfo
  {journal} {Nuclear Fusion}\ }\textbf {\bibinfo {volume} {57}},\ \bibinfo
  {pages} {114004} (\bibinfo {year} {2017})}\BibitemShut {NoStop}%
\bibitem [{\citenamefont {Walkden}\ \emph
  {et~al.}(2017{\natexlab{a}})\citenamefont {Walkden}, \citenamefont {Wynn},
  \citenamefont {Militello}, \citenamefont {Lipschultz}, \citenamefont
  {Matthews}, \citenamefont {Guillemaut}, \citenamefont {Harrison},
  \citenamefont {Moulton},\ and\ \citenamefont {{JET Contributors}}}]{walkden}%
  \BibitemOpen
  \bibfield  {author} {\bibinfo {author} {\bibfnamefont {N.~R.}\ \bibnamefont
  {Walkden}}, \bibinfo {author} {\bibfnamefont {A.}~\bibnamefont {Wynn}},
  \bibinfo {author} {\bibfnamefont {F.}~\bibnamefont {Militello}}, \bibinfo
  {author} {\bibfnamefont {B.}~\bibnamefont {Lipschultz}}, \bibinfo {author}
  {\bibfnamefont {G.}~\bibnamefont {Matthews}}, \bibinfo {author}
  {\bibfnamefont {C.}~\bibnamefont {Guillemaut}}, \bibinfo {author}
  {\bibfnamefont {J.}~\bibnamefont {Harrison}}, \bibinfo {author}
  {\bibfnamefont {D.}~\bibnamefont {Moulton}}, \ and\ \bibinfo {author}
  {\bibnamefont {{JET Contributors}}},\ }\bibfield  {title} {\enquote {\bibinfo
  {title} {Statistical analysis of the ion flux to the {JET} outer wall},}\
  }\href@noop {} {\bibfield  {journal} {\bibinfo  {journal} {Nuclear Fusion}\
  }\textbf {\bibinfo {volume} {57}},\ \bibinfo {pages} {036016} (\bibinfo
  {year} {2017}{\natexlab{a}})}\BibitemShut {NoStop}%
\bibitem [{\citenamefont {Walkden}\ \emph
  {et~al.}(2017{\natexlab{b}})\citenamefont {Walkden}, \citenamefont {Wynn},
  \citenamefont {Militello}, \citenamefont {Lipschultz}, \citenamefont
  {Matthews}, \citenamefont {Guillemaut}, \citenamefont {Harrison},
  \citenamefont {Moulton},\ and\ \citenamefont {{C}ontributors}}]{walkden2017}%
  \BibitemOpen
  \bibfield  {author} {\bibinfo {author} {\bibfnamefont {N.~R.}\ \bibnamefont
  {Walkden}}, \bibinfo {author} {\bibfnamefont {A.}~\bibnamefont {Wynn}},
  \bibinfo {author} {\bibfnamefont {F.}~\bibnamefont {Militello}}, \bibinfo
  {author} {\bibfnamefont {B.}~\bibnamefont {Lipschultz}}, \bibinfo {author}
  {\bibfnamefont {G.}~\bibnamefont {Matthews}}, \bibinfo {author}
  {\bibfnamefont {C.}~\bibnamefont {Guillemaut}}, \bibinfo {author}
  {\bibfnamefont {J.}~\bibnamefont {Harrison}}, \bibinfo {author}
  {\bibfnamefont {D.}~\bibnamefont {Moulton}}, \ and\ \bibinfo {author}
  {\bibfnamefont {J.}~\bibnamefont {{C}ontributors}},\ }\bibfield  {title}
  {\enquote {\bibinfo {title} {Interpretation of scrape-off layer profile
  evolution and first-wall ion flux statistics on {JET} using a stochastic
  framework based on filamentary motion},}\ }\href@noop {} {\bibfield
  {journal} {\bibinfo  {journal} {Plasma Physics and Controlled Fusion}\
  }\textbf {\bibinfo {volume} {59}},\ \bibinfo {pages} {085009} (\bibinfo
  {year} {2017}{\natexlab{b}})}\BibitemShut {NoStop}%
\bibitem [{\citenamefont {Kube}\ \emph {et~al.}(2018)\citenamefont {Kube},
  \citenamefont {Garcia}, \citenamefont {Theodorsen}, \citenamefont {Brunner},
  \citenamefont {Kuang}, \citenamefont {La{B}ombard},\ and\ \citenamefont
  {Terry}}]{kube2018intermittent}%
  \BibitemOpen
  \bibfield  {author} {\bibinfo {author} {\bibfnamefont {R.}~\bibnamefont
  {Kube}}, \bibinfo {author} {\bibfnamefont {O.~E.}\ \bibnamefont {Garcia}},
  \bibinfo {author} {\bibfnamefont {A.}~\bibnamefont {Theodorsen}}, \bibinfo
  {author} {\bibfnamefont {D.}~\bibnamefont {Brunner}}, \bibinfo {author}
  {\bibfnamefont {A.~Q.}\ \bibnamefont {Kuang}}, \bibinfo {author}
  {\bibfnamefont {B.}~\bibnamefont {La{B}ombard}}, \ and\ \bibinfo {author}
  {\bibfnamefont {J.~L.}\ \bibnamefont {Terry}},\ }\bibfield  {title} {\enquote
  {\bibinfo {title} {Intermittent electron density and temperature fluctuations
  and associated fluxes in the {A}lcator {C}-{M}od scrape-off layer},}\
  }\href@noop {} {\bibfield  {journal} {\bibinfo  {journal} {Plasma Physics and
  Controlled Fusion}\ }\textbf {\bibinfo {volume} {60}},\ \bibinfo {pages}
  {065002} (\bibinfo {year} {2018})}\BibitemShut {NoStop}%
\bibitem [{\citenamefont {Garcia}\ \emph {et~al.}(2018)\citenamefont {Garcia},
  \citenamefont {Kube}, \citenamefont {Theodorsen}, \citenamefont {LaBombard},\
  and\ \citenamefont {Terry}}]{garcia2018intermittent}%
  \BibitemOpen
  \bibfield  {author} {\bibinfo {author} {\bibfnamefont {O.~E.}\ \bibnamefont
  {Garcia}}, \bibinfo {author} {\bibfnamefont {R.}~\bibnamefont {Kube}},
  \bibinfo {author} {\bibfnamefont {A.}~\bibnamefont {Theodorsen}}, \bibinfo
  {author} {\bibfnamefont {B.}~\bibnamefont {LaBombard}}, \ and\ \bibinfo
  {author} {\bibfnamefont {J.~L.}\ \bibnamefont {Terry}},\ }\bibfield  {title}
  {\enquote {\bibinfo {title} {Intermittent fluctuations in the {A}lcator
  {C}-{M}od scrape-off layer for ohmic and high confinement mode plasmas},}\
  }\href@noop {} {\bibfield  {journal} {\bibinfo  {journal} {Physics of
  Plasmas}\ }\textbf {\bibinfo {volume} {25}},\ \bibinfo {pages} {056103}
  (\bibinfo {year} {2018})}\BibitemShut {NoStop}%
\bibitem [{\citenamefont {Theodorsen}\ \emph {et~al.}(2018)\citenamefont
  {Theodorsen}, \citenamefont {Garcia}, \citenamefont {Kube}, \citenamefont
  {La{B}ombard},\ and\ \citenamefont {Terry}}]{theodorsen2018universality}%
  \BibitemOpen
  \bibfield  {author} {\bibinfo {author} {\bibfnamefont {A.}~\bibnamefont
  {Theodorsen}}, \bibinfo {author} {\bibfnamefont {O.~E.}\ \bibnamefont
  {Garcia}}, \bibinfo {author} {\bibfnamefont {R.}~\bibnamefont {Kube}},
  \bibinfo {author} {\bibfnamefont {B.}~\bibnamefont {La{B}ombard}}, \ and\
  \bibinfo {author} {\bibfnamefont {J.~L.}\ \bibnamefont {Terry}},\ }\bibfield
  {title} {\enquote {\bibinfo {title} {Universality of {P}oisson-driven plasma
  fluctuations in the {A}lcator {C}-{M}od scrape-off layer},}\ }\href@noop {}
  {\bibfield  {journal} {\bibinfo  {journal} {Physics of Plasmas}\ }\textbf
  {\bibinfo {volume} {25}},\ \bibinfo {pages} {122309} (\bibinfo {year}
  {2018})}\BibitemShut {NoStop}%
\bibitem [{\citenamefont {Bencze}\ \emph {et~al.}(2019)\citenamefont {Bencze},
  \citenamefont {Berta}, \citenamefont {Buz{\'a}s}, \citenamefont {Hacek},
  \citenamefont {Krbec}, \citenamefont {Szuty{\'a}nyi},\ and\ \citenamefont
  {{the COMPASS Team}}}]{benze}%
  \BibitemOpen
  \bibfield  {author} {\bibinfo {author} {\bibfnamefont {A.}~\bibnamefont
  {Bencze}}, \bibinfo {author} {\bibfnamefont {M.}~\bibnamefont {Berta}},
  \bibinfo {author} {\bibfnamefont {A.}~\bibnamefont {Buz{\'a}s}}, \bibinfo
  {author} {\bibfnamefont {P.}~\bibnamefont {Hacek}}, \bibinfo {author}
  {\bibfnamefont {J.}~\bibnamefont {Krbec}}, \bibinfo {author} {\bibfnamefont
  {M.}~\bibnamefont {Szuty{\'a}nyi}}, \ and\ \bibinfo {author} {\bibnamefont
  {{the COMPASS Team}}},\ }\bibfield  {title} {\enquote {\bibinfo {title}
  {Characterization of edge and scrape-off layer fluctuations using the fast
  {Li-BES} system on compass},}\ }\href@noop {} {\bibfield  {journal} {\bibinfo
   {journal} {Plasma Physics and Controlled Fusion}\ }\textbf {\bibinfo
  {volume} {61}},\ \bibinfo {pages} {085014} (\bibinfo {year}
  {2019})}\BibitemShut {NoStop}%
\bibitem [{\citenamefont {Kube}\ \emph {et~al.}(2019)\citenamefont {Kube},
  \citenamefont {Garcia}, \citenamefont {Theodorsen}, \citenamefont {Kuang},
  \citenamefont {LaBombard}, \citenamefont {Terry},\ and\ \citenamefont
  {Brunner}}]{kube2019statistical}%
  \BibitemOpen
  \bibfield  {author} {\bibinfo {author} {\bibfnamefont {R.}~\bibnamefont
  {Kube}}, \bibinfo {author} {\bibfnamefont {O.~E.}\ \bibnamefont {Garcia}},
  \bibinfo {author} {\bibfnamefont {A.}~\bibnamefont {Theodorsen}}, \bibinfo
  {author} {\bibfnamefont {A.~Q.}\ \bibnamefont {Kuang}}, \bibinfo {author}
  {\bibfnamefont {B.}~\bibnamefont {LaBombard}}, \bibinfo {author}
  {\bibfnamefont {J.~L.}\ \bibnamefont {Terry}}, \ and\ \bibinfo {author}
  {\bibfnamefont {D.}~\bibnamefont {Brunner}},\ }\bibfield  {title} {\enquote
  {\bibinfo {title} {Statistical properties of the plasma fluctuations and
  turbulent cross-field fluxes in the outboard mid-plane scrape-off layer of
  {A}lcator {C}-{M}od},}\ }\href@noop {} {\bibfield  {journal} {\bibinfo
  {journal} {Nuclear Materials and Energy}\ }\textbf {\bibinfo {volume} {18}},\
  \bibinfo {pages} {193} (\bibinfo {year} {2019})}\BibitemShut {NoStop}%
\bibitem [{\citenamefont {Kuang}\ \emph {et~al.}(2019)\citenamefont {Kuang},
  \citenamefont {La{B}ombard}, \citenamefont {Brunner}, \citenamefont {Garcia},
  \citenamefont {Kube},\ and\ \citenamefont {Theodorsen}}]{kuang}%
  \BibitemOpen
  \bibfield  {author} {\bibinfo {author} {\bibfnamefont {A.}~\bibnamefont
  {Kuang}}, \bibinfo {author} {\bibfnamefont {B.}~\bibnamefont {La{B}ombard}},
  \bibinfo {author} {\bibfnamefont {D.}~\bibnamefont {Brunner}}, \bibinfo
  {author} {\bibfnamefont {O.~E.}\ \bibnamefont {Garcia}}, \bibinfo {author}
  {\bibfnamefont {R.}~\bibnamefont {Kube}}, \ and\ \bibinfo {author}
  {\bibfnamefont {A.}~\bibnamefont {Theodorsen}},\ }\bibfield  {title}
  {\enquote {\bibinfo {title} {Plasma fluctuations in the scrape-off layer and
  at the divertor target in {A}lcator {C}-{M}od and their relationship to
  divertor collisionality and density shoulder formation},}\ }\href@noop {}
  {\bibfield  {journal} {\bibinfo  {journal} {Nuclear Materials and Energy}\
  }\textbf {\bibinfo {volume} {19}},\ \bibinfo {pages} {295} (\bibinfo {year}
  {2019})}\BibitemShut {NoStop}%
\bibitem [{\citenamefont {Kube}\ \emph {et~al.}(2020)\citenamefont {Kube},
  \citenamefont {Theodorsen}, \citenamefont {Garcia}, \citenamefont {Brunner},
  \citenamefont {LaBombard},\ and\ \citenamefont
  {Terry}}]{kube2020intermittent}%
  \BibitemOpen
  \bibfield  {author} {\bibinfo {author} {\bibfnamefont {R.}~\bibnamefont
  {Kube}}, \bibinfo {author} {\bibfnamefont {A.}~\bibnamefont {Theodorsen}},
  \bibinfo {author} {\bibfnamefont {O.~E.}\ \bibnamefont {Garcia}}, \bibinfo
  {author} {\bibfnamefont {D.}~\bibnamefont {Brunner}}, \bibinfo {author}
  {\bibfnamefont {B.}~\bibnamefont {LaBombard}}, \ and\ \bibinfo {author}
  {\bibfnamefont {J.~L.}\ \bibnamefont {Terry}},\ }\bibfield  {title} {\enquote
  {\bibinfo {title} {Comparison between mirror {L}angmuir probe and gas-puff
  imaging measurements of intermittent fluctuations in the {A}lcator {C}-{M}od
  scrape-off layer},}\ }\href@noop {} {\bibfield  {journal} {\bibinfo
  {journal} {Journal of Plasma Physics}\ }\textbf {\bibinfo {volume} {86}},\
  \bibinfo {pages} {905860519} (\bibinfo {year} {2020})}\BibitemShut {NoStop}%
\bibitem [{\citenamefont {Garcia}(2012)}]{garcia2012stochastic}%
  \BibitemOpen
  \bibfield  {author} {\bibinfo {author} {\bibfnamefont {O.~E.}\ \bibnamefont
  {Garcia}},\ }\bibfield  {title} {\enquote {\bibinfo {title} {Stochastic
  modeling of intermittent scrape-off layer plasma fluctuations},}\ }\href@noop
  {} {\bibfield  {journal} {\bibinfo  {journal} {Physical Review Letters}\
  }\textbf {\bibinfo {volume} {108}},\ \bibinfo {pages} {265001} (\bibinfo
  {year} {2012})}\BibitemShut {NoStop}%
\bibitem [{\citenamefont {Kube}\ and\ \citenamefont
  {Garcia}(2015)}]{kube2015pop}%
  \BibitemOpen
  \bibfield  {author} {\bibinfo {author} {\bibfnamefont {R.}~\bibnamefont
  {Kube}}\ and\ \bibinfo {author} {\bibfnamefont {O.~E.}\ \bibnamefont
  {Garcia}},\ }\bibfield  {title} {\enquote {\bibinfo {title} {Convergence of
  statistical moments of particle density time series in scrape-off layer
  plasmas},}\ }\href@noop {} {\bibfield  {journal} {\bibinfo  {journal}
  {Physics of Plasmas}\ }\textbf {\bibinfo {volume} {22}},\ \bibinfo {pages}
  {012502} (\bibinfo {year} {2015})}\BibitemShut {NoStop}%
\bibitem [{\citenamefont {Theodorsen}\ and\ \citenamefont
  {Garcia}(2016)}]{theodorsen2016pop}%
  \BibitemOpen
  \bibfield  {author} {\bibinfo {author} {\bibfnamefont {A.}~\bibnamefont
  {Theodorsen}}\ and\ \bibinfo {author} {\bibfnamefont {O.~E.}\ \bibnamefont
  {Garcia}},\ }\bibfield  {title} {\enquote {\bibinfo {title} {Level crossings,
  excess times, and transient plasma–wall interactions in fusion plasmas},}\
  }\href@noop {} {\bibfield  {journal} {\bibinfo  {journal} {Physics of
  Plasmas}\ }\textbf {\bibinfo {volume} {23}},\ \bibinfo {pages} {040702}
  (\bibinfo {year} {2016})}\BibitemShut {NoStop}%
\bibitem [{\citenamefont {Garcia}\ \emph {et~al.}(2016)\citenamefont {Garcia},
  \citenamefont {Kube}, \citenamefont {Theodorsen},\ and\ \citenamefont
  {P{\'e}cseli}}]{garcia2016stochastic}%
  \BibitemOpen
  \bibfield  {author} {\bibinfo {author} {\bibfnamefont {O.~E.}\ \bibnamefont
  {Garcia}}, \bibinfo {author} {\bibfnamefont {R.}~\bibnamefont {Kube}},
  \bibinfo {author} {\bibfnamefont {A.}~\bibnamefont {Theodorsen}}, \ and\
  \bibinfo {author} {\bibfnamefont {H.~L.}\ \bibnamefont {P{\'e}cseli}},\
  }\bibfield  {title} {\enquote {\bibinfo {title} {Stochastic modelling of
  intermittent fluctuations in the scrape-off layer: {C}orrelations,
  distributions, level crossings, and moment estimation},}\ }\href@noop {}
  {\bibfield  {journal} {\bibinfo  {journal} {Physics of Plasmas}\ }\textbf
  {\bibinfo {volume} {23}},\ \bibinfo {pages} {052308} (\bibinfo {year}
  {2016})}\BibitemShut {NoStop}%
\bibitem [{\citenamefont {Militello}\ and\ \citenamefont
  {Omotani}(2016{\natexlab{a}})}]{militello2016}%
  \BibitemOpen
  \bibfield  {author} {\bibinfo {author} {\bibfnamefont {F.}~\bibnamefont
  {Militello}}\ and\ \bibinfo {author} {\bibfnamefont {J.~T.}\ \bibnamefont
  {Omotani}},\ }\bibfield  {title} {\enquote {\bibinfo {title} {Scrape off
  layer profiles interpreted with filament dynamics},}\ }\href@noop {}
  {\bibfield  {journal} {\bibinfo  {journal} {Nuclear Fusion}\ }\textbf
  {\bibinfo {volume} {56}},\ \bibinfo {pages} {104004} (\bibinfo {year}
  {2016}{\natexlab{a}})}\BibitemShut {NoStop}%
\bibitem [{\citenamefont {Militello}\ and\ \citenamefont
  {Omotani}(2016{\natexlab{b}})}]{mo2016}%
  \BibitemOpen
  \bibfield  {author} {\bibinfo {author} {\bibfnamefont {F.}~\bibnamefont
  {Militello}}\ and\ \bibinfo {author} {\bibfnamefont {J.~T.}\ \bibnamefont
  {Omotani}},\ }\bibfield  {title} {\enquote {\bibinfo {title} {On the relation
  between non-exponential scrape off layer profiles and the dynamics of
  filaments},}\ }\href@noop {} {\bibfield  {journal} {\bibinfo  {journal}
  {Nuclear Fusion}\ }\textbf {\bibinfo {volume} {58}},\ \bibinfo {pages}
  {125004} (\bibinfo {year} {2016}{\natexlab{b}})}\BibitemShut {NoStop}%
\bibitem [{\citenamefont {Theodorsen}\ and\ \citenamefont
  {Garcia}(2017)}]{theodorsen2017ps}%
  \BibitemOpen
  \bibfield  {author} {\bibinfo {author} {\bibfnamefont {A.}~\bibnamefont
  {Theodorsen}}\ and\ \bibinfo {author} {\bibfnamefont {O.~E.}\ \bibnamefont
  {Garcia}},\ }\bibfield  {title} {\enquote {\bibinfo {title} {Statistical
  properties of a filtered {P}oisson process with additive random noise:
  distributions, correlations and moment estimation},}\ }\href@noop {}
  {\bibfield  {journal} {\bibinfo  {journal} {Physica Scripta}\ }\textbf
  {\bibinfo {volume} {92}},\ \bibinfo {pages} {054002} (\bibinfo {year}
  {2017})}\BibitemShut {NoStop}%
\bibitem [{\citenamefont {Garcia}\ and\ \citenamefont
  {Theodorsen}(2017{\natexlab{a}})}]{garcia2017pop}%
  \BibitemOpen
  \bibfield  {author} {\bibinfo {author} {\bibfnamefont {O.~E.}\ \bibnamefont
  {Garcia}}\ and\ \bibinfo {author} {\bibfnamefont {A.}~\bibnamefont
  {Theodorsen}},\ }\bibfield  {title} {\enquote {\bibinfo {title}
  {Auto-correlation function and frequency spectrum due to a super-position of
  uncorrelated exponential pulses},}\ }\href@noop {} {\bibfield  {journal}
  {\bibinfo  {journal} {Physics of Plasmas}\ }\textbf {\bibinfo {volume}
  {24}},\ \bibinfo {pages} {032309} (\bibinfo {year}
  {2017}{\natexlab{a}})}\BibitemShut {NoStop}%
\bibitem [{\citenamefont {Theodorsen}\ and\ \citenamefont
  {Garcia}(2018{\natexlab{a}})}]{theodorsen2018pre}%
  \BibitemOpen
  \bibfield  {author} {\bibinfo {author} {\bibfnamefont {A.}~\bibnamefont
  {Theodorsen}}\ and\ \bibinfo {author} {\bibfnamefont {O.~E.}\ \bibnamefont
  {Garcia}},\ }\bibfield  {title} {\enquote {\bibinfo {title} {Level crossings
  and excess times due to a superposition of uncorrelated exponential
  pulses},}\ }\href@noop {} {\bibfield  {journal} {\bibinfo  {journal}
  {Physical Review E}\ }\textbf {\bibinfo {volume} {97}},\ \bibinfo {pages}
  {012110} (\bibinfo {year} {2018}{\natexlab{a}})}\BibitemShut {NoStop}%
\bibitem [{\citenamefont {Militello}\ \emph {et~al.}(2018)\citenamefont
  {Militello}, \citenamefont {Farley}, \citenamefont {Mukhi}, \citenamefont
  {Walkden},\ and\ \citenamefont {Omotani}}]{militello2018}%
  \BibitemOpen
  \bibfield  {author} {\bibinfo {author} {\bibfnamefont {F.}~\bibnamefont
  {Militello}}, \bibinfo {author} {\bibfnamefont {T.}~\bibnamefont {Farley}},
  \bibinfo {author} {\bibfnamefont {K.}~\bibnamefont {Mukhi}}, \bibinfo
  {author} {\bibfnamefont {N.}~\bibnamefont {Walkden}}, \ and\ \bibinfo
  {author} {\bibfnamefont {J.~T.}\ \bibnamefont {Omotani}},\ }\bibfield
  {title} {\enquote {\bibinfo {title} {A two-dimensional statistical framework
  connecting thermodynamic profiles with filaments in the scrape off layer and
  application to experiments},}\ }\href@noop {} {\bibfield  {journal} {\bibinfo
   {journal} {Physics of Plasmas}\ }\textbf {\bibinfo {volume} {25}},\ \bibinfo
  {pages} {056112} (\bibinfo {year} {2018})}\BibitemShut {NoStop}%
\bibitem [{\citenamefont {Theodorsen}\ and\ \citenamefont
  {Garcia}(2018{\natexlab{b}})}]{theodorsen2018ppcf}%
  \BibitemOpen
  \bibfield  {author} {\bibinfo {author} {\bibfnamefont {A.}~\bibnamefont
  {Theodorsen}}\ and\ \bibinfo {author} {\bibfnamefont {O.~E.}\ \bibnamefont
  {Garcia}},\ }\bibfield  {title} {\enquote {\bibinfo {title} {Probability
  distribution functions for intermittent scrape-off layer plasma
  fluctuations},}\ }\href@noop {} {\bibfield  {journal} {\bibinfo  {journal}
  {Plasma Physics and Controlled Fusion}\ }\textbf {\bibinfo {volume} {60}},\
  \bibinfo {pages} {034006} (\bibinfo {year} {2018}{\natexlab{b}})}\BibitemShut
  {NoStop}%
\bibitem [{\citenamefont {Terry}\ \emph {et~al.}(2008)\citenamefont {Terry},
  \citenamefont {Greenwald}, \citenamefont {Leboeuf}, \citenamefont {McKee},
  \citenamefont {Mikkelsen}, \citenamefont {Nevins}, \citenamefont {Newman},
  \citenamefont {Stotler}, \citenamefont {{Task {G}roup on {V}erification and
  {V}alidation, U.S. {B}urning {P}lasma {O}rganization}},\ and\ \citenamefont
  {{U.S. {T}ransport {T}ask {F}orce}}}]{terry}%
  \BibitemOpen
  \bibfield  {author} {\bibinfo {author} {\bibfnamefont {P.~W.}\ \bibnamefont
  {Terry}}, \bibinfo {author} {\bibfnamefont {M.}~\bibnamefont {Greenwald}},
  \bibinfo {author} {\bibfnamefont {J.-N.}\ \bibnamefont {Leboeuf}}, \bibinfo
  {author} {\bibfnamefont {G.~R.}\ \bibnamefont {McKee}}, \bibinfo {author}
  {\bibfnamefont {D.~R.}\ \bibnamefont {Mikkelsen}}, \bibinfo {author}
  {\bibfnamefont {W.~M.}\ \bibnamefont {Nevins}}, \bibinfo {author}
  {\bibfnamefont {D.~E.}\ \bibnamefont {Newman}}, \bibinfo {author}
  {\bibfnamefont {D.~P.}\ \bibnamefont {Stotler}}, \bibinfo {author}
  {\bibnamefont {{Task {G}roup on {V}erification and {V}alidation, U.S.
  {B}urning {P}lasma {O}rganization}}}, \ and\ \bibinfo {author} {\bibnamefont
  {{U.S. {T}ransport {T}ask {F}orce}}},\ }\bibfield  {title} {\enquote
  {\bibinfo {title} {Validation in fusion research: {T}owards guidelines and
  best practices},}\ }\href@noop {} {\bibfield  {journal} {\bibinfo  {journal}
  {Physics of Plasmas}\ }\textbf {\bibinfo {volume} {15}},\ \bibinfo {pages}
  {062503} (\bibinfo {year} {2008})}\BibitemShut {NoStop}%
\bibitem [{\citenamefont {Greenwald}(2010)}]{greenwald}%
  \BibitemOpen
  \bibfield  {author} {\bibinfo {author} {\bibfnamefont {M.}~\bibnamefont
  {Greenwald}},\ }\bibfield  {title} {\enquote {\bibinfo {title} {Verification
  and validation for magnetic fusion},}\ }\href@noop {} {\bibfield  {journal}
  {\bibinfo  {journal} {Plasma Physics and Controlled Fusion}\ }\textbf
  {\bibinfo {volume} {17}},\ \bibinfo {pages} {058101} (\bibinfo {year}
  {2010})}\BibitemShut {NoStop}%
\bibitem [{\citenamefont {Holland}(2016)}]{holland}%
  \BibitemOpen
  \bibfield  {author} {\bibinfo {author} {\bibfnamefont {C.}~\bibnamefont
  {Holland}},\ }\bibfield  {title} {\enquote {\bibinfo {title} {Validation
  metrics for turbulent plasma transport},}\ }\href@noop {} {\bibfield
  {journal} {\bibinfo  {journal} {Physics of Plasmas}\ }\textbf {\bibinfo
  {volume} {23}},\ \bibinfo {pages} {060901} (\bibinfo {year}
  {2016})}\BibitemShut {NoStop}%
\bibitem [{\citenamefont {Decristoforo}, \citenamefont {Theodorsen},\ and\
  \citenamefont {Garcia}(2020)}]{decristoforo2020intermittent}%
  \BibitemOpen
  \bibfield  {author} {\bibinfo {author} {\bibfnamefont {G.}~\bibnamefont
  {Decristoforo}}, \bibinfo {author} {\bibfnamefont {A.}~\bibnamefont
  {Theodorsen}}, \ and\ \bibinfo {author} {\bibfnamefont {O.~E.}\ \bibnamefont
  {Garcia}},\ }\bibfield  {title} {\enquote {\bibinfo {title} {Intermittent
  fluctuations due to {L}orentzian pulses in turbulent thermal convection},}\
  }\href@noop {} {\bibfield  {journal} {\bibinfo  {journal} {Physics of
  Fluids}\ }\textbf {\bibinfo {volume} {32}},\ \bibinfo {pages} {085102}
  (\bibinfo {year} {2020})}\BibitemShut {NoStop}%
\bibitem [{\citenamefont {Pogutse}\ \emph {et~al.}(1994)\citenamefont
  {Pogutse}, \citenamefont {Kerner}, \citenamefont {Gribkov}, \citenamefont
  {Bazdenkov},\ and\ \citenamefont {Osipenko}}]{pogutse}%
  \BibitemOpen
  \bibfield  {author} {\bibinfo {author} {\bibfnamefont {O.}~\bibnamefont
  {Pogutse}}, \bibinfo {author} {\bibfnamefont {W.}~\bibnamefont {Kerner}},
  \bibinfo {author} {\bibfnamefont {V.}~\bibnamefont {Gribkov}}, \bibinfo
  {author} {\bibfnamefont {S.}~\bibnamefont {Bazdenkov}}, \ and\ \bibinfo
  {author} {\bibfnamefont {M.}~\bibnamefont {Osipenko}},\ }\bibfield  {title}
  {\enquote {\bibinfo {title} {The resistive interchange convection in the edge
  of tokamak plasmas},}\ }\href@noop {} {\bibfield  {journal} {\bibinfo
  {journal} {Plasma Physics and Controlled Fusion}\ }\textbf {\bibinfo {volume}
  {36}},\ \bibinfo {pages} {1963} (\bibinfo {year} {1994})}\BibitemShut
  {NoStop}%
\bibitem [{\citenamefont {Sugama}\ and\ \citenamefont {Horton}(1995)}]{sugama}%
  \BibitemOpen
  \bibfield  {author} {\bibinfo {author} {\bibfnamefont {H.}~\bibnamefont
  {Sugama}}\ and\ \bibinfo {author} {\bibfnamefont {W.}~\bibnamefont
  {Horton}},\ }\bibfield  {title} {\enquote {\bibinfo {title} {L-{H}
  confinement mode dynamics in three-dimensional state space},}\ }\href@noop {}
  {\bibfield  {journal} {\bibinfo  {journal} {Plasma Physics and Controlled
  Fusion}\ }\textbf {\bibinfo {volume} {37}},\ \bibinfo {pages} {345} (\bibinfo
  {year} {1995})}\BibitemShut {NoStop}%
\bibitem [{\citenamefont {Beyer}\ and\ \citenamefont
  {Spatschek}(1996)}]{beyer}%
  \BibitemOpen
  \bibfield  {author} {\bibinfo {author} {\bibfnamefont {P.}~\bibnamefont
  {Beyer}}\ and\ \bibinfo {author} {\bibfnamefont {K.~H.}\ \bibnamefont
  {Spatschek}},\ }\bibfield  {title} {\enquote {\bibinfo {title} {Center
  manifold theory for the dynamics of the {L}–{H}-transition},}\ }\href@noop
  {} {\bibfield  {journal} {\bibinfo  {journal} {Physics of Plasmas}\ }\textbf
  {\bibinfo {volume} {3}},\ \bibinfo {pages} {995} (\bibinfo {year}
  {1996})}\BibitemShut {NoStop}%
\bibitem [{\citenamefont {Horton}, \citenamefont {Hu},\ and\ \citenamefont
  {Laval}(1996)}]{horton}%
  \BibitemOpen
  \bibfield  {author} {\bibinfo {author} {\bibfnamefont {W.}~\bibnamefont
  {Horton}}, \bibinfo {author} {\bibfnamefont {G.}~\bibnamefont {Hu}}, \ and\
  \bibinfo {author} {\bibfnamefont {G.}~\bibnamefont {Laval}},\ }\bibfield
  {title} {\enquote {\bibinfo {title} {Turbulent transport in mixed states of
  convective cells and sheared flows},}\ }\href@noop {} {\bibfield  {journal}
  {\bibinfo  {journal} {Physics of Plasmas}\ }\textbf {\bibinfo {volume} {3}},\
  \bibinfo {pages} {2912} (\bibinfo {year} {1996})}\BibitemShut {NoStop}%
\bibitem [{\citenamefont {Berning}\ and\ \citenamefont
  {Spatschek}(2000)}]{berning}%
  \BibitemOpen
  \bibfield  {author} {\bibinfo {author} {\bibfnamefont {M.}~\bibnamefont
  {Berning}}\ and\ \bibinfo {author} {\bibfnamefont {K.~H.}\ \bibnamefont
  {Spatschek}},\ }\bibfield  {title} {\enquote {\bibinfo {title} {Bifurcations
  and transport barriers in the resistive-g paradigm},}\ }\href@noop {}
  {\bibfield  {journal} {\bibinfo  {journal} {Physical Review E}\ }\textbf
  {\bibinfo {volume} {62}},\ \bibinfo {pages} {1162} (\bibinfo {year}
  {2000})}\BibitemShut {NoStop}%
\bibitem [{\citenamefont {Garcia}\ \emph
  {et~al.}(2003{\natexlab{a}})\citenamefont {Garcia}, \citenamefont {Bian},
  \citenamefont {Paulsen}, \citenamefont {Benkadda},\ and\ \citenamefont
  {Rypdal}}]{garcia2003rb}%
  \BibitemOpen
  \bibfield  {author} {\bibinfo {author} {\bibfnamefont {O.~E.}\ \bibnamefont
  {Garcia}}, \bibinfo {author} {\bibfnamefont {N.~H.}\ \bibnamefont {Bian}},
  \bibinfo {author} {\bibfnamefont {J.-V.}\ \bibnamefont {Paulsen}}, \bibinfo
  {author} {\bibfnamefont {S.}~\bibnamefont {Benkadda}}, \ and\ \bibinfo
  {author} {\bibfnamefont {K.}~\bibnamefont {Rypdal}},\ }\bibfield  {title}
  {\enquote {\bibinfo {title} {Confinement and bursty transport in a
  flux-driven convection model with sheared flows},}\ }\href@noop {} {\bibfield
   {journal} {\bibinfo  {journal} {Plasma Physics and Controlled Fusion}\
  }\textbf {\bibinfo {volume} {45}},\ \bibinfo {pages} {919} (\bibinfo {year}
  {2003}{\natexlab{a}})}\BibitemShut {NoStop}%
\bibitem [{\citenamefont {Garcia}\ and\ \citenamefont
  {Bian}(2003)}]{garciabian2003}%
  \BibitemOpen
  \bibfield  {author} {\bibinfo {author} {\bibfnamefont {O.~E.}\ \bibnamefont
  {Garcia}}\ and\ \bibinfo {author} {\bibfnamefont {N.~H.}\ \bibnamefont
  {Bian}},\ }\bibfield  {title} {\enquote {\bibinfo {title} {Bursting and
  large-scale intermittency in turbulent convection with differential
  rotation},}\ }\href@noop {} {\bibfield  {journal} {\bibinfo  {journal}
  {Physical Review E}\ }\textbf {\bibinfo {volume} {68}},\ \bibinfo {pages}
  {047301} (\bibinfo {year} {2003})}\BibitemShut {NoStop}%
\bibitem [{\citenamefont {Garcia}\ \emph
  {et~al.}(2003{\natexlab{b}})\citenamefont {Garcia}, \citenamefont {Bian},
  \citenamefont {Naulin}, \citenamefont {Nielsen},\ and\ \citenamefont
  {Rasmussen}}]{garcia2003ps}%
  \BibitemOpen
  \bibfield  {author} {\bibinfo {author} {\bibfnamefont {O.~E.}\ \bibnamefont
  {Garcia}}, \bibinfo {author} {\bibfnamefont {N.~H.}\ \bibnamefont {Bian}},
  \bibinfo {author} {\bibfnamefont {V.}~\bibnamefont {Naulin}}, \bibinfo
  {author} {\bibfnamefont {A.~H.}\ \bibnamefont {Nielsen}}, \ and\ \bibinfo
  {author} {\bibfnamefont {J.~J.}\ \bibnamefont {Rasmussen}},\ }\bibfield
  {title} {\enquote {\bibinfo {title} {Two-dimensional convection and
  interchange motions in fluids and magnetized plasmas},}\ }\href@noop {}
  {\bibfield  {journal} {\bibinfo  {journal} {Physica Scripta}\ }\textbf
  {\bibinfo {volume} {T122}},\ \bibinfo {pages} {104} (\bibinfo {year}
  {2003}{\natexlab{b}})}\BibitemShut {NoStop}%
\bibitem [{\citenamefont {Wilczynski}\ \emph {et~al.}(2019)\citenamefont
  {Wilczynski}, \citenamefont {Hughes}, \citenamefont {Loo}, \citenamefont
  {Arter},\ and\ \citenamefont {Militello}}]{wilczynski}%
  \BibitemOpen
  \bibfield  {author} {\bibinfo {author} {\bibfnamefont {F.}~\bibnamefont
  {Wilczynski}}, \bibinfo {author} {\bibfnamefont {D.~W.}\ \bibnamefont
  {Hughes}}, \bibinfo {author} {\bibfnamefont {S.~V.}\ \bibnamefont {Loo}},
  \bibinfo {author} {\bibfnamefont {W.}~\bibnamefont {Arter}}, \ and\ \bibinfo
  {author} {\bibfnamefont {F.}~\bibnamefont {Militello}},\ }\bibfield  {title}
  {\enquote {\bibinfo {title} {Stability of scrape-off layer plasma: {A}
  modified {R}ayleigh–{B}{\'e}nard problem},}\ }\href@noop {} {\bibfield
  {journal} {\bibinfo  {journal} {Physics of Plasmas}\ }\textbf {\bibinfo
  {volume} {26}},\ \bibinfo {pages} {022510} (\bibinfo {year}
  {2019})}\BibitemShut {NoStop}%
\bibitem [{\citenamefont {Benkadda}, \citenamefont {Garbet},\ and\
  \citenamefont {Verma}(1994)}]{benkadda}%
  \BibitemOpen
  \bibfield  {author} {\bibinfo {author} {\bibfnamefont {S.}~\bibnamefont
  {Benkadda}}, \bibinfo {author} {\bibfnamefont {X.}~\bibnamefont {Garbet}}, \
  and\ \bibinfo {author} {\bibfnamefont {A.}~\bibnamefont {Verma}},\ }\bibfield
   {title} {\enquote {\bibinfo {title} {Interchange instability turbulence
  model in edge tokamak plasma},}\ }\href@noop {} {\bibfield  {journal}
  {\bibinfo  {journal} {Contributions to Plasma Physics}\ }\textbf {\bibinfo
  {volume} {34}},\ \bibinfo {pages} {247} (\bibinfo {year} {1994})}\BibitemShut
  {NoStop}%
\bibitem [{\citenamefont {Garcia}(2001)}]{garcia-2001}%
  \BibitemOpen
  \bibfield  {author} {\bibinfo {author} {\bibfnamefont {O.~E.}\ \bibnamefont
  {Garcia}},\ }\bibfield  {title} {\enquote {\bibinfo {title} {Two-field
  transport models for magnetized plasmas},}\ }\href@noop {} {\bibfield
  {journal} {\bibinfo  {journal} {Journal of Plasma Physics}\ }\textbf
  {\bibinfo {volume} {65}},\ \bibinfo {pages} {81} (\bibinfo {year}
  {2001})}\BibitemShut {NoStop}%
\bibitem [{\citenamefont {Sarazin}\ and\ \citenamefont
  {Ghendrih}(1998)}]{sarazin1998theoretical}%
  \BibitemOpen
  \bibfield  {author} {\bibinfo {author} {\bibfnamefont {Y.}~\bibnamefont
  {Sarazin}}\ and\ \bibinfo {author} {\bibfnamefont {P.}~\bibnamefont
  {Ghendrih}},\ }\bibfield  {title} {\enquote {\bibinfo {title} {Intermittent
  particle transport in two-dimensional edge turbulence},}\ }\href@noop {}
  {\bibfield  {journal} {\bibinfo  {journal} {Physics of Plasmas}\ }\textbf
  {\bibinfo {volume} {5}},\ \bibinfo {pages} {4214} (\bibinfo {year}
  {1998})}\BibitemShut {NoStop}%
\bibitem [{\citenamefont {Ghendrih}\ \emph {et~al.}(2003)\citenamefont
  {Ghendrih}, \citenamefont {Sarazin}, \citenamefont {Attuel}, \citenamefont
  {Benkadda}, \citenamefont {Beyer}, \citenamefont {Falchetto}, \citenamefont
  {Figarella}, \citenamefont {Garbet}, \citenamefont {Grandgirard},\ and\
  \citenamefont {Ottaviani}}]{ghendrih2003}%
  \BibitemOpen
  \bibfield  {author} {\bibinfo {author} {\bibfnamefont {P.}~\bibnamefont
  {Ghendrih}}, \bibinfo {author} {\bibfnamefont {Y.}~\bibnamefont {Sarazin}},
  \bibinfo {author} {\bibfnamefont {G.}~\bibnamefont {Attuel}}, \bibinfo
  {author} {\bibfnamefont {S.}~\bibnamefont {Benkadda}}, \bibinfo {author}
  {\bibfnamefont {P.}~\bibnamefont {Beyer}}, \bibinfo {author} {\bibfnamefont
  {G.}~\bibnamefont {Falchetto}}, \bibinfo {author} {\bibfnamefont
  {C.}~\bibnamefont {Figarella}}, \bibinfo {author} {\bibfnamefont
  {X.}~\bibnamefont {Garbet}}, \bibinfo {author} {\bibfnamefont
  {V.}~\bibnamefont {Grandgirard}}, \ and\ \bibinfo {author} {\bibfnamefont
  {M.}~\bibnamefont {Ottaviani}},\ }\bibfield  {title} {\enquote {\bibinfo
  {title} {Theoretical analysis of the influence of external biasing on long
  range turbulent transport in the scrape-off layer},}\ }\href@noop {}
  {\bibfield  {journal} {\bibinfo  {journal} {Nuclear Fusion}\ }\textbf
  {\bibinfo {volume} {43}},\ \bibinfo {pages} {1013} (\bibinfo {year}
  {2003})}\BibitemShut {NoStop}%
\bibitem [{\citenamefont {Sarazin}\ \emph {et~al.}(2003)\citenamefont
  {Sarazin}, \citenamefont {Ghendrih}, \citenamefont {Attuel}, \citenamefont
  {Cl{\'e}ment}, \citenamefont {Garbet}, \citenamefont {Grandgirard},
  \citenamefont {Ottaviani}, \citenamefont {Benkadda}, \citenamefont {Beyer},
  \citenamefont {Bian},\ and\ \citenamefont
  {Figarella}}]{sarazin2003theoretical}%
  \BibitemOpen
  \bibfield  {author} {\bibinfo {author} {\bibfnamefont {Y.}~\bibnamefont
  {Sarazin}}, \bibinfo {author} {\bibfnamefont {P.}~\bibnamefont {Ghendrih}},
  \bibinfo {author} {\bibfnamefont {G.}~\bibnamefont {Attuel}}, \bibinfo
  {author} {\bibfnamefont {C.}~\bibnamefont {Cl{\'e}ment}}, \bibinfo {author}
  {\bibfnamefont {X.}~\bibnamefont {Garbet}}, \bibinfo {author} {\bibfnamefont
  {V.}~\bibnamefont {Grandgirard}}, \bibinfo {author} {\bibfnamefont
  {M.}~\bibnamefont {Ottaviani}}, \bibinfo {author} {\bibfnamefont
  {S.}~\bibnamefont {Benkadda}}, \bibinfo {author} {\bibfnamefont
  {P.}~\bibnamefont {Beyer}}, \bibinfo {author} {\bibfnamefont
  {N.}~\bibnamefont {Bian}}, \ and\ \bibinfo {author} {\bibfnamefont
  {C.}~\bibnamefont {Figarella}},\ }\bibfield  {title} {\enquote {\bibinfo
  {title} {Theoretical understanding of turbulent transport in the {SOL}},}\
  }\href@noop {} {\bibfield  {journal} {\bibinfo  {journal} {Journal of Nuclear
  Materials}\ }\textbf {\bibinfo {volume} {313}},\ \bibinfo {pages} {796}
  (\bibinfo {year} {2003})}\BibitemShut {NoStop}%
\bibitem [{\citenamefont {Garcia}\ \emph {et~al.}(2004)\citenamefont {Garcia},
  \citenamefont {Naulin}, \citenamefont {Nielsen},\ and\ \citenamefont
  {Rasmussen}}]{garcia-esel-prl}%
  \BibitemOpen
  \bibfield  {author} {\bibinfo {author} {\bibfnamefont {O.~E.}\ \bibnamefont
  {Garcia}}, \bibinfo {author} {\bibfnamefont {V.}~\bibnamefont {Naulin}},
  \bibinfo {author} {\bibfnamefont {A.~H.}\ \bibnamefont {Nielsen}}, \ and\
  \bibinfo {author} {\bibfnamefont {J.~J.}\ \bibnamefont {Rasmussen}},\
  }\bibfield  {title} {\enquote {\bibinfo {title} {Computations of intermittent
  transport in scrape-off layer plasmas},}\ }\href@noop {} {\bibfield
  {journal} {\bibinfo  {journal} {Physical Review Letters}\ }\textbf {\bibinfo
  {volume} {92}},\ \bibinfo {pages} {165003} (\bibinfo {year}
  {2004})}\BibitemShut {NoStop}%
\bibitem [{\citenamefont {Garcia}\ \emph
  {et~al.}(2005{\natexlab{b}})\citenamefont {Garcia}, \citenamefont {Naulin},
  \citenamefont {Nielsen},\ and\ \citenamefont {Rasmussen}}]{garcia-esel-php}%
  \BibitemOpen
  \bibfield  {author} {\bibinfo {author} {\bibfnamefont {O.~E.}\ \bibnamefont
  {Garcia}}, \bibinfo {author} {\bibfnamefont {V.}~\bibnamefont {Naulin}},
  \bibinfo {author} {\bibfnamefont {A.~H.}\ \bibnamefont {Nielsen}}, \ and\
  \bibinfo {author} {\bibfnamefont {J.~J.}\ \bibnamefont {Rasmussen}},\
  }\bibfield  {title} {\enquote {\bibinfo {title} {Turbulence and intermittent
  transport at the boundary of magnetized plasmas},}\ }\href@noop {} {\bibfield
   {journal} {\bibinfo  {journal} {Physics of Plasmas}\ }\textbf {\bibinfo
  {volume} {12}},\ \bibinfo {pages} {062309} (\bibinfo {year}
  {2005}{\natexlab{b}})}\BibitemShut {NoStop}%
\bibitem [{\citenamefont {Garcia}\ \emph {et~al.}(2006)\citenamefont {Garcia},
  \citenamefont {Naulin}, \citenamefont {Nielsen},\ and\ \citenamefont
  {Rasmussen}}]{garcia-esel-ps}%
  \BibitemOpen
  \bibfield  {author} {\bibinfo {author} {\bibfnamefont {O.~E.}\ \bibnamefont
  {Garcia}}, \bibinfo {author} {\bibfnamefont {V.}~\bibnamefont {Naulin}},
  \bibinfo {author} {\bibfnamefont {A.~H.}\ \bibnamefont {Nielsen}}, \ and\
  \bibinfo {author} {\bibfnamefont {J.~J.}\ \bibnamefont {Rasmussen}},\
  }\bibfield  {title} {\enquote {\bibinfo {title} {Turbulence simulations of
  blob formation and radial propagation in toroidally magnetized plasmas},}\
  }\href@noop {} {\bibfield  {journal} {\bibinfo  {journal} {Physica Scripta}\
  }\textbf {\bibinfo {volume} {T122}},\ \bibinfo {pages} {89} (\bibinfo {year}
  {2006})}\BibitemShut {NoStop}%
\bibitem [{\citenamefont {Myra}, \citenamefont {Russell},\ and\ \citenamefont
  {D’{I}ppolito}(2008)}]{myra2008}%
  \BibitemOpen
  \bibfield  {author} {\bibinfo {author} {\bibfnamefont {J.~R.}\ \bibnamefont
  {Myra}}, \bibinfo {author} {\bibfnamefont {D.~A.}\ \bibnamefont {Russell}}, \
  and\ \bibinfo {author} {\bibfnamefont {D.~A.}\ \bibnamefont
  {D’{I}ppolito}},\ }\bibfield  {title} {\enquote {\bibinfo {title}
  {Transport of perpendicular edge momentum by drift-interchange turbulence and
  blobs},}\ }\href@noop {} {\bibfield  {journal} {\bibinfo  {journal} {Physics
  of Plasmas}\ }\textbf {\bibinfo {volume} {15}},\ \bibinfo {pages} {032304}
  (\bibinfo {year} {2008})}\BibitemShut {NoStop}%
\bibitem [{\citenamefont {Russell}, \citenamefont {Myra},\ and\ \citenamefont
  {D’{I}ppolito}(2009)}]{russell}%
  \BibitemOpen
  \bibfield  {author} {\bibinfo {author} {\bibfnamefont {D.~A.}\ \bibnamefont
  {Russell}}, \bibinfo {author} {\bibfnamefont {J.~R.}\ \bibnamefont {Myra}}, \
  and\ \bibinfo {author} {\bibfnamefont {D.~A.}\ \bibnamefont
  {D’{I}ppolito}},\ }\bibfield  {title} {\enquote {\bibinfo {title}
  {Saturation mechanisms for edge turbulence},}\ }\href@noop {} {\bibfield
  {journal} {\bibinfo  {journal} {Physics of Plasmas}\ }\textbf {\bibinfo
  {volume} {16}},\ \bibinfo {pages} {122304} (\bibinfo {year}
  {2009})}\BibitemShut {NoStop}%
\bibitem [{\citenamefont {Myra}\ \emph {et~al.}(2013)\citenamefont {Myra},
  \citenamefont {Davis}, \citenamefont {D’{I}ppolito}, \citenamefont
  {La{B}ombard}, \citenamefont {Russell1}, \citenamefont {Terry},\ and\
  \citenamefont {Zweben}}]{myra2013}%
  \BibitemOpen
  \bibfield  {author} {\bibinfo {author} {\bibfnamefont {J.~R.}\ \bibnamefont
  {Myra}}, \bibinfo {author} {\bibfnamefont {W.~M.}\ \bibnamefont {Davis}},
  \bibinfo {author} {\bibfnamefont {D.~A.}\ \bibnamefont {D’{I}ppolito}},
  \bibinfo {author} {\bibfnamefont {B.}~\bibnamefont {La{B}ombard}}, \bibinfo
  {author} {\bibfnamefont {D.~A.}\ \bibnamefont {Russell1}}, \bibinfo {author}
  {\bibfnamefont {J.~L.}\ \bibnamefont {Terry}}, \ and\ \bibinfo {author}
  {\bibfnamefont {S.~J.}\ \bibnamefont {Zweben}},\ }\bibfield  {title}
  {\enquote {\bibinfo {title} {Edge sheared flows and the dynamics of
  blob-filaments},}\ }\href@noop {} {\bibfield  {journal} {\bibinfo  {journal}
  {Nuclear Fusion}\ }\textbf {\bibinfo {volume} {53}},\ \bibinfo {pages}
  {073013} (\bibinfo {year} {2013})}\BibitemShut {NoStop}%
\bibitem [{\citenamefont {Bisai}\ \emph {et~al.}(2004)\citenamefont {Bisai},
  \citenamefont {Das}, \citenamefont {Deshpande}, \citenamefont {Jha},
  \citenamefont {Kaw}, \citenamefont {Sen},\ and\ \citenamefont
  {Singh}}]{bisai2004}%
  \BibitemOpen
  \bibfield  {author} {\bibinfo {author} {\bibfnamefont {N.}~\bibnamefont
  {Bisai}}, \bibinfo {author} {\bibfnamefont {A.}~\bibnamefont {Das}}, \bibinfo
  {author} {\bibfnamefont {S.}~\bibnamefont {Deshpande}}, \bibinfo {author}
  {\bibfnamefont {R.}~\bibnamefont {Jha}}, \bibinfo {author} {\bibfnamefont
  {P.}~\bibnamefont {Kaw}}, \bibinfo {author} {\bibfnamefont {A.}~\bibnamefont
  {Sen}}, \ and\ \bibinfo {author} {\bibfnamefont {R.}~\bibnamefont {Singh}},\
  }\bibfield  {title} {\enquote {\bibinfo {title} {Simulation of plasma
  transport by coherent structures in scrape-off-layer tokamak plasmas},}\
  }\href@noop {} {\bibfield  {journal} {\bibinfo  {journal} {Physics of
  Plasmas}\ }\textbf {\bibinfo {volume} {11}},\ \bibinfo {pages} {4018}
  (\bibinfo {year} {2004})}\BibitemShut {NoStop}%
\bibitem [{\citenamefont {Bisai}\ \emph
  {et~al.}(2005{\natexlab{a}})\citenamefont {Bisai}, \citenamefont {Das},
  \citenamefont {Deshpande}, \citenamefont {Jha}, \citenamefont {Kaw},
  \citenamefont {Sen},\ and\ \citenamefont {Singh}}]{bisai2005edge}%
  \BibitemOpen
  \bibfield  {author} {\bibinfo {author} {\bibfnamefont {N.}~\bibnamefont
  {Bisai}}, \bibinfo {author} {\bibfnamefont {A.}~\bibnamefont {Das}}, \bibinfo
  {author} {\bibfnamefont {S.}~\bibnamefont {Deshpande}}, \bibinfo {author}
  {\bibfnamefont {R.}~\bibnamefont {Jha}}, \bibinfo {author} {\bibfnamefont
  {P.}~\bibnamefont {Kaw}}, \bibinfo {author} {\bibfnamefont {A.}~\bibnamefont
  {Sen}}, \ and\ \bibinfo {author} {\bibfnamefont {R.}~\bibnamefont {Singh}},\
  }\bibfield  {title} {\enquote {\bibinfo {title} {Edge and scrape-off layer
  tokamak plasma turbulence simulation using two-field fluid model},}\
  }\href@noop {} {\bibfield  {journal} {\bibinfo  {journal} {Physics of
  Plasmas}\ }\textbf {\bibinfo {volume} {12}},\ \bibinfo {pages} {072520}
  (\bibinfo {year} {2005}{\natexlab{a}})}\BibitemShut {NoStop}%
\bibitem [{\citenamefont {Bisai}\ \emph
  {et~al.}(2005{\natexlab{b}})\citenamefont {Bisai}, \citenamefont {Das},
  \citenamefont {Deshpande}, \citenamefont {Jha}, \citenamefont {Kaw},
  \citenamefont {Sen},\ and\ \citenamefont {Singh}}]{bisai2005edge2}%
  \BibitemOpen
  \bibfield  {author} {\bibinfo {author} {\bibfnamefont {N.}~\bibnamefont
  {Bisai}}, \bibinfo {author} {\bibfnamefont {A.}~\bibnamefont {Das}}, \bibinfo
  {author} {\bibfnamefont {S.}~\bibnamefont {Deshpande}}, \bibinfo {author}
  {\bibfnamefont {R.}~\bibnamefont {Jha}}, \bibinfo {author} {\bibfnamefont
  {P.}~\bibnamefont {Kaw}}, \bibinfo {author} {\bibfnamefont {A.}~\bibnamefont
  {Sen}}, \ and\ \bibinfo {author} {\bibfnamefont {R.}~\bibnamefont {Singh}},\
  }\bibfield  {title} {\enquote {\bibinfo {title} {Formation of a density blob
  and its dynamics in the edge and the scrape-off layer of a tokamak plasma},}\
  }\href@noop {} {\bibfield  {journal} {\bibinfo  {journal} {Physics of
  Plasmas}\ }\textbf {\bibinfo {volume} {12}},\ \bibinfo {pages} {102515}
  (\bibinfo {year} {2005}{\natexlab{b}})}\BibitemShut {NoStop}%
\bibitem [{\citenamefont {Nielsen}\ \emph {et~al.}(2015)\citenamefont
  {Nielsen}, \citenamefont {Xu}, \citenamefont {Madsen}, \citenamefont
  {Naulin}, \citenamefont {Rasmussen},\ and\ \citenamefont
  {Wan}}]{nielsen2015}%
  \BibitemOpen
  \bibfield  {author} {\bibinfo {author} {\bibfnamefont {A.~H.}\ \bibnamefont
  {Nielsen}}, \bibinfo {author} {\bibfnamefont {G.~S.}\ \bibnamefont {Xu}},
  \bibinfo {author} {\bibfnamefont {J.}~\bibnamefont {Madsen}}, \bibinfo
  {author} {\bibfnamefont {V.}~\bibnamefont {Naulin}}, \bibinfo {author}
  {\bibfnamefont {J.~J.}\ \bibnamefont {Rasmussen}}, \ and\ \bibinfo {author}
  {\bibfnamefont {B.~N.}\ \bibnamefont {Wan}},\ }\bibfield  {title} {\enquote
  {\bibinfo {title} {Simulation of transition dynamics to high confinement in
  fusion plasmas},}\ }\href@noop {} {\bibfield  {journal} {\bibinfo  {journal}
  {Physics Letters A}\ }\textbf {\bibinfo {volume} {379}},\ \bibinfo {pages}
  {3097} (\bibinfo {year} {2015})}\BibitemShut {NoStop}%
\bibitem [{\citenamefont {Nielsen}\ \emph {et~al.}(2017)\citenamefont
  {Nielsen}, \citenamefont {Rasmussen}, \citenamefont {Madsen}, \citenamefont
  {Xu}, \citenamefont {Naulin}, \citenamefont {Olsen}, \citenamefont
  {L{\o}iten}, \citenamefont {Hansen}, \citenamefont {Yan}, \citenamefont
  {Toph{\o}j},\ and\ \citenamefont {Wan}}]{nielsen2017}%
  \BibitemOpen
  \bibfield  {author} {\bibinfo {author} {\bibfnamefont {A.~H.}\ \bibnamefont
  {Nielsen}}, \bibinfo {author} {\bibfnamefont {J.~J.}\ \bibnamefont
  {Rasmussen}}, \bibinfo {author} {\bibfnamefont {J.}~\bibnamefont {Madsen}},
  \bibinfo {author} {\bibfnamefont {G.~S.}\ \bibnamefont {Xu}}, \bibinfo
  {author} {\bibfnamefont {V.}~\bibnamefont {Naulin}}, \bibinfo {author}
  {\bibfnamefont {J.~M.~B.}\ \bibnamefont {Olsen}}, \bibinfo {author}
  {\bibfnamefont {M.}~\bibnamefont {L{\o}iten}}, \bibinfo {author}
  {\bibfnamefont {S.~K.}\ \bibnamefont {Hansen}}, \bibinfo {author}
  {\bibfnamefont {N.}~\bibnamefont {Yan}}, \bibinfo {author} {\bibfnamefont
  {L.}~\bibnamefont {Toph{\o}j}}, \ and\ \bibinfo {author} {\bibfnamefont
  {B.~N.}\ \bibnamefont {Wan}},\ }\bibfield  {title} {\enquote {\bibinfo
  {title} {Numerical simulations of blobs with ion dynamics},}\ }\href@noop {}
  {\bibfield  {journal} {\bibinfo  {journal} {Plasma Physics and Controlled
  Fusion}\ }\textbf {\bibinfo {volume} {59}},\ \bibinfo {pages} {025012}
  (\bibinfo {year} {2017})}\BibitemShut {NoStop}%
\bibitem [{\citenamefont {Olsen}\ \emph {et~al.}(2018)\citenamefont {Olsen},
  \citenamefont {Nielsen}, \citenamefont {Rasmussen}, \citenamefont {Madsen},
  \citenamefont {Eich}, \citenamefont {Sieglin},\ and\ \citenamefont
  {Naulin}}]{olsen2018}%
  \BibitemOpen
  \bibfield  {author} {\bibinfo {author} {\bibfnamefont {J.}~\bibnamefont
  {Olsen}}, \bibinfo {author} {\bibfnamefont {A.~H.}\ \bibnamefont {Nielsen}},
  \bibinfo {author} {\bibfnamefont {J.~J.}\ \bibnamefont {Rasmussen}}, \bibinfo
  {author} {\bibfnamefont {J.}~\bibnamefont {Madsen}}, \bibinfo {author}
  {\bibfnamefont {T.}~\bibnamefont {Eich}}, \bibinfo {author} {\bibfnamefont
  {B.}~\bibnamefont {Sieglin}}, \ and\ \bibinfo {author} {\bibfnamefont
  {V.}~\bibnamefont {Naulin}},\ }\bibfield  {title} {\enquote {\bibinfo {title}
  {Scrape-off layer power fall-off length from turbulence simulations of
  {ASDEX} {U}pgrade {L}-mode},}\ }\href@noop {} {\bibfield  {journal} {\bibinfo
   {journal} {Plasma Physics and Controlled Fusion}\ }\textbf {\bibinfo
  {volume} {60}},\ \bibinfo {pages} {085018} (\bibinfo {year}
  {2018})}\BibitemShut {NoStop}%
\bibitem [{\citenamefont {Dudson}\ \emph {et~al.}(2009)\citenamefont {Dudson},
  \citenamefont {Umansky}, \citenamefont {Xu}, \citenamefont {Snyder},\ and\
  \citenamefont {Wilson}}]{dudson2009bout++}%
  \BibitemOpen
  \bibfield  {author} {\bibinfo {author} {\bibfnamefont {B.}~\bibnamefont
  {Dudson}}, \bibinfo {author} {\bibfnamefont {M.}~\bibnamefont {Umansky}},
  \bibinfo {author} {\bibfnamefont {X.}~\bibnamefont {Xu}}, \bibinfo {author}
  {\bibfnamefont {P.}~\bibnamefont {Snyder}}, \ and\ \bibinfo {author}
  {\bibfnamefont {H.}~\bibnamefont {Wilson}},\ }\bibfield  {title} {\enquote
  {\bibinfo {title} {{BOUT}++: {A} framework for parallel plasma fluid
  simulations},}\ }\href@noop {} {\bibfield  {journal} {\bibinfo  {journal}
  {Computer Physics Communications}\ }\textbf {\bibinfo {volume} {180}},\
  \bibinfo {pages} {1467} (\bibinfo {year} {2009})}\BibitemShut {NoStop}%
\bibitem [{\citenamefont {Militello}\ \emph {et~al.}(2017)\citenamefont
  {Militello}, \citenamefont {Dudson}, \citenamefont {Easy}, \citenamefont
  {Kirk},\ and\ \citenamefont {Naylor}}]{militello2017interaction}%
  \BibitemOpen
  \bibfield  {author} {\bibinfo {author} {\bibfnamefont {F.}~\bibnamefont
  {Militello}}, \bibinfo {author} {\bibfnamefont {B.}~\bibnamefont {Dudson}},
  \bibinfo {author} {\bibfnamefont {L.}~\bibnamefont {Easy}}, \bibinfo {author}
  {\bibfnamefont {A.}~\bibnamefont {Kirk}}, \ and\ \bibinfo {author}
  {\bibfnamefont {P.}~\bibnamefont {Naylor}},\ }\bibfield  {title} {\enquote
  {\bibinfo {title} {On the interaction of scrape off layer filaments},}\
  }\href@noop {} {\bibfield  {journal} {\bibinfo  {journal} {Plasma Physics and
  Controlled Fusion}\ }\textbf {\bibinfo {volume} {59}},\ \bibinfo {pages}
  {125013} (\bibinfo {year} {2017})}\BibitemShut {NoStop}%
\bibitem [{\citenamefont {Byrne}\ and\ \citenamefont
  {Hindmarsh}(1999)}]{byrne1999pvode}%
  \BibitemOpen
  \bibfield  {author} {\bibinfo {author} {\bibfnamefont {G.~D.}\ \bibnamefont
  {Byrne}}\ and\ \bibinfo {author} {\bibfnamefont {A.~C.}\ \bibnamefont
  {Hindmarsh}},\ }\bibfield  {title} {\enquote {\bibinfo {title} {{PVODE}, an
  {ODE} solver for parallel computers},}\ }\href@noop {} {\bibfield  {journal}
  {\bibinfo  {journal} {The International Journal of High Performance Computing
  Applications}\ }\textbf {\bibinfo {volume} {13}},\ \bibinfo {pages} {354}
  (\bibinfo {year} {1999})}\BibitemShut {NoStop}%
\bibitem [{\citenamefont {Maggs}\ and\ \citenamefont
  {Morales}(2011)}]{maggs2011}%
  \BibitemOpen
  \bibfield  {author} {\bibinfo {author} {\bibfnamefont {J.~E.}\ \bibnamefont
  {Maggs}}\ and\ \bibinfo {author} {\bibfnamefont {G.~J.}\ \bibnamefont
  {Morales}},\ }\bibfield  {title} {\enquote {\bibinfo {title} {Generality of
  deterministic chaos, exponential spectra, and {L}orentzian pulses in
  magnetically confined plasmas},}\ }\href@noop {} {\bibfield  {journal}
  {\bibinfo  {journal} {Physical Review Letters}\ }\textbf {\bibinfo {volume}
  {107}},\ \bibinfo {pages} {185003} (\bibinfo {year} {2011})}\BibitemShut
  {NoStop}%
\bibitem [{\citenamefont {Garcia}\ and\ \citenamefont
  {Theodorsen}(2017{\natexlab{b}})}]{garcia2017l}%
  \BibitemOpen
  \bibfield  {author} {\bibinfo {author} {\bibfnamefont {O.~E.}\ \bibnamefont
  {Garcia}}\ and\ \bibinfo {author} {\bibfnamefont {A.}~\bibnamefont
  {Theodorsen}},\ }\bibfield  {title} {\enquote {\bibinfo {title} {Power law
  spectra and intermittent fluctuations due to uncorrelated {L}orentzian
  pulses},}\ }\href@noop {} {\bibfield  {journal} {\bibinfo  {journal} {Physics
  of Plasmas}\ }\textbf {\bibinfo {volume} {24}},\ \bibinfo {pages} {020704}
  (\bibinfo {year} {2017}{\natexlab{b}})}\BibitemShut {NoStop}%
\bibitem [{\citenamefont {Garcia}\ and\ \citenamefont
  {Theodorsen}(2018{\natexlab{a}})}]{garcia2018l}%
  \BibitemOpen
  \bibfield  {author} {\bibinfo {author} {\bibfnamefont {O.~E.}\ \bibnamefont
  {Garcia}}\ and\ \bibinfo {author} {\bibfnamefont {A.}~\bibnamefont
  {Theodorsen}},\ }\bibfield  {title} {\enquote {\bibinfo {title} {Skewed
  {L}orentzian pulses and exponential frequency power spectra},}\ }\href@noop
  {} {\bibfield  {journal} {\bibinfo  {journal} {Physics of Plasmas}\ }\textbf
  {\bibinfo {volume} {25}},\ \bibinfo {pages} {014503} (\bibinfo {year}
  {2018}{\natexlab{a}})}\BibitemShut {NoStop}%
\bibitem [{\citenamefont {Garcia}\ and\ \citenamefont
  {Theodorsen}(2018{\natexlab{b}})}]{garcia2018l2}%
  \BibitemOpen
  \bibfield  {author} {\bibinfo {author} {\bibfnamefont {O.~E.}\ \bibnamefont
  {Garcia}}\ and\ \bibinfo {author} {\bibfnamefont {A.}~\bibnamefont
  {Theodorsen}},\ }\bibfield  {title} {\enquote {\bibinfo {title} {Intermittent
  fluctuations due to uncorrelated {L}orentzian pulses},}\ }\href@noop {}
  {\bibfield  {journal} {\bibinfo  {journal} {Physics of Plasmas}\ }\textbf
  {\bibinfo {volume} {25}},\ \bibinfo {pages} {014506} (\bibinfo {year}
  {2018}{\natexlab{b}})}\BibitemShut {NoStop}%
\bibitem [{\citenamefont {Decristoforo}\ \emph {et~al.}(2020)\citenamefont
  {Decristoforo}, \citenamefont {Militello}, \citenamefont {Nicholas},
  \citenamefont {Omotani}, \citenamefont {Marsden}, \citenamefont {Walkden},\
  and\ \citenamefont {Garcia}}]{decristoforo2020blob}%
  \BibitemOpen
  \bibfield  {author} {\bibinfo {author} {\bibfnamefont {G.}~\bibnamefont
  {Decristoforo}}, \bibinfo {author} {\bibfnamefont {F.}~\bibnamefont
  {Militello}}, \bibinfo {author} {\bibfnamefont {T.}~\bibnamefont {Nicholas}},
  \bibinfo {author} {\bibfnamefont {J.}~\bibnamefont {Omotani}}, \bibinfo
  {author} {\bibfnamefont {C.}~\bibnamefont {Marsden}}, \bibinfo {author}
  {\bibfnamefont {N.}~\bibnamefont {Walkden}}, \ and\ \bibinfo {author}
  {\bibfnamefont {O.~E.}\ \bibnamefont {Garcia}},\ }\bibfield  {title}
  {\enquote {\bibinfo {title} {Blob interactions in {2D} scrape-off layer
  simulations},}\ }\href@noop {} {\bibfield  {journal} {\bibinfo  {journal}
  {Physics of Plasmas}\ }\textbf {\bibinfo {volume} {27}},\ \bibinfo {pages}
  {122301} (\bibinfo {year} {2020})}\BibitemShut {NoStop}%
\bibitem [{\citenamefont {Halpern}\ \emph {et~al.}(2016)\citenamefont
  {Halpern}, \citenamefont {Ricci}, \citenamefont {Jolliet}, \citenamefont
  {Loizu}, \citenamefont {Morales}, \citenamefont {Mosetto}, \citenamefont
  {Musil}, \citenamefont {Riva}, \citenamefont {Tran},\ and\ \citenamefont
  {Wersal}}]{halpern}%
  \BibitemOpen
  \bibfield  {author} {\bibinfo {author} {\bibfnamefont {F.~D.}\ \bibnamefont
  {Halpern}}, \bibinfo {author} {\bibfnamefont {P.}~\bibnamefont {Ricci}},
  \bibinfo {author} {\bibfnamefont {S.}~\bibnamefont {Jolliet}}, \bibinfo
  {author} {\bibfnamefont {J.}~\bibnamefont {Loizu}}, \bibinfo {author}
  {\bibfnamefont {J.}~\bibnamefont {Morales}}, \bibinfo {author} {\bibfnamefont
  {A.}~\bibnamefont {Mosetto}}, \bibinfo {author} {\bibfnamefont
  {F.}~\bibnamefont {Musil}}, \bibinfo {author} {\bibfnamefont
  {F.}~\bibnamefont {Riva}}, \bibinfo {author} {\bibfnamefont {T.~M.}\
  \bibnamefont {Tran}}, \ and\ \bibinfo {author} {\bibfnamefont
  {C.}~\bibnamefont {Wersal}},\ }\bibfield  {title} {\enquote {\bibinfo {title}
  {The {GBS} code for tokamak scrape-off layer simulations},}\ }\href@noop {}
  {\bibfield  {journal} {\bibinfo  {journal} {Journal of Computational
  Physics}\ }\textbf {\bibinfo {volume} {315}},\ \bibinfo {pages} {388}
  (\bibinfo {year} {2016})}\BibitemShut {NoStop}%
\bibitem [{\citenamefont {Tamain}\ \emph {et~al.}(2016)\citenamefont {Tamain},
  \citenamefont {Bufferand}, \citenamefont {Ciraoloa}, \citenamefont {Colin},
  \citenamefont {Galassi}, \citenamefont {Ghendrih},\ and\ \citenamefont
  {Schwander}}]{tamain}%
  \BibitemOpen
  \bibfield  {author} {\bibinfo {author} {\bibfnamefont {P.}~\bibnamefont
  {Tamain}}, \bibinfo {author} {\bibfnamefont {H.}~\bibnamefont {Bufferand}},
  \bibinfo {author} {\bibfnamefont {G.}~\bibnamefont {Ciraoloa}}, \bibinfo
  {author} {\bibfnamefont {C.}~\bibnamefont {Colin}}, \bibinfo {author}
  {\bibfnamefont {D.}~\bibnamefont {Galassi}}, \bibinfo {author} {\bibfnamefont
  {P.}~\bibnamefont {Ghendrih}}, \ and\ \bibinfo {author} {\bibfnamefont
  {F.}~\bibnamefont {Schwander}},\ }\bibfield  {title} {\enquote {\bibinfo
  {title} {The {TOKAM3X} code for edge turbulence fluid simulations of tokamak
  plasmas in versatile magnetic geometries},}\ }\href@noop {} {\bibfield
  {journal} {\bibinfo  {journal} {Journal of Computational Physics}\ }\textbf
  {\bibinfo {volume} {321}},\ \bibinfo {pages} {606} (\bibinfo {year}
  {2016})}\BibitemShut {NoStop}%
\bibitem [{\citenamefont {Dudson}\ and\ \citenamefont {Leddy}(2017)}]{dudson}%
  \BibitemOpen
  \bibfield  {author} {\bibinfo {author} {\bibfnamefont {B.~D.}\ \bibnamefont
  {Dudson}}\ and\ \bibinfo {author} {\bibfnamefont {J.}~\bibnamefont {Leddy}},\
  }\bibfield  {title} {\enquote {\bibinfo {title} {Hermes: global plasma edge
  fluid turbulence simulations},}\ }\href@noop {} {\bibfield  {journal}
  {\bibinfo  {journal} {Plasma Physics and Controlled Fusion}\ }\textbf
  {\bibinfo {volume} {59}},\ \bibinfo {pages} {054010} (\bibinfo {year}
  {2017})}\BibitemShut {NoStop}%
\bibitem [{\citenamefont {Held}\ \emph {et~al.}(2018)\citenamefont {Held},
  \citenamefont {Wiesenberger}, \citenamefont {Kube},\ and\ \citenamefont
  {Kendl}}]{held}%
  \BibitemOpen
  \bibfield  {author} {\bibinfo {author} {\bibfnamefont {M.}~\bibnamefont
  {Held}}, \bibinfo {author} {\bibfnamefont {M.}~\bibnamefont {Wiesenberger}},
  \bibinfo {author} {\bibfnamefont {R.}~\bibnamefont {Kube}}, \ and\ \bibinfo
  {author} {\bibfnamefont {A.}~\bibnamefont {Kendl}},\ }\bibfield  {title}
  {\enquote {\bibinfo {title} {Non-{O}berbeck–{B}oussinesq zonal flow
  generation},}\ }\href@noop {} {\bibfield  {journal} {\bibinfo  {journal}
  {Nuclear Fusion}\ }\textbf {\bibinfo {volume} {58}},\ \bibinfo {pages}
  {104001} (\bibinfo {year} {2018})}\BibitemShut {NoStop}%
\bibitem [{\citenamefont {Wiesenberger}\ \emph {et~al.}(2019)\citenamefont
  {Wiesenberger}, \citenamefont {Einkemmer}, \citenamefont {Held},
  \citenamefont {Gutierrez-Milla}, \citenamefont {S{\'a}ez},\ and\
  \citenamefont {Iakymchuk}}]{wiesenberger}%
  \BibitemOpen
  \bibfield  {author} {\bibinfo {author} {\bibfnamefont {M.}~\bibnamefont
  {Wiesenberger}}, \bibinfo {author} {\bibfnamefont {L.}~\bibnamefont
  {Einkemmer}}, \bibinfo {author} {\bibfnamefont {M.}~\bibnamefont {Held}},
  \bibinfo {author} {\bibfnamefont {A.}~\bibnamefont {Gutierrez-Milla}},
  \bibinfo {author} {\bibfnamefont {X.}~\bibnamefont {S{\'a}ez}}, \ and\
  \bibinfo {author} {\bibfnamefont {R.}~\bibnamefont {Iakymchuk}},\ }\bibfield
  {title} {\enquote {\bibinfo {title} {Reproducibility, accuracy and
  performance of the {FELTOR} code and library on parallel computer
  architectures},}\ }\href@noop {} {\bibfield  {journal} {\bibinfo  {journal}
  {Computer Physics Communications}\ }\textbf {\bibinfo {volume} {238}},\
  \bibinfo {pages} {145} (\bibinfo {year} {2019})}\BibitemShut {NoStop}%
\bibitem [{\citenamefont {Stegmeir}\ \emph {et~al.}(2019)\citenamefont
  {Stegmeir}, \citenamefont {Ross}, \citenamefont {Body}, \citenamefont
  {Francisquez}, \citenamefont {Zholobenko}, \citenamefont {Coster},
  \citenamefont {Jenko}, \citenamefont {Rogers},\ and\ \citenamefont
  {Kang}}]{stegmeir}%
  \BibitemOpen
  \bibfield  {author} {\bibinfo {author} {\bibfnamefont {A.}~\bibnamefont
  {Stegmeir}}, \bibinfo {author} {\bibfnamefont {A.}~\bibnamefont {Ross}},
  \bibinfo {author} {\bibfnamefont {T.}~\bibnamefont {Body}}, \bibinfo {author}
  {\bibfnamefont {M.}~\bibnamefont {Francisquez}}, \bibinfo {author}
  {\bibfnamefont {W.}~\bibnamefont {Zholobenko}}, \bibinfo {author}
  {\bibfnamefont {D.}~\bibnamefont {Coster}}, \bibinfo {author} {\bibfnamefont
  {F.}~\bibnamefont {Jenko}}, \bibinfo {author} {\bibfnamefont {B.~N.}\
  \bibnamefont {Rogers}}, \ and\ \bibinfo {author} {\bibfnamefont {K.~S.}\
  \bibnamefont {Kang}},\ }\bibfield  {title} {\enquote {\bibinfo {title}
  {Global turbulence simulations of the tokamak edge region with {GRILLIX}},}\
  }\href@noop {} {\bibfield  {journal} {\bibinfo  {journal} {Physics of
  Plasmas}\ }\textbf {\bibinfo {volume} {26}},\ \bibinfo {pages} {052517}
  (\bibinfo {year} {2019})}\BibitemShut {NoStop}%
\bibitem [{\citenamefont {Zhang}\ \emph {et~al.}(2019)\citenamefont {Zhang},
  \citenamefont {Chen}, \citenamefont {Xu},\ and\ \citenamefont {Xia}}]{zhang}%
  \BibitemOpen
  \bibfield  {author} {\bibinfo {author} {\bibfnamefont {D.~R.}\ \bibnamefont
  {Zhang}}, \bibinfo {author} {\bibfnamefont {Y.~P.}\ \bibnamefont {Chen}},
  \bibinfo {author} {\bibfnamefont {X.~Q.}\ \bibnamefont {Xu}}, \ and\ \bibinfo
  {author} {\bibfnamefont {T.~Y.}\ \bibnamefont {Xia}},\ }\bibfield  {title}
  {\enquote {\bibinfo {title} {Self-consistent simulation of transport and
  turbulence in tokamak edge plasma by coupling {SOLPS-ITER} and {BOUT++}},}\
  }\href@noop {} {\bibfield  {journal} {\bibinfo  {journal} {Physics of
  Plasmas}\ }\textbf {\bibinfo {volume} {26}},\ \bibinfo {pages} {012508}
  (\bibinfo {year} {2019})}\BibitemShut {NoStop}%
\bibitem [{\citenamefont {Russell}, \citenamefont {Myra},\ and\ \citenamefont
  {Stotler}(2019)}]{rms}%
  \BibitemOpen
  \bibfield  {author} {\bibinfo {author} {\bibfnamefont {D.~A.}\ \bibnamefont
  {Russell}}, \bibinfo {author} {\bibfnamefont {J.~R.}\ \bibnamefont {Myra}}, \
  and\ \bibinfo {author} {\bibfnamefont {D.~P.}\ \bibnamefont {Stotler}},\
  }\bibfield  {title} {\enquote {\bibinfo {title} {A reduced model of
  neutral-plasma interactions in the edge and scrape-off-layer: {V}erification
  comparisons with kinetic {M}onte {C}arlo simulations},}\ }\href@noop {}
  {\bibfield  {journal} {\bibinfo  {journal} {Physics of Plasmas}\ }\textbf
  {\bibinfo {volume} {26}},\ \bibinfo {pages} {022304} (\bibinfo {year}
  {2019})}\BibitemShut {NoStop}%
\bibitem [{\citenamefont {Giacomin}\ and\ \citenamefont
  {Ricci}(2020)}]{giacomin}%
  \BibitemOpen
  \bibfield  {author} {\bibinfo {author} {\bibfnamefont {M.}~\bibnamefont
  {Giacomin}}\ and\ \bibinfo {author} {\bibfnamefont {P.}~\bibnamefont
  {Ricci}},\ }\bibfield  {title} {\enquote {\bibinfo {title} {Investigation of
  turbulent transport regimes in the tokamak edge by using two-fluid
  simulations},}\ }\href@noop {} {\bibfield  {journal} {\bibinfo  {journal}
  {Journal of Plasma Physics}\ }\textbf {\bibinfo {volume} {86}},\ \bibinfo
  {pages} {905860502} (\bibinfo {year} {2020})}\BibitemShut {NoStop}%
\end{thebibliography}%

\end{document}